# Microscopic Origin of Regeneration Noise in Relaxation Oscillator and its Macroscopic Circuit Manifestation

Y. Ng, B. Leung, M. Kononenko, S. Safavi-Naeini
ECE Department, University of Waterloo, Waterloo, Ontario, Canada

***Abstract*— This paper augments the existing macroscopic circuit noise model for phase noise in relaxation oscillators by showing the microscopic origins of the noise and explains temperature dependency. The noise arises from fluctuation of the energy accompanying the excess carriers on device (transistors) capacitors in the oscillator. Such fluctuation has its physical origin from the noise of such carriers, which, microscopically, are distributed across the energy levels (Fermi-Dirac). Furthermore this energy can be interpreted, circuit-wise, such that its gradient, with respect to circuit state variables, correspond to time evolution of current and voltage i.e. the oscillator dynamics. Three methods: potential energy based (macroscopic), free energy based (microscopic), Langevin equation based, are used to develop the noise model. The model's temperature variation over range of 77K-300K was compared to measured results on oscillators fabricated in $\mathbf{0.13}\ \boldsymbol{\mu m}$ CMOS technology. The trend agree reasonably well, where above a crossover temperature, the phase noise is a monotonic increasing function of temperature, while below the crossover temperature, the phase noise stays relatively constant and an explanation based on Langevin equation, extended to quantum regime, is offered.**

## I. Introduction

CIRCUITS exhibiting metastability and regeneration, such as Schmitt triggers, are important components in many electronic circuits like relaxation oscillator, comparators, sense amplifier and flip-flops [1]-[6]. Among them, relaxation oscillator [1], already used in traditional applications such as phase locked loop for clock recovery, recently also find applications in time to digital converters (TDCs) [7], [8], [9]. Reference [1] developed a noise model of relaxation oscillator, focusing on the noise spike arising from regeneration. This allowed the oscillator to operate at high frequency (eq. 37 in [10]) where the loop gain $g_m R$ approached unity, when noise spike appears. For TDC [7], this increases the oversampling frequency, hence resolution.

TABLE 1 A COMPARISON OF THE NOISE PROBLEMS TO BE SOLVED BY THE NOISE MODELS AMONG DIFFERENT WORK

| Relation among different works | | |
|---|---|---|
| Noise Problems trying to tackle | Description of method | |
| No noise spike | Switching model [11], ISF [12], frequency domain [13] | |
| Noise spike at room temperature (300 $K$) | Mapped thermodynamics system [1], large deviation [15] | |
| Noise spike variation as Temp varies 300 $K$ to 77 $K$ | This work | Method 1: potential energy based Section II, applicable in high temperature range (macroscopic) |
| | | Method 2: free energy based Section III, applicable in high temperature range, with temperature dependency consistent with Method 1 (microscopic) |
| | | Method 3: Langevin equation based), applicable in high temperature (trend consistent with Methods 1 and 2) and low temperature range Section IV |

Past noise models [8], [11], [12], [13] (see Table 1) do not include any noise spike in the noise process calculation. While [1] does include noise spike, it focused the investigation predominantly to room temperature. Given some of the recent applications in TDCs [9], [14], which involve a larger temperature range, the temperature dependency merits further investigation. As [1]'s temperature dependency arises from a model that is based on mapping the nonlinear dynamics equation of motion of the circuit to the corresponding equation of state in a thermodynamics system, as far as temperature dependency is concerned, it is natural to look at the thermodynamics systems of the circuit itself, rather than the mapped thermodynamics system.

As for the study of the circuit's thermodynamics property, it is most directly related to the circuit's energy, rather than its individual node voltage and branch current. Recently the use of energy centric methodologies towards study circuit property has been attempted [8], [16]. Similarly, this work tries to look at the circuit from an energy viewpoint.

Three methods, employing energy-based approach, summarized in Table 2, are proposed, which extends the predictive power of the model to larger range of operating condition.

In the first method, rather than identifying the potential $E$ as gradient of current flow in the circuit, as in [1], it is now identified as the potential energy of the circuit, specifically that of electrostatic energy stored in gate to source capacitance $C_{gs}$ of the two cross coupled transistors. Consequently, instead of plotting $E$ vs normalized current $z$, potential energy $E$ is plotted against normalized differential current, now interpreted as order parameter. Subsequent development of noise model follows [1], via thermodynamic argument, but with the advantage of the present approach in having a more intuitive explanation of $E$ as the energy. The concept of temperature is introduced into the noise model by adding thermal fluctuation to the current charging $C_{gs}$ (arising from current noise in the resistor $R$, having current noise power spectral density (PSD) proportional to $\frac{4kT}{R}$), resulting in fluctuation along the potential energy curve.

Meanwhile even though this sets the stage of looking at $E$ as energy, having the temperature introduced as a macroscopic concept via thermodynamic argument, does not capture microscopic details e.g. the microscopic behavior of excess charges on $C_{gs}$ of the two cross coupled transistors (such as fluctuation in occupancy of energy levels, arising from thermal energy of excess charges, acquired in $R$, before they flow from resistor to $C_{gs}$) is ignored. The statistical mechanics (microscopic) nature in the fluctuation of the excess charges ensemble is not called on to clarify the thermodynamics (macroscopic) fluctuation of current charging $C_{gs}$, and hence resulting fluctuation in $E$. Therefore, the next method (Method 2) is developed, in which we identify $E$ as free energy and address this limitation. Following the principle of statistical mechanics, we start by developing the Hamiltonian of the ensemble, with interaction energy $J$, as sum over ensemble's occupancy over energy levels. This is used to calculate partition function, and then free energy, which captures the temperature effect. The noise fluctuation so derived agrees with method 1 and depends on $kT$ as well. Note that, with microscopic details, method 2 explicitly calculates the expression of free energy and hence noise, whereas method 1 adds thermal noise to the system with potential energy $E$ and then uses the mapped thermodynamic system, via thermodynamic argument (essentially in a roundabout way), to obtain noise.

As temperature decreases, past publications [14], [17], [18] on the behavior of circuit noise have explained its temperature dependence. Thermal noise sources dependency on $kT$ is one factor. In addition, specific to particular circuits considered, control circuit for LC oscillator based PLL's temperature variation [17], $Q$ factor of LC oscillator's temperature dependency is also offered [18] as other explanations. This paper attempts to offer another explanation, specific to the relaxation oscillator circuit under consideration: the fast regeneration of the relaxation oscillator. This explanation, being specific to relaxation oscillator (since our circuit has no control circuit, nor LC tank with the corresponding $Q$ factor, while in [17] and [18] the LC oscillator does not experience regeneration), is attempt on explaining low temperature noise dependency in relaxation oscillator arising from quantum noise. Thus, Method 2 is hard to apply, as free energy formulation in quantum mechanics for relaxation oscillator is difficult. A full description using free energy is not necessary for noise calculation and so we propose Method 3.

In the third method, we revert to Method 1, where noise is added to the deterministic part of circuit, represented in a macroscopic manner, but now with the noise source generalized so that the expression includes both thermal and quantum fluctuation, then subsequent application of Method 1 is complicated[1].

Instead we use Langevin dynamics [19], [20], which represents the deterministic part of circuit in a macroscopic manner, and also have noise source includes both thermal and quantum fluctuation, but this time, find the equation of motion of the deterministic part by linearizing circuit equation in Method 1, and obtain the noise by calculating how noise source modulates this deterministic trajectory. The result approaches that of Methods 1 and 2 when quantum fluctuations are small compared to thermal fluctuations (e.g. at room temperature), but at low temperature, Method 3 accounts for quantum fluctuation, and extends the range of the model.

[1] It should be noted, whereas in Method 1, with thermal noise alone present in the noise source, thermodynamic argument leads to a closed form solution (eq. 43 and 49 of [1]), here having also quantum noise present, coefficient in equation of state, namely a, b in eq. 42 of [1], no longer has the simple form following the equations,(it now has dependency on quantum effect, see pg. 282, eq. 27 of [19]) and the subsequent derivation is complicated.

The organization of the paper is as follows. Section II reviews previous model and Method 1. Section III covers Method 2. Section IV covers method3. Experimental results are presented in section V and compared with model. The low temperature (semi-microscopic) parameters (e.g. $g_m$) are obtained from Eldo simulation and fit into model. Conclusions are drawn in Section VI.

## II. Method 1: Potential Energy Based

### A. Previous Model: 'E', its Gradient as Current and Voltage

Figure 1 shows the application of a Schmitt trigger in the design of a relaxation oscillator. [1][2] The Schmitt trigger governs the regeneration process in the switching of states for a relaxation oscillator. All noise in the circuit is lumped and represented by a single noise current source $I_n$. The noise spike from this regeneration process is the focus of this paper, and comparison with previous noise paper on relaxation oscillator is given in Table 1. For simplification, $R_1 = R_2 = R$ and $k_1 = k_2 = k$. Previously, [1] ignored $C_{gs2}$ and encapsulated the parasitic effects of $C_{gs1}$ as $\varepsilon = \frac{RC_{gs1}}{I_0}$. During the regeneration process, the Schmitt trigger is approximated by a first order differential equation in normalized current, $z = \frac{2i_{d1} - I_0}{I_0}$ as shown in (1).

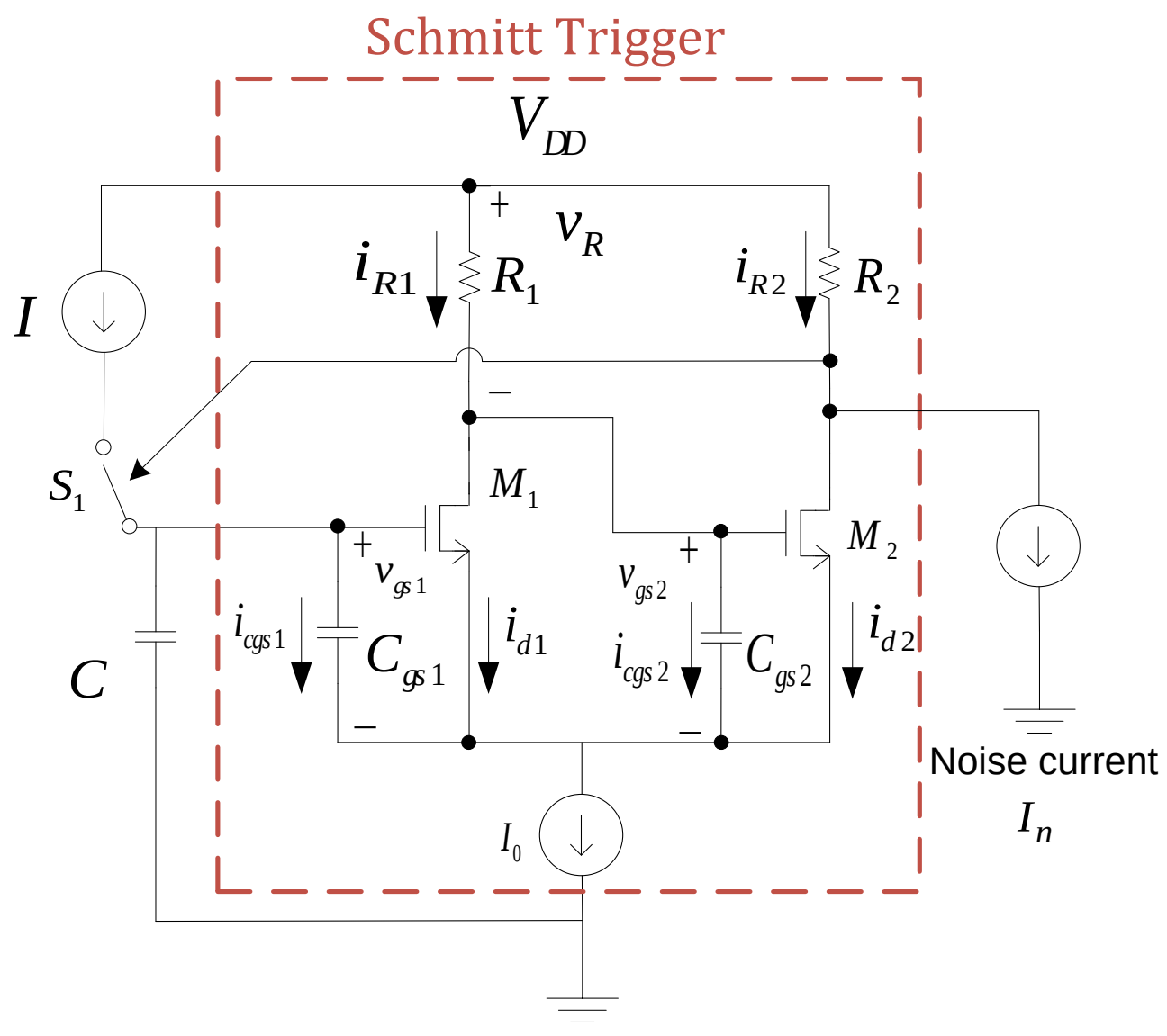

Figure 1. Ground Capacitor Relaxation Oscillator Model. $S_1$ is a current-controlled switch that flips to fill $C_{gs1}$ when $M_2$ is on.

$$\frac{dz}{dt} = \frac{1}{\varepsilon}\left(\left(\sqrt{2}(1 - g_m R)\right) z + \frac{1}{8} z^3\right) \tag{1}$$

Reference [1] assumed (1) describes a first order system with a potential function [21], (denoted in [1] as $E$) such that:

$$\frac{dz}{dt} = \frac{dE}{dz} \tag{2}$$

By substituting (2) into (1), and integrating with respect to $z$, a first order differential equation (ODE) in $E$, with $z$, rather $t$, as variable is obtained.

$$E(z) = \frac{\sqrt{2}}{\varepsilon}\left(\frac{1}{2}(1 - g_m R) z^2 + \frac{1}{32} z^4\right) \tag{3}$$

[2] Reference [19] elaborates on the equivalence to a floating capacitor based relaxation oscillator, the one described in [1].

TABLE 2 RELATION BETWEEN DIFFERENT APPROACHES ON CALCULATING NOISE SPIKE OF RELAXATION OSCILLATOR FROM THE UNDERLYING BISTABLE CIRCUIT/SCHMITT TRIGGER

| Method and relation among methods | Conditions where Method is applicable | Limitations of methods |
|---|---|---|
| Method 1 (macroscopic):<br>• Develop equation of motion and find the energy of which the equation of motion is the gradient.<br>• Add noise current to circuit originating from a single source<br>• Map circuit to equivalent thermodynamic system such as a van der Waals gas or Ising Magnet<br>• Obtain fluctuation and map back. | • Higher Temperature Range<br>• When Equivalent Thermodynamic System Exists | • Candidate thermodynamic system may not exist |
| Method 2 (microscopic)<br>• Find interaction energy between charges in the ensemble<br>• Find the Hamiltonian of the system<br>• Obtain the partition function<br>• Find free energy of charges from partition function<br>• Obtain fluctuation from partition function<br><br>Relation to Method 1:<br>• Noise expression consistent with Method 1, but shows microscopic origin of temperature dependency<br>• Does not require equivalent thermodynamic system, unlike Method 1 | • Higher Temperature Range | • Free energy not easy to calculate<br>• Free energy particularly hard to calculate if Hamiltonian has to be calculated including quantum mechanics, e.g. if quantum fluctuation has to be included |
| Method 3 (macroscopic)<br>• Formulate deterministic part of Langevin equation about metastable point for circuit by using an approximate linear circuit and take Fourier transform<br>• Find correlation function for noise current source in frequency domain<br>• Multiply noise source and deterministic part to get energy fluctuation and then relate to current fluctuation.<br>Relation to Method 1:<br>• Properly linearize Method 1 in in deterministic part calculation so that noise expression approaches that at high temperature limit, while in noise source incorporate quantum noise source<br>Relation to Method 2:<br>• Noise expressions are consistent at high-temperature limit | • When noise source consists of one degree of freedom and one can linearize the deterministic part of a system | • Langevin equation is limited to solving the energy fluctuation, and not the other thermodynamic properties.<br>• Works only on linear or linearizable deterministic systems (unlike method 1: nonlinear, particularly any bistable system with thermodynamic phase change). |

Reference [1] identifies the first order ODE, as a gradient system and interprets $E$ as "potential", in the sense of its gradient $\frac{dE}{dz}$ giving the time evolution of current $\frac{dz}{dt}$, which gives the time evolution of the normalized currents and voltages. The particular form of (3) ensures that when $g_m R > 1$, the gradient has one metastable state between two stable states. At the metastable state, the circuit regenerates into one of the two stable states. If we turn on the noise by adding $I_n$ into (1), the resulting fluctuation maps into the fluctuation of a corresponding thermodynamic system. This system is the van der Waals gas. It then calculates the fluctuation, and maps back to noise in the relaxation oscillator. As expected, the fluctuation has a noise spike just like the van der Waals gas. This is given as first entry in Table 2.

However, the temperature dependency in the fluctuation formula is obtained via mapping and does not arise from physical first principles. To improve understanding of this dependency, it is therefore natural to identify the physical meaning of $E$, (with dimension Joule), in the form of some sort of energy.

## B. The Physical Interpretation of $E$ as Potential Energy in the Schmitt Trigger

Method 1 gives a first attempt to identify the physics of $E$ as energy, while leaving the subsequent mapping to thermodynamic system and fluctuation as has been presented in [1], with the view that the fluctuation expression so obtained remains consistent with [1]**.** As will be shown, this will lead to Method 2, which further establishes the energy connotation of $E$.

Towards this goal, we start with (3) and obtain the physical interpretation of $E$ by analyzing the dynamics of the Schmitt trigger using circuit theory as shown in [21]. Following [22] and [23], we transform the coordinate, such that in this transformed coordinate, the resulting system, $E$ has physical meaning as energy, and it behaves as a 2-level system (with stable and metastable states, as in Figure 1, see also Appendix A).

Equation (4) comes from applying Kirchhoff's Current Law (KCL) at the drain nodes of both $M_1$ and $M_2$.

$$i_{cgs1} = i_{R2} - i_{d2} \quad i_{cgs2} = i_{R1} - i_{d1} \tag{4}$$

Applying Kirchhoff's Voltage Law (KVL) around the resistor loop $R_1$, $R_2$ and transistors $M_1$, $M_2$, and setting $R_1 = R_2 = R$ yields (5).

$$i_{R1} - i_{R2} = \frac{v_{gs2} - v_{gs1}}{R} \tag{5}$$

Due to the tail current source $I_0$, the common mode current is $\frac{I_0}{2}$. Equations (6) and (7) give the currents $i_{R1}$ and $i_{R2}$ with respect to the common mode current.

$$i_{R1} = \frac{I_0}{2} + \frac{v_{gs2} - v_{gs1}}{2R} \tag{6}$$

$$i_{R2} = \frac{I_0}{2} - \frac{v_{gs2} - v_{gs1}}{2R} \tag{7}$$

Equation (8) relates the capacitor currents $i_{Cgs1}$ and $i_{Cgs2}$ to the capacitor voltages $v_{gs1}$ and $v_{gs2}$.

$$i_{cgs1} = C_{gs1}\frac{dv_{gs1}}{dt} \qquad i_{cgs2} = C_{gs2}\frac{dv_{gs2}}{dt} \tag{8}$$

Finally, (9) relates the drain currents $i_{d1}$ and $i_{d2}$ to $v_{gs1}$ and $v_{gs2}$ using the transistor device equation. Ignore the Early effect.

$$v_{gs1} = \sqrt{\frac{2i_{d1}}{k_n}} + v_t \quad v_{gs2} = \sqrt{\frac{2i_{d2}}{k_n}} + v_t \tag{9}$$

Combining (4)-(9), setting $C_{gs1} = C_{gs2} = C_{gs}$, and using $g_m$ yields (10). Equation (10) gives the time evolution of the differential current through the oscillator[3].

$$\frac{C_{gs}}{g_m}\frac{d(i_{d1}-i_{d2})}{dt} = \frac{1}{R}\left(\sqrt{\frac{2i_{d2}}{k_n}} - \sqrt{\frac{2i_{d1}}{k_n}}\right) + i_{d1} - i_{d2} \tag{10}$$

The normalized differential mode signal is (11). Use this as a transformed co-ordinate.

$$\Delta = \frac{i_{d\,1} - i_{d\,2}}{I_0} \tag{11}$$

Substituting (11) into (9) gives the time evolution of $\Delta$ as shown in (12).

$$\frac{C_{gs}}{g_m}\frac{d\Delta}{dt} = -\frac{1}{g_m R}\left(\sqrt{1+\Delta} - \sqrt{1-\Delta}\right) + \Delta \tag{12}$$

Expanding the square roots in (12) about $\Delta = 0$ using (13) yields (14); a first-order equation of motion with a cubic nonlinear term.

$$\sqrt{1+\Delta} - \sqrt{1-\Delta} = \Delta + \frac{1}{8}\Delta^3 + O(\Delta^5) \tag{13}$$

$$RC_{gs}\frac{d\Delta}{dt} = \frac{1}{2}(Rg_m - 1)\Delta - \frac{1}{8}\Delta^3 \tag{14}$$

The energy is the integral on the right-hand side (RHS) of (14), and upon scaling of $C_{gs}$ and $\frac{I_o}{g_m}$, has units of energy. Thus, in terms of transformed coordinate, $E$ does have dimension of energy. The physical interpretation of $E$ is the potential energy (electrostatic potential energy) of charges on two capacitors $C_{gs1}$ and $C_{gs2}$ and is indeed energy[3]. Upon normalization this is (15).

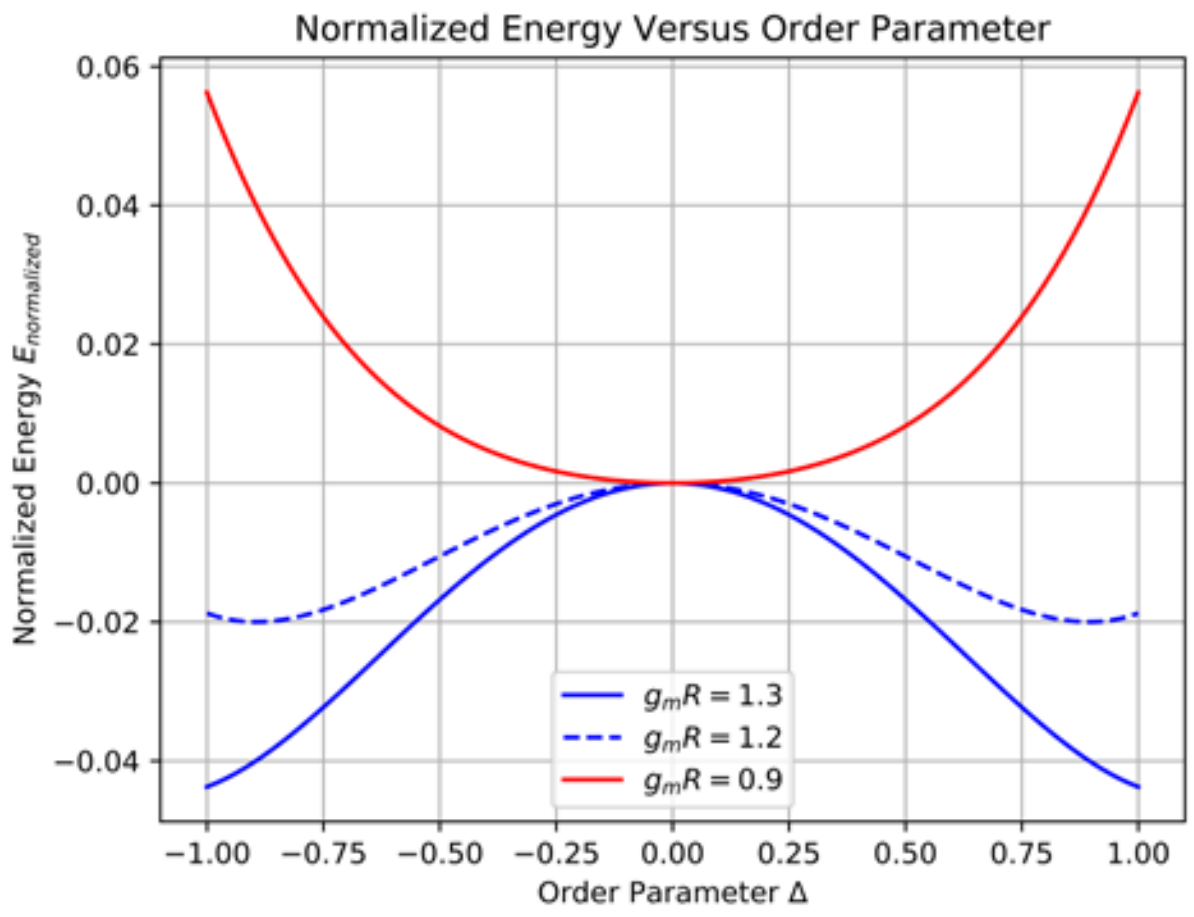

Figure 2. POTENTIAL ENERGY VS. ORDER PARAMETER FOR DIFFERENT LOOP GAINS

$$E_{normalized} \;=\; -\left(\frac{(Rg_m - 1)}{4}\Delta^2 - \frac{1}{32}\Delta^4\right) + Const \tag{15}$$

Figure 2 shows $E_{normalized}$ as a function of $\Delta$ for three different values of $g_mR$. When $g_mR > 1$, $E_{normalized}$ has two local minima separated by a local maximum at $\Delta = 0$. The two local minima are the stable states of a two-level system, and $\Delta = 0$ is a metastable state. The stable states correspond to the two stable states of the Schmitt Trigger in Figure 1. This plot is similar to Figure 7 of [1].

Subsequent development of noise model follows [1], via thermodynamic argument. Temperature is introduced into the noise model by turning on $I_n$. This adds thermal fluctuations to the current charging $C_{gs2}$ with current noise power spectral density proportional to $kT$. This results in fluctuation along the potential energy curve, which is also given by the noise formula in eq. 49 of [1]. The resulting fluctuation is no longer directly proportional to $kT$, but it is monotonic increasing with $T$.

Meanwhile, having the temperature introduced as a macroscopic concept via thermodynamic argument, does not capture microscopic details such as the microscopic behavior of excess charges on $C_{gs}$ of the two cross coupled transistors, such as fluctuation in occupancy of energy levels, arising from thermal energy of excess charge fluctuation in occupancy of energy levels, arising from thermal energy of excess charges.

On the other hand, connecting the circuit dynamics to a thermodynamic system like the van der Waals gas invites the use of Hamiltonian dynamics to describe the circuit. This applies particularly to microscopic details of the circuit such as the energy level distribution.

In Section II, we include microscopic effects by considering the different configurations in which these charges are distributed over energy levels, which is a function of temperature. Then we plot the resulting energy, now the free energy, which has temperature dependency (note $E$ has no temperature dependency). The two-level system is also present in the free energy plot. This is summarized in Table 2 Row 2.[4]

## III. METHOD 2: FREE ENERGY BASED

Method 2 is conceptually demanding since instead of the circuit (macroscopic) concepts involving more familiar variables such as current and voltage, it now involves device (microscopic) concepts, like energy levels. To simplify, we will only explain the concepts most connected to temperature dependency (like distribution of ensemble over energy levels) with microscopic details, while concepts with secondary connection to temperature dependency, (like transfer of excess carriers of the ensemble across cross coupled $M_1$-$M_2$, via loop gain $g_mR$) remains to be explained at the macroscopic level. Figure 3 summarizes the process of deriving the current

[3] Remember these charges are coupled via transistors $M_1$ and $M_2$, and so the energy includes the energy involves coupling.

[4] It should be noted that, for simplification, we focus on energy (microscopic aspect) arising from interaction part of the system, while leaving the description of other aspect (such as transport), using macroscopic (phenomenological) description (e.g. $g_m$)

fluctuation, what the goal of the section is, whether it involves a macroscopic or microscopic description of the circuit, and how the parts of the derivation are connected.

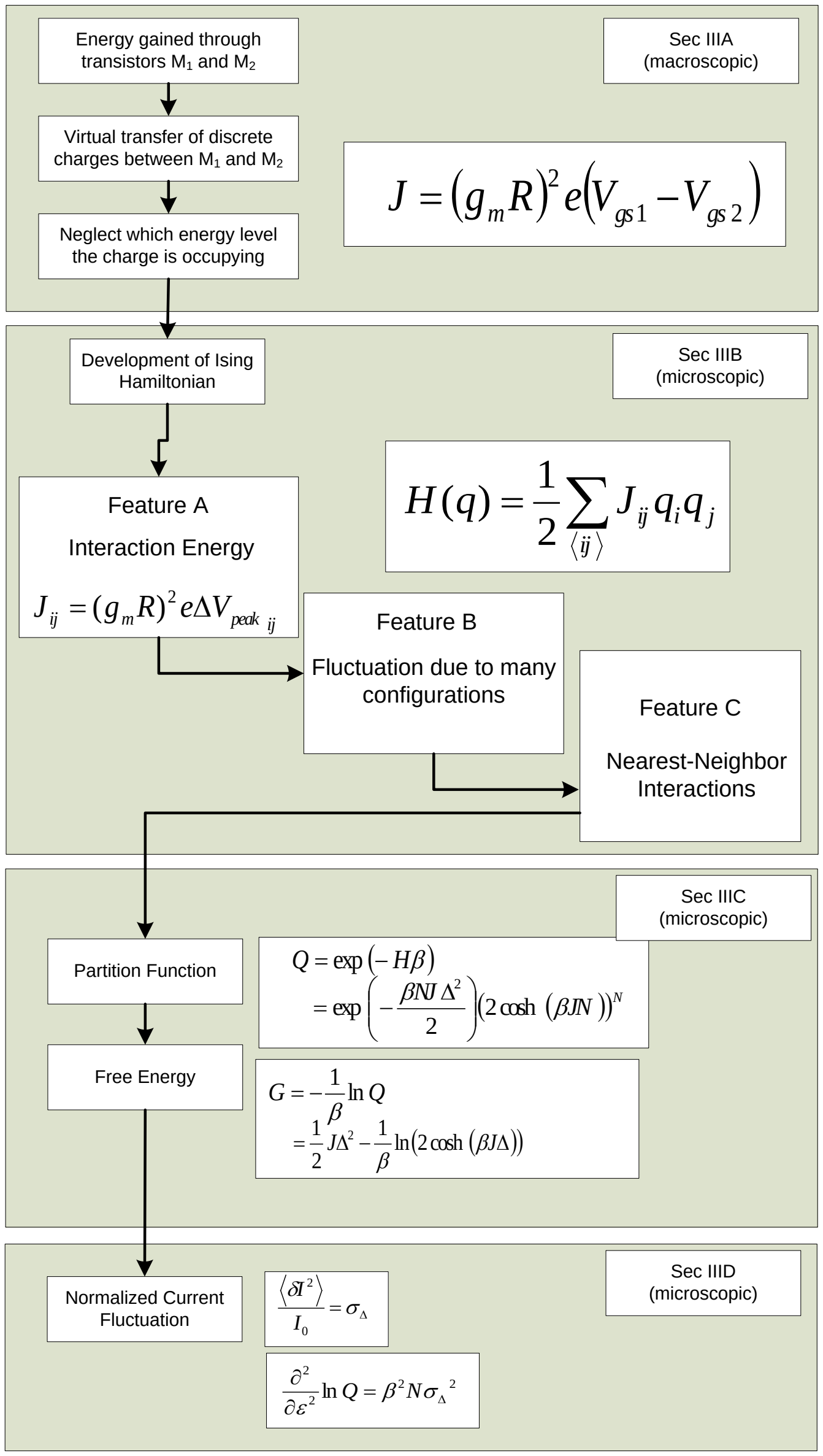

Figure 3. AN OUTLINE FOR DERIVING THE CURRENT FLUCTUATION USING METHOD 2

## A. Interaction Energy between Each Pair of Coupled Charges with Identical Energy

We start by looking at energy due to coupling between each pair of charges, assuming the pair is at the same energy level. Therefore occupancy is not an issue and temperature dependency is not explicit. There is already some microscopic aspect, because unlike Method 1, which deals with current (continuum of charge), the ensemble in Method 2 consists of discrete charges. Figure 1 is therefore redrawn to show this aspect. Specifically the Schmitt Trigger is redrawn in Figure 4a, where the charges (electrons in NMOS) in the inversion layer of the transistors and also $C_{gs}$ of the cross coupled transistors. We assume all the charge occupies same energy level. $M_{1\text{-}2}$ is redrawn as a symmetrical construction of transistor $M_1$, $M_2$ to highlight the cross coupling. In Figure 4a, the positive charges are represented by crosses, while the negative charges are represented by dashes.

Now, we can represent the current flowing through the channel in $M_1$ to be proportional to the number of charges in the channel, i.e. $i_1 = WQ_I v_d$ with $Q_1$ the total inversion layer charges per unit area. [5] Since $i_1 + i_2 = I_0$, the total number of charges in both channels is constant.

Having a net negative charge in the channel between drain and source of the transistor causes the same net positive charge to collect on the gate of the transistor. Since the electrons in the channel of both transistors are conserved due to the current summing into a constant tail current source, the total charges on top of both parasitic capacitors top plates are conserved. Equation (16) gives the conservation relation.

$$N = N_1 + N_2 \tag{16}$$

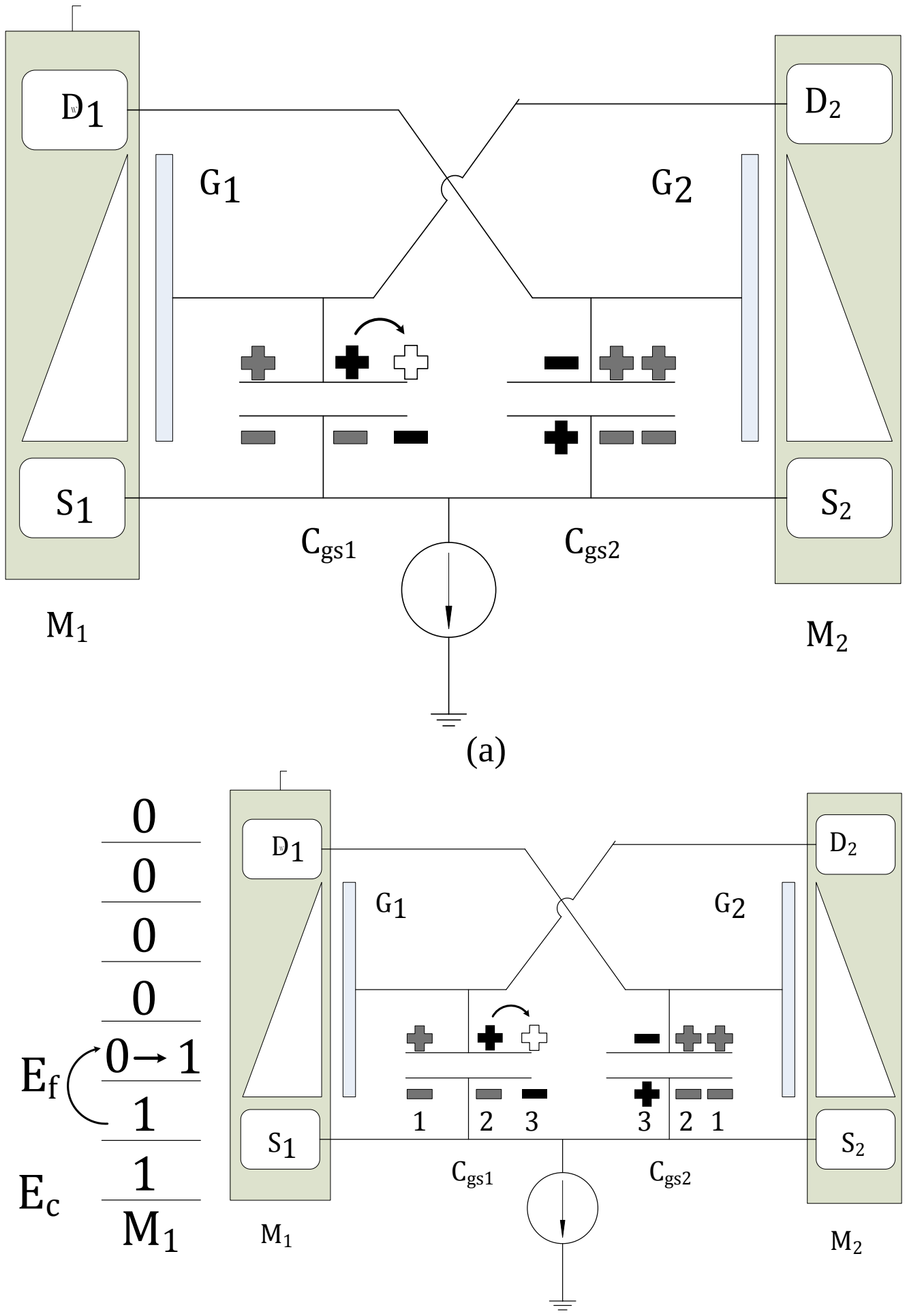

Figure 4. A MICROSCOPIC MODEL OF THE ENSEMBLE OF CHARGES ON THE SCHMITT TRIGGER. (A) SHOWS A SPATIAL DISTRIBUTION OF CHARGES OVER THE PARASTIIC CAPACITORS, AND (B) ADDS AN ENERGY LEVEL DISTRIBUTION INTO THE SPATIAL DISTRIBUTION

In equation (16), $N_1$ is the number of charges on the top plate of $C_{gs1}$ and $N_2$ is the number of charges on the top plate of $C_{gs2}$. Specifically, for our thermodynamic system, the Schmitt trigger, to identify the interaction energy, we find the change of energy by adding an electron in the channel of $M_1$ and removing a charge from the channel of $M_2$.

To calculate the interaction energy, which is due to transfer of charge between $M_1$ and $M_2$, consider the change of charges in transistor $M_1$ and $M_2$, involved in this transfer. Equation (17) gives the total number of excess charges on the top plate of $M_1$.

$$\begin{aligned} Q_1 &= C_{ox}(WL)V_{gs1} \\ &= C_{ox}(WL)V_{ds2} \end{aligned} \tag{17}$$

[5] Using this description Δ in (11) can be shown to be comparable to Δ in (42)

In equation (17), $C_{ox}$ is gate capacitance per unit area and $W$, $L$ are width and length of $M_1$, and $V_{gs1}$ is the gate-to-source voltage across transistor $M_2$. In the second line this is made the same as $V_{ds2}$, since the cross coupling connects the gate of $M_2$ to the drain of $M_1$. Assume that $\delta Q$ charges were transferred during regeneration. The resulting change in voltage is given by equation (18) where the resulting change is $\delta V_{gs}$.

$$\begin{aligned} \delta V_{gs1} &= \frac{1}{C_{ox}(WL)} \delta Q_1 \\ \delta Q_2 &= C_{ox}\,(WL)\,\delta V_{gs2} \end{aligned} \tag{18}$$

Conservation of charge[6] implies equation (19) and the resulting change in energy as given in equation (20).

$$\delta Q_1 - \delta Q_2 = 0 \tag{19}$$

$$\delta E = \delta V_{ds1} \delta Q_1 \tag{20}$$

The symmetric construction of the Schmitt trigger circuit allows $\delta V_{ds1}$ to be related to $\delta Q_2$ as shown in (21).

$$\begin{aligned} \delta V_{ds1} &= -(g_m R)^2 \delta V_{gs2} \\ &= (g_m R)^2 \frac{\delta Q_2}{C_{ox} WL} \end{aligned} \tag{21}$$

By substituting (21) to equation (20), the change in energy is given by equation (22).

$$\Delta E = \frac{(g_m R)^2 \delta Q_2 \delta Q_1}{C_{ox}(WL)} \tag{22}$$

Note in the Schmitt trigger circuit this change in energy, is coupled between the charges of the 2 ensembles, on two individual capacitors. The charges are

i) not physically transferred; rather one is delivered from ground via the current source, while an equal amount is removed to ground by the same current source

ii) the charge was originally provided through the resistor, or heat reservoir.

In the present case, considering one pair of charge, we have $\delta Q = e$, and this change of energy in the interaction operation, is interpreted as the interaction energy per pair. For introduction of an electron in transistor $M_1$ and removal of an electron in transistor$M_2$, $J$ is given as equation (23).

$$J = \frac{(g_m R)^2 e^2}{C_{ox}(WL)} \tag{23}$$

This change results in an interaction involving two neighboring electrons on $M_1$ ($V_{gs1}$), but coupled through electron on $M_2$ ($V_{gs2}$). The associated change is given by $e\left(V_{gs2} - V_{gs1}\right)$. Equivalently, starting with $e$ (on $M_1$), it goes through a voltage change of $V_{gs2} - V_{gs1}$ (via $M_2$) to get its next charge of, and so $\frac{e}{C_{ox} WL}$ can be identified with $V_{gs2} - V_{gs1}$. For simplicity we approximate this by a constant voltage, denoted as $\Delta V_{peak}$.[7] This leads to equation (24).

$$\begin{aligned} J &= (g_m R)^2 e (V_{gs1} - V_{gs2}) \\ &= (g_m R)^2 e V_{peak} \end{aligned} \tag{24}$$

[6] This is assuming a) $i_{tail} = i_{d1} + i_{d2} + i_{cgs1} + i_{cgs2} \sim= i_{d1} + i_{d2}$ (neglecting $i_{cgs1}, i_{cgs2}$ which equal $C\frac{dv_{gs1}}{dt}$, $C\frac{dv_{gs2}}{dt}$, as $i_{d1}, i_{d2}$ changes slowly at beginning of regeneration(from simulation), $C\frac{dv_{gs1}}{dt}$, $C\frac{dv_{gs2}}{dt} \sim 0$. Then $i_{d1}$ = -$i_{d2}$ or $\delta Q_1$ = - $\delta Q_2$, with the presumption that the $\delta Q_1$, $\delta Q_2$, happens slowly, so that it is consistent with $C\frac{dv_{gs1}}{dt}$, $C\frac{dv_{gs2}}{dt} \sim 0$

b) charge, being drain current id1, id2 NOT displacement current $i_{cgs1}$, $i_{cgs2}$, has charge travel through and experience potential $v_{ds}$(i.e. drain to source), which due to cross coupling equal $v_{gs}$ i.e. from current source, to source, to drain to gate of other side transistor, or top plate of capacitor

c) Meanwhile the other transistor, a charge is pulled from its substrate to the channel/inversion layer of M2.)

[7] $V_{peak}$ in [1] (eq. 52) is the peak voltage across capacitor C. At metastable point, voltage across capacitor, from symmetry, is given by $V_{gs2} - V_{gs1}$. We reuse the same symbol $V_{peak}$ here.

Hence, interaction energy per pair has been identified with the macroscopic parameters $g_m$, $R$[8], as shown in Figure 4a. Summing over all pairs should give us interaction energy. Note, in order to highlight the microscopic feature, we now interpret $(g_m R)^2$ in (23) as energy gain (microscopic feature) of each pair of electrons through coupling; rather than as voltage gain squared of $\delta V_{gs2}$ (macroscopic) through coupling, as in (21). However, this implicitly assumes every pair has same interaction energy, and ignores microscopic details such as individual charge occupying different energy level.

B. Interaction Including Distribution of Charges over Energy Levels

### 1. Microscopic Energy Level Distribution for a Single Transistor Incorporating Temperature Effects

We will next go into coupling incorporating different energy levels that each pair of charges occupies. In (24) we assumed every pair of electron goes through the same voltage change ($V_{gs2} - V_{gs1}$), illustrated using Figure 4a, when, in reality, electrons are distributed across energy levels. Figure 4a is redrawn in Figure 4b to highlight that. Following quantum mechanics, these energy levels are the eigenenergies of the cross coupled system.

Specifically in Figure 4b, a simple example is shown, with microscopic features. This includes conduction band energy $E_c$, the Fermi level $E_f$, the eigenenergies $\{E_1, E_2, \ldots, E_7\}$, and the occupancies. Figure 4b shows a change in occupancy where the number of electrons on $M_1$ increases from 2 to 3. Appendix B elaborates on the microscopic picture, where, starting from the simple single MOS transistor (no cross coupling i.e. no interaction), we begin with the more familiar E-x of MOS transistor, a representation is introduced, where E-k diagram at different x locations is also shown (essentially Figure 4b is like Figure 24d, but only $E_c$ is shown). Appendix B, also shows that with this E-x/E-k representation (microscopic), the macroscopic I-V characteristics can be recovered. This then allows us to retain easily the concept of $(g_m R)^2$ in (23), when moving to microscopic description. This representation is the usual way in texts that handle I-V characteristics in the microscopic details (e.g. fig.7.3 in [24], fig.1.7.2 in [25]). In such representation, at any $x$, the ensemble of mobile/excess charges on capacitor is shown both spatially (top plate of capacitor) and energetically (occupying energy levels).

Figure 4b shows the case when all the lowest energy levels are occupied (shown as 1). This corresponds to the case when temperature $T = 0$. Now summing over the pairs, as discussed after (24), will give us interaction energy (denoted as $E_{int}$, but this time we have taken distribution into consideration i.e. $\sum q_i q_j$ (if no distribution, but just sum, we will apply (23), which is $3[e(-e)]$, since each pair(from (23)) gives us $e(-e)$, and from Figure 4a, there are 3 electrons, and so 3 pairs(all have identical interaction energy).

When the temperature is above absolute zero, some of the electrons will be excited to a higher energy level, leaving some of the lower levels empty. The probability of a given energy level being occupied depends on $T$, and follows Fermi-Dirac statistics. Section III.B.4.b considers temperature effects in greater detail, discussing multiple configurations as well. It is necessary to sum over the product of all occupancies $q_i q_j$ because the interaction energy per pair described by $J_{ij}$ depends on $q_i$ and $q_j$. When the temperature is above absolute zero, there exist multiple configurations of the system for a given total energy, obtained by evaluating the Hamiltonian. The probability of the system being in a given configuration depends on the total energy of that configuration.

### 2. Effect of Coupling Transistors

Next, we discuss interaction due to coupling. First let us start with $T = 0$ i.e. Figure 4a. Repeating footnote 8, but this time with charges distribution over energy levels, charge is now further identified with energy level it is associated with. Thus, whereas with footnote 8, pertaining to Figure 4a, which starts with red positive charge on M1, now at Figure 4b, starting with this same charge, we need to identified with the energy level, which we assume starts at 2. Therefore we called the charge $q_{2_{M_1\,top}}(black)$. Because of cross-coupling between $M_1$

8 Physically, starting initially$q_{2_{M1\,\mathrm{top}}}$ $(black)$, via cross coupling, puts on $M_2$, $q_{3_{M_2\,top}}$ $(black)$. Because of inversion from transistor M1, positive charge, $q_{2_{M_1\,top}}$ $(black)$, results in negative charge, $q_{3_{M_2\,top}}$ $(black)$, on the top plate of $M_2$. This results in positive charge, $q_{3_{M_2\,bot}}$ $(black)$, at the bottom plate of M2. Because of charge conservation due to the constant current source, a negative charge $q_{3_{M_1\,bot}}$ $(black)$ is deposited on the bottom plate of $C_{gs1}$. Because of charge neutrality, negative charge, $q_{3_{M_1\,bot}}$ $(black)$, results in positive charge, $q_{3_{M_1\,top}}$ $(white)$, on top plate of $C_{gs1}$. Thus this highlights the coupling of two charges on top plate capacitor of M1, spatially (horizontal), actually arises from the feedback coupling across M1-2 via coupling of two charges on top plate capacitor of M1, energetically (vertical). Contrast this to standalone M1, where two charges on plates have no coupling (spatially nor energetically). Note, the feedback coupling between $q_{3_{M_2\,top}}$ $(black)$, $q_{3_{M_2\,bot}}$ $(black)$, and $q_{3_{M_1\,bot}}$ $(black)$, the interaction is $g_m R$. From symmetry we go from M2 to M1 and find the interaction also as $g_m R$. If we repeat by going through the opposite direction, where there is also a factor of $g_m R$, and this accounts for $(g_m R)^2$, or J part in (27).

and $M_2$, $q_{2_{M_1\ top}}\ (black)$, puts a positive charge on $M_2$. We now again need to identify the energy level. Because of Pauli Exclusion Principle, this needed to be put on the next level $E_3$, i.e. $q_{3_{M_2\ top}}\ (black)$. Hence in this new description we have:

Starting initially $q_{2_{M_1\ top}}(black)$, via cross coupling, puts $q_{3_{M_2\ top}}(black)$ on $M_2$. Because of inversion from transistor $M_1$, positive charge, $q_{2_{M_1\ top}}(black)$, results in negative charge, $q_{3_{M_2\ top}}(black)$, on top plate of $M_2$. This results in positive charge $q_{3_{M_2\ bot}}(black)$ at the bottom plate of $M_2$. Because of charge conservation due to the constant current source, a negative charge $q_{3_{M_1\ bot}}(black)$ is deposited on the bottom plate of $C_{gs1}$. Because of charge neutrality, negative charge $q_{3_{M_1\ bot}}(black)$ results in positive charge $q_{3_{M_1\ top}}(white)$ on top plate of $C_{gs1}$. Thus this highlights the coupling of two charges on top plate capacitor of $M_1$, spatially (horizontal), actually arises from the feedback coupling across $M_1$ and $M_2$ via coupling of two charges on top plate capacitor of $M_1$, energetically (vertical).

At this point to highlight the role of coupling, contrast this to standalone $M_1$, where two charges on plates have no coupling (spatially nor energetically). Note, the feedback coupling, via $q_{3_{M_2\ top}}(black)$, $q_{3_{M_2\ bot}}(black)$, and $q_{3_{M_1\ bot}}(black)$, the interaction is $g_m R$. From symmetry we go from $M_2$ to $M_1$ and find the interaction also as $g_m R$. If we repeat by going through the opposite direction, where there is also a factor of $g_m R$, and this accounts for $(g_m R)^2$, or $J$ part in (27).

### 3. Ensemble of Excess Carriers/Interaction Energy $J$ Arising from Coupling and Hamiltonian $H$

Now that we have described the excess carriers as distributed throughout the energy levels, we describe this ensemble of carriers over energy eigenvalues/levels, which arise from the underlying Hamiltonian (which in turn depends on the potential energy (i.e. interaction energy, as a result of coupling). It should be noted, due to coupling the energy level is not the same as in the standalone case. [9, 10]

Next we discuss the main features of $J$. Using Figure 4b, and starting from $M_1$, $C_{gs1}$, let us look at the ensemble of positive charges in the low-temperature limit.

The occupancy of energy level is shown (at $T = 0$, where the lowest levels are occupied). Starting with the two lowest levels i.e. $i = 1, 2$, the first level for excess charge is practically $E_c$, as pointed out in Appendix B. This sets up the capacitor voltage, $V_{gs1}$ and bias $M_1$. Through cross coupling via $M_2$ this will put more electrons on $M_1$. The next such electron ($q_3$) is on $i = 3$ because if we add an electron, due to Pauli Exclusion Principle, it will occupy $i = 3$. The Pauli Exclusion Principle also shows why this model corresponds to the Ising model; two members of the ensemble do not share an eigenstate, or equivalently, site. The resulting change in voltage is as given by (22) and the coupling strength $J$, as given by (24) will, if we consider only the nearest neighbor (like Ising)), will lead to Hamiltonian $H$ in (25). The occupancy of charges at energy level is denoted by 1.

### 4. Hamiltonian $H$ and Corresponding $J$'s Microscopic Origin

Having shown that microscopic details are necessary to explain interaction energy when coupling depends on distribution over microscopic energy levels and that this leads to probability of such distribution depends on interaction energy, we elaborate on how to calculate the Hamiltonian.

In order to calculate the Hamiltonian, we capture the microscopic nature of the energy level occupied by the ensemble of charges, or equivalently the system of charges. Then its energy is given as the sum of energy from

---

[9] Using the $J$ discussed below, for interaction potential among the electron ensembles on the capacitor in cross coupled (Schmitt Trigger) system, we can describe the Hamiltonian of this ensemble (discussed below). As discussed in the introduction, Table 2, in this section, we are content just to discuss the classical Hamiltonian (like Ising), and derive the resulting fluctuation, which of course is just valid in the classical, or high temperature regime. The $J$ we discussed in II.A, rooted in a potential that is dependent on 'distance' between energy level, is taken as a scalar, and therefore, when we write out the Hamiltonian, it is also treated as a scalar. Strictly speaking, potential is a function of a quantum-mechanical state vector, and is itself a vector operator. We simply take the energy level (eigenenergy, rooted in quantum mechanics) of the quantum state, but itself a scalar, to represent potential operator, pretty much in classical, the phase space (the position/momentum) (scalar) representation is used instead of the full quantum state vector [[19], pg. 281]. In the later sections, when the more complete quantum mechanical treatment (involving time evolution is required), care has to be exercised.

[10] With interaction, the Hamiltonian has a potential energy term. The eigenenergies are much more complicated, and typically can be solved only by methods such as the Hartree-Fock formalism [16]. In this paper, we do not need to solve such energy, only try to make some observation on the features.

.

individual charges i.e. a Hamiltonian $H$. Assume that potential energy arising from interaction energy is the dominant term in $H$

$$H = \underbrace{\mathcal{E}\left(\sum_{i(M_1)} q_i - \sum_{i(M_2)} q_i\right)}_{KE} + \underbrace{\frac{1}{2}\sum_{\langle ij\rangle} J_{ij} q_i q_j}_{PE} \tag{25}$$

In (25), $q_i$ and $q_j$ are the corresponding discrete charges, which, occupy different energy levels, and so different from the situation in (23). Because of this, $J$ becomes $J_{ij}$, so that in (24), the potential $V_{peak}$ becomes $V_{peak_{ij}}$, as the energy from one charge to the other depends on energy level it occupies. Thus, the potential energy arises from pairwise interaction energy between electrons in the ensemble, discussed in last section, but now, in the example above, considering electron $q_2$, the next such electron ($q_3$) is on $i = 3$, as shown, because if we add electron, it follows Pauli exclusion principle.

Meanwhile, from last section, which considers one such pair, we now consider summing over all the possible pairs. With energy level concept, we can take the sum over $N$ charges, to be the sum over all occupied energy levels. Thus let us define $q_i$ as the occupancy of charges at energy level $i$ and $q_j$ as the occupancy of charges at energy level $j$. For any level, the occupancy has a value of 1 if it is occupied and is 0 when it is not occupied. The sum is then given as (26). The term $\langle ij\rangle$ denotes summation over nearest neighbors (e.g. 4 and 5, 5 and 6)

$$\frac{1}{2}\sum_{\langle \mathrm{ij}\rangle} J_{ij} q_i q_j = \frac{1}{2}\sum_{\langle\, ij\,\rangle} (g_m R)^2 e V_{peak_{ij}} q_i q_j \tag{26}$$

The factor of $\frac{1}{2}$ accounts for double counting when charges are swapped. Whereas $i, j$ in (26) index individual charges on $C_{gs1}$ and $C_{gs2}$, so that the sum is over all excess charges on $C_{gs1}$ and $C_{gs2}$, $i, j$ now index occupancy of individual levels on $C_{gs1}$ and $C_{gs2}$, so that the sum is over all energy levels. From (26), the left-hand side is $H$. The summand in the right-hand side is now interaction energy per energy level pair. What (26) says is that in microscopic representation, the summand depends on distribution, and further the resulting sum is the Hamiltonian of the microscopic ensemble. To highlight this we write $H$ as $H(q)$, where $q$ stands for the occupancy of the ensemble distribution, leading to (27). Define $q_i$ as the occupancy of charges at energy level $i$ and $q_j$ as the occupancy of charges at energy level $j$. For any level, the occupancy has a value of 1 if it is occupied and 0 when it is not occupied. Note that the $q_i q_j$ factor depends on microscopic distribution and vary with temperature.

$$H(q) = \frac{1}{2}\sum_{\langle ij\rangle} J_{ij}\, q_i\, q_j = \frac{1}{2}\sum_{\langle ij\rangle} (g_m R)^2 e V_{peak_{ij}} q_i q_j \tag{27}$$

Thus the Hamiltonian, $H(q)$, where $q$ stands for the occupancy of the ensemble, has the same form as (25), but $q_i$ and $q_j$ can vary continuously from 0 to 1. There are three features on $J_{ij}$, as described in sections III.B.4.a, III.B.4.b, and III.B.4.c .

$$\begin{aligned} H(q) &= \mathcal{E}\left(\sum_{i(M_1)} q_i - \sum_{i(M_2)} q_i\right) + \frac{1}{2}\sum_{\langle ij\rangle} J_{ij} q_i q_j \\ &= \mathcal{E}\left(\sum_{i(M_1)} q_i - \sum_{i(M_2)} q_i\right) + \frac{1}{2}\sum_{\langle ij\rangle} (g_m R)^2\, e V_{peak_{\mathrm{ij}}} q_i q_j \end{aligned} \tag{28}$$

***a. Feature A: Coupling Energy***

We now elaborate on $V_{peak_{ij}}$ in (28) further (cap potential differential as $V_{peak}$ and energy level difference as $ij$) and also make the $ij$ elaboration consistent with $ij$ in $\sum q_i q_j$.

Let us illustrate via an example. Start by referring to Figure 4b, i.e. referring to charge on capacitor plate (horizontal). On $M_1$ we want to find the interaction between $q_{2_{M_1}top}(black)$ and $q_{3_{M_1}top}(white)$. Between negative charge $q_{3_{M_1}bot}(black)$ and positive charge $q_{3_{M_1}top}(white)$, as well as negative charge $q_{3_{M_2\,top}}(black)$, and positive charge $q_{3_{M_2\,bot}}(black)$, the electrostatic energy (contributing to $V_{peak}$ part of $V_{peak_{ij}}$) can be then be calculated, by following eq. 2.36 of [26]. The result is (29) where $q_i$ is the test charge, and $V(P_i)$ is the voltage or potential $q_i$ experienced, at position $P_i$.

$$W = \sum q_i V(P_i) \tag{29}$$

This, together with energy imparted through cross coupling, gives rise to interaction energy, given in (28) as equation (30).

$$\frac{1}{2}\sum_{\langle ij\rangle}^{N} J_{ij} q_i q_j = \frac{1}{2}\sum_{\langle ij\rangle}(g_m R)^2 e\, V_{peak_{ij}}\, q_i q_j \tag{30}$$

In (30), $N$ is the number of excess charges on each capacitor. Specifically, every test charge $q_{i,}$ on top plate of $C_{gs1}$ of $M_1$, when goes through $M_1$, picks up energy that is proportional to $(g_m R)^2$ ($\frac{1}{2}$ $(g_m R)^2$ is power gain of $M_1$). The resulting charge, at drain on $M_1$, resides on top plate $C_{gs2}$ of $M_2$. Following last paragraph, it induces charge on bottom plate of $C_{gs2}$, then bottom plate of $C_{gs1}$, then another charge on top plate of $C_{gs1}$. In the process it picks up electrostatic energy as given by (29), proportional to potential across the capacitors ($V_{gs1}, V_{gs2}$), and since there are two capacitors, to their difference ($V_{peak}$). Thus we have the correspondence shown in (31).

$$\begin{gathered} W \leftrightarrow \sum_i \sum_j J_{ij} q_i q_j \\ q_i \leftrightarrow e\, q_i \\ \sum V(P_i) \leftrightarrow \sum (g_m R)^2 V_{peak} q_i \end{gathered} \tag{31}$$

To put this in words, we have $q_i$ is the test charge (on top plate of capacitor $C_{gs}$ of $M_1$), and so we consider $eq_i$ also as test charge ($q_i$ now is the occupancy of the energy level $i$; so if $q_i = 1$implies that the $i^{th}$ level is occupied, and there is a charge $e$, which is the test charge $q_i$).

$V(P_i)$ is the voltage or potential experienced(contributed by charges at position $P_i$). Here, charges are 'rest' of charges on bottom plate of $M_1$ capacitors. Now instead of characterizing the potential from the 'rest' of the charges using their position, here, all the 'rest' of the charges have the same potential (neglecting energy level for time being). This potential can be viewed as the work done. Since the charge has to be moved via the cross coupling transistor $M_1, M_2$ (contributing the factor $(g_m R)^2$), and then across the two capacitors with potential difference $V_{peak}$. Therefore work done or potential experienced is $(g_m R)^2 V_{peak}$

Meanwhile in (29), the rest of charges with potential $V_{gs1}$ (in bottom plate of $M_1$) all occupy an energy level $E_j$, or, due to charge neutrality, all have a corresponding charge (in top plate of $M_1$), occupying level $E_j$ i.e. having an equivalent occupancy of $q_j$.[11] Thus, the interaction energy per test charge is (32).[12]

$$\sum_i V(Pi) = \sum_i (g_m R)^2 V_{peak} q_j \tag{32}$$

Next taking energy level into consideration, $q_{2_{M_1\,top}}(black)$ begets $q_{3_{M_1\,top}}(white)$ and the potential difference is not just $V_{peak}$, but difference between $V_{gs1}$ offset by energy at $i = 2$ and $V_{gs2}$ offset by energy at $i = 3$, or $V_{peak}$ plus energy difference between level 2 and 3 i.e $ij$. We denote this as $V_{peak_{ij}}$.

Finally to get the overall interaction energy from interaction energy per test charge (given in (32)), we sum up all the test charges $q_i$ on the top plate of $C_{gs1}$ (again using the occupancy representation and across the energy level) i.e. (32) is summed to give (33).

$$\sum_i \sum_j J_{ij} q_i q_j = \sum_i q_i \sum_j (g_m R)^2 V_{peak_{ij}} q_j$$

---

[11] Note, coupling two capacitors give horizontal coupling, which, as explained before, translate into vertical coupling i.e. energy levels $E_i, E_j$, or $q_i, q_j$.

[12] Equation (37) indicates the sum of charges (in the range of $kT$, centered around $E_f$) that participates in the differential mode of current flow [30], and hence represents the current, which is then consistent with the definition of $\Delta$ given in (11) (macroscopic, also given in terms of current) of II.B.

(33)

Equation (33) can be re-written as (36) by following equations (34) and (35).

$$\sum_i \sum_j J_{ij} q_i q_j = \sum_i q_i \sum_j \left(g_m R\right)^2 V_{peak_{ij}} q_j \quad (34)$$

$$= \sum_i \sum_j (g_m R)^2 V_{peak_{ij}} q_i q_j \quad (35)$$

$$= \frac{1}{2} \sum_{ij} (g_m R)^2 V_{peak_{ij}} q_i q_j \quad (36)$$

Equation (36) is the same as (32), but now with explicit explanation on where the $V_{peak}$ and $ij$ part in $V_{peak_{ij}}$ in (32) originates.

In (36), $q_i$ and $q_j$ refer to physical charges, whereas in (26), $q_i$ and $q_j$ refer to occupancies of energy level $i, j$. They are indices to do bookkeeping, or counting (e.g. occupied states have $q_i = 1$, unoccupied states have $q_i = 0$). In this way the sum over all energy levels will count only those contributions when energy level is occupied, i.e. equivalently to $\sum_{all\ charges} q_i q_j$, where $q_i$ and $q_j$ are the $i^{th}$ and $j^{th}$ charges.

There is still the relevance of energy level in index/energy level "co-ordinate"; but that shows up in the vertical description, as for example, in feature C (see (40) and III.B.4.c); here, in feature A, it is mainly on top/bottom plate.

### *b. Feature B: Fluctuation and Existence of Many Configurations with Different Probabilities*

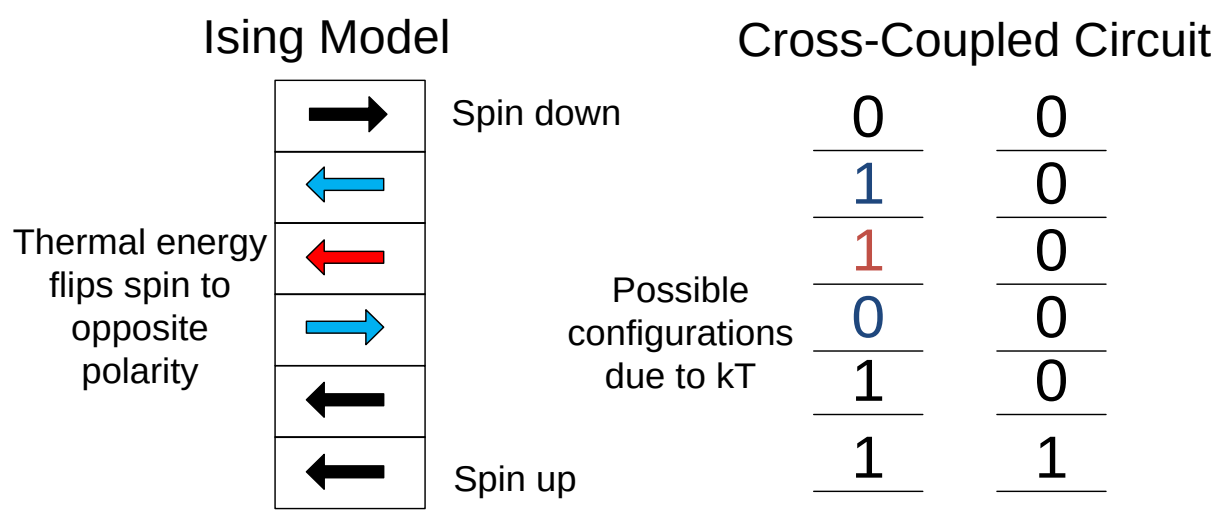

Figure 5. COMPARISON OF THE OCCUPANCIES ON THE CROSS-COUPLED CIRCUIT WITH A CHAIN OF MAGNETIC MOMENTS IN THE ISING MODEL.

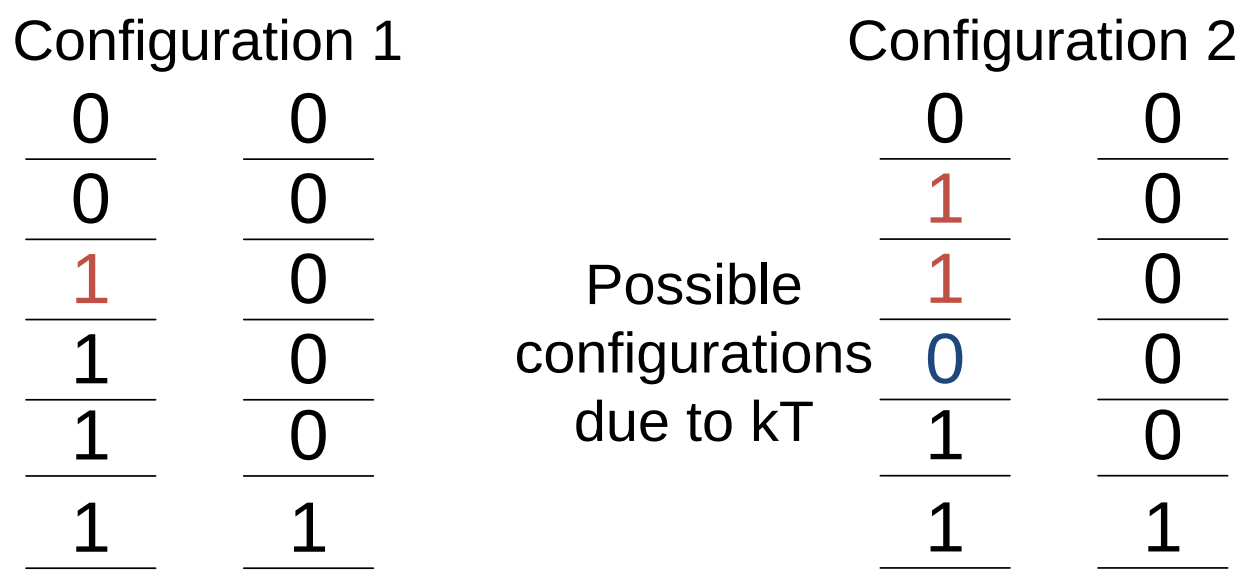

Figure 6. A COMPARISON BETWEEN TWO CONFIGURATIONS SHOWING THE EFFECT OF TEMPERATURE ON THE DISTRIBUTION OF EXCESS CHARGES.

Consider the ensemble in a metastable state, just like Ising model in a disorder state, where disorder/metastable results from fluctuation (see Figure 5). With fluctuation there are [1 0] pairs. Without fluctuation, the ensemble settles into order/stable state i.e. $M_1$ is all [1 1], $M_2$ all [0 0], or vice versa. Let us consider two such configurations, shown in Figure 6.

Both configurations have 4 positive charges at M1 and 1 positive charge at $M_2$, so have the same order parameter Δ, given as (37). [13] The order parameter is the difference in the number of positive charges on $C_{gs1}$ and

[13] Note, when evaluating the sum, on the LHS, only those $i^{th}$ positions having charges contribute. Equivalently, on the RHS, only those energy levels occupied (i.e. $q_i = 1$) contribute.

$C_{gs2}$. Its value is 1 if all charges are on $C_{gs1}$ and $-1$ if all charges are on $C_{gs2}$. Changing the order parameter is equivalent to phase change in the Schmitt Trigger.

$$\Delta = \frac{\sum_{i(M_1)} q_i - \sum_{i(M_2)} q_i}{\sum_{i(M_1)} q_i + \sum_{i(M_2)} q_i} \tag{37}$$

Meanwhile Configuration 1 has different pairs of [1 1], [0 0], [1 0][14]), than Configuration 2. This means Configuration 1 has a different energy $H$ than Configuration 2. $H$ is $\frac{1}{2}\sum_{\langle ij\rangle} J_{ij} q_i q_j$ (refer to (27)). This determines the probability of each configuration, as discussed next.

We now discuss in detail the energy and probability of configurations (degrees of freedom). As discussed, the ensembles, unlike in section III.B which gives one degree of representation for the entire ensemble, resulting in a potential $E$, now is going to possess many degrees of freedom, each one for each configuration. As an example, using one-dimensional representation, with an ensemble of three members ($N = 3$), the configuration is designated as $q = (q_1, q_2, \dots, q_N) = (q_1, q_2, q_3)$, distributed over all the energy states available. Appendix B describes the eigenenergy and $E - k$ diagram for the excess charge.

We designate $q_1$ as the occupancy of charge in energy level 1 ($q_i = 0$ is not occupied and 1 when occupied) and $q_j$ to be the occupancy of charge in the next higher energy level $j$. Note we are recycling notation of $q_i$ and $q_j$. These charges are all going to be occupying starting from $E_c$[15], which is shifted by source voltage $V_s$, i.e. $qV_s$ (assuming we neglect $V_{ds}$ drop for now) [16]. So, the first charge occupies the state $E_c$, and the next charge, by Pauli Exclusion Principle (assume no degeneracy), occupies the next higher energy level.

Figure 4 is now redrawn, starting with the case at metastable state i.e. $\Delta = 0$. An example for the configurations of ensemble with $N = 6$ and $\Delta = 0$ is shown in Figure 7, where $i = 1$ is the conduction band energy $E_c$. When $\Delta = 0$, both transistors are on, and are biased at the same voltage. The number of charges is the same for both transistors, meaning $N_1 = N_2$. At temperature $T = 0$, the charges occupy the lowest energy levels possible and there is only one possible configuration. At temperature, $T > 0$, there are many possible configurations with different energy level occupancy of the charges. The three configurations of $N_1$ shown in Figure 7 (The distributions over the three $M_1$'s (or left hand column) are as following: Configuration 1: $q = (1, 0, 0, 0, 1, 1)$; Configuration 2: $q = (0, 0, 1, 1, 1, 0)$; and Configuration 3: $q = (0, 1, 0, 1, 0, 1)$. Each configuration corresponds to a particular value of the order parameter (order parameter, denoted as difference of occupancy, this being sum of all the 1's in the configuration, where again 1 means the particular energy level is occupied, in our case). For example, configuration 1, 2, 3's order parameter (from occupancy) is $0$.

Now we go to the energy, or $H$ of the configuration, as defined in (28). Configuration $(\nu)$ 1 and 2 have different energy $E_\nu$, and so when exponentiated, it refers to different probabilities (Boltzmann probabilities). To calculate these probabilities, we make use of (28). For example, (assuming $J = 1$) Configuration 1 energy is $q_5 q_6$ because there are 2 consecutive 1's and so $J$ between 2 of them. $q_1$ is neglected because, as noted, the interaction energy between charges of non-adjacent energy level is not taken into account when summing $Jq_i q_j$. This is because we neglect this energy, which is not adjacent. Section III.B.4.c gives more detail. Meanwhile, Configuration 2 is $q_3 q_4 q_5$ because there are three consecutive 1s. Hence energy is $q_3 q_4 + q_4 q_5$ and this gives 2. In summary, Configurations 1 and 2 have different energies, and so when weighed by $\beta$ and taking exponential, result in different probabilities of being occupied.

---

[14] Comment on [1 0] vs [1 1] or [0 0] pair:

- There is inversion of $M_1$ and $M_2$ with $1e$ in and $1e$ out. This due to inversion in the interaction path (explained as interaction path in feature C (see III.B.4.c(3)).
- This inversion normally has adjacent pairs' configuration as in Figure 5, but [1 0], but not normally as Figure 6 i.e. [1 1] or [0 0]
- Due to thermal energy, [1 1] or [0 0]: in Figure 6 can exist. This explains the origin of different configurations.
- This is similar to spin up/down configuration as Ising description in Figure 5.

[15] It is only the excess charge, induced by capacitor voltage, that forms the inversion layer, that is primarily contributing to the ensemble, as it is only these charges that flow (via source/drain) to resistor, the heat reservoir, or dissipation source, maintain thermal equilibrium via the temperature established, and therefore qualifies to be denoted as the ensemble. Any other electrons, originally in the conduction band, apart from the inversion charge induced through capacitor (and hence also flows), which strictly speaking contributes to the ensemble, is tiny, as the material is $p$ type i.e. it is neglected.

[16] We are considering bottom plate. For top plate, there is a shift due to capacitor voltage, i.e. no offset of $\frac{1}{2}CV^2$ from $E_c$ or $qV_s$

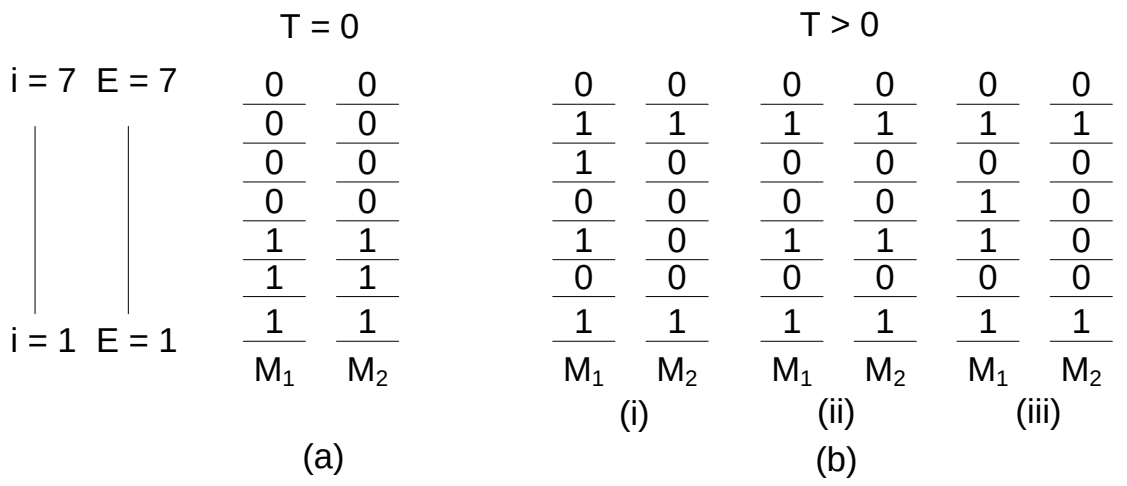

Figure 7. VARIOUS CONFIGURATIONS AT METASTABLE STATE FOR SIX CHARGES.

Another example configuration, for the case at stable state, of ensemble with $N = 6$ and $\Delta = 1$ is shown in Figure 8. With $\Delta = 1$, transistor $M_1$ is on and transistor $M_2$ is off. Thus, the Schmitt trigger is at stable state. The number of configurations for metastable state is larger than the configurations for stable state. In the case metastable state, both transistors are on, meaning in Figure 7 both $M_1$ and $M_2$ are on, and that the charges are free to distribute themselves among these levels (we show 3 possibilities/configurations). In the case of stable state, one transistor is on and the other off, meaning, for the off transistor, there is no charge, or no distribution. Thus, the resulting possibilities/configurations (with charges on the 'on' transistor distributing) is fewer (we show one such possibility in Figure 8).

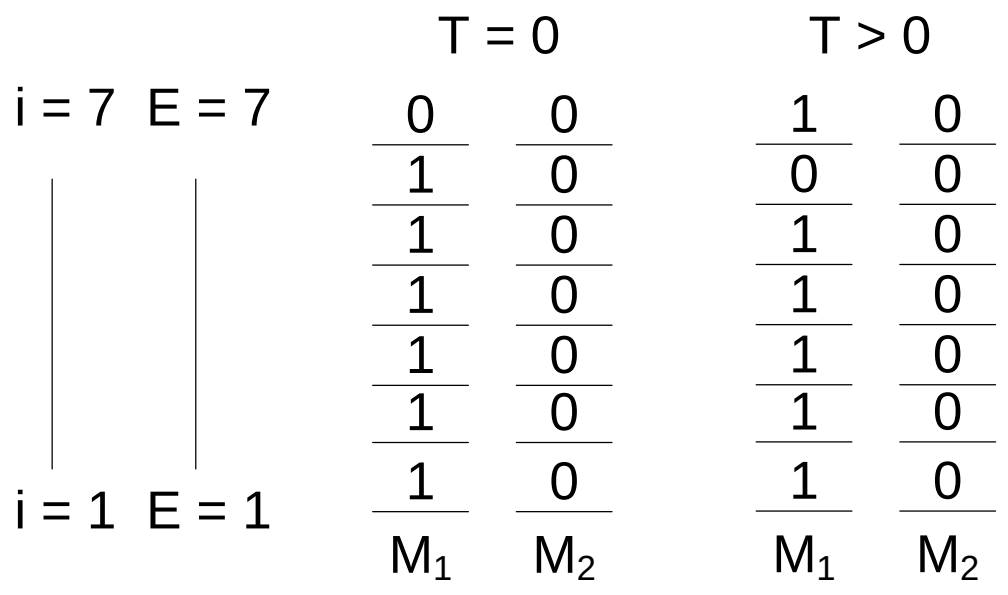

Figure 8. VARIOUS CONFIGURATIONS AT STABLE STATE FOR SIX CHARGES

Now we can combine the examples from Figure 6 which cover going from $\Delta(M)$ to number of configuration) and the example in Figure 7, which covers the probability of each of such configuration, and get the free energy. But before we do that we cover one more feature (feature C) of the ensemble.

***c. Feature C: Modeling the Schmitt Trigger using only Nearest Neighbor Interactions in the 'Ising'-like Hamiltonian***

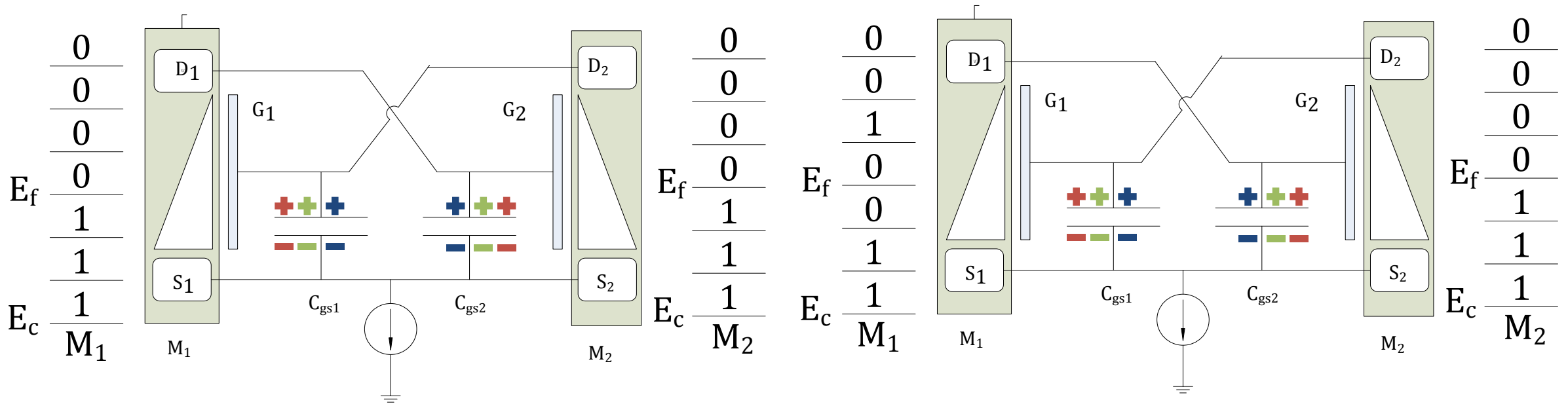

Figure 9. MICROSCOPIC EXAMPLE OF INTERACTION ENERGY MODEL: NEAREST NEIGHBOR

**(1) Qualitative Discussion on Independence of Interaction Energy on Spatial Distribution and Ising Behavior from Interaction Energy**

As illustrated above, the interaction energy accompanying the transfer 'virtual' of electrons is independent of the physical location or placement of the charges (the charges are spread throughout the top plate of the capacitor in an arbitrary fashion; a particular charge needs not be spatially identified with specific 'site' on the top plate, unlike spin, where each spin is situated on a specific site[17]). Therefore, whereas, in Ising model, interaction energy is dependent on the distance (spatial) between sites), here, rather, the coupling, or interaction energy, between 2 charges (occupying two energy levels; vertical coupling) is dependent on the distance (energy difference) between the 2 energy levels occupied (shown in (31)).

**(2) Formula for Calculating Interaction Energy Based on Energy Index**

Mathematically the interaction energy, from (28), is $\frac{1}{2}\sum_{\langle ij\rangle} J_{ij} q_i q_j$. Looking at the change in total energy from one state to another, the $\frac{1}{2}\sum_{\langle ij\rangle} J_{ij} q_i q_j$ can be interpreted as proportional to the sum of changing $\Delta v$, which is consistent with (28) as each term $q_i q_j$ in the sum in (28) is interpreted as adding a change in energy or $\Delta v$. Basically (28) is static, as you are given a configuration (spin or charge) and then for each pair you calculate energy as a product (again spin or charge) and then you sum all the pairs (adjacent or not) for that configuration i.e. $\frac{1}{2}\sum_{\langle ij\rangle} q_i q_j$. This will be elaborated in the explanation of interaction path below, using Figure 10.

**(3) Justification of Neglecting Next-Nearest Neighbor Interactions via Interaction Energy Formula From III.B.4.c(2)**

To show interaction $J$ is not constant (and eventually based on energy index), specifically now we have to make expression for $J$ more sophisticated. First we look macroscopically. Its dependency on biasing condition:

Qualitatively as biasing is changed: (Referring to Section II and Appendix A), and starting from metastable i.e. $M_1$, $M_2$ on, e.g. by biasing current $I_{d1}$ (nominally $\frac{I_0}{2}$, i.e. half of current source) from $M_1$ to $M_2$ (or vice versa) to stable state ($M_1$ or $M_2$ off), i.e. $i_{d1} = I_0$, $i_{d2} = 0$. This results in coupling becoming weaker as the feedback loop is weakened. Specifically the turning on of $M_1$ is counteracted by turning off of $M_2$, resulting in $V_{gs} - V_t$ for both transistors giving the respective currents that sum to a constant common mode current, as explained in Method 1 (Section II) in (6) and (7).

Equation (24) is rewritten as (38), with interaction energy being expressed as $J_{ij}$. As the electron goes through different energy levels (i.e. $i, j$) $I_d$ changes, $g_m$ changes and loop gain changes. This is shown in (38) by making $g_{m1}$ and $g_{m2}$ functions of Δ.

$$J_{ij} = g_{m1}(\Delta) g_{m2}(\Delta) R_1 R_2\, eV_{peak} \tag{38}$$

The loop gain $g_{m1} g_{m2} R_1 R_2$ depends on the biasing condition of $M_1, M_2$ in the stable versus metastable state (see II and Appendix A). For example as $i_{d1}$ increases and $i_{d2}$ decreases, the loop gain changes i.e. loop gain is also a function of Δ as defined in equation (38). The order parameter Δ is defined in (37).

Second, we want to show how $J_{ij}$ is related to the energy indices $i, j$ in a microscopic fashion. We do this by identifying the energy occupancy at a metastable versus stable state. From microscopic argument, having different energy levels occupied means different biasing voltage $V_{gs} - V_t$ on the capacitor. This results in different drain current $I_d$. Equation (39) relates the biasing voltage to the distribution of excess carriers on the top plate of a transistor. It holds for $M_1$ and $M_2$.[18] [19] [20]

$$V_{gs} - V_t \propto \frac{\sum_k q_k E_k}{\sum_k q_k}$$

---

[17] On the other hand, like the Ising model, where one and only one spin occupies per site, here, assuming no degeneracy, one and only one electron occupies a level. It should be noted there are not too many electrons (these are excess electrons, due to charge on cap, and for our case is on the order a few hundreds to thousands) than levels (correspond to levels in the conduction band of a doped silicon, the top plate of the capacitor)

[18] The macroscopic expression for the voltage of a capacitor $V = \frac{Q}{C}$ is related to the microscopic description $V = \frac{\sum_i q_i E_i}{\sum_i q_i}$. However, the microscopic description takes the distribution of charges into account whereas the macroscopic description assumes all charges have the same energy. To derive the macroscopic expression, let $E_i = \frac{\sum_i q_i}{C}$, and let $Q = \sum_i q_i$. $E_i$ no longer depends on the index directly, and so it will factor out of the sum in the microscopic description. Referring to Figure 24d in Appendix B, there is a correspondence between $E_f - E_C$ and $V_{gs} - V_t$ due to the correspondence between the microscopic and macroscopic description of charge. $E_f - E_c$ is microscopic and $V_{gs} - V_t$ is macroscopic. The offset of $V_T$ is required to compensate for the charge in the transistor inversion layer. Similarly, in Figure 9 three electrons span from $E_c$ to $E_f$, with their occupancy being $q_i = (1\ 1\ 1\ 0\ 0\ 0)$. This allows one to apply (31) to calculate $V$, relating the microscopic charge distribution to the macroscopic $V_{gs} - V_t$.

[19] To see how electron level affects macroscopic quantities such as $V_{gs1}$, $V_{gs2}$, follows E-K diagram, see Appendix B.

[20] $V_{peak}$ has more to do with electrostatic energy transferred as charge is moved from $C_{gs2}$ to $C_{gs1}$, as such is average (not individual energy level); and does not get factored in microscopic determination of $J$.

(39)

For illustration purposes, we follow the energy level distribution of $M_1$ as shown in Figure 10ai. Figure 10ai depicts the energy distribution for $M_2$ to the right of $M_1$. As the circuit goes from metastable to stable (i.e. Figure 10ai to Figure 10aii), the coupling from $M_1$ to $M_2$ decreases. For a particular electron occupying that energy level, we can again associate the $J_{ij}$ for that index as the energy needed to put or remove the electron at that energy level, as explained in (i) (macroscopic), but this time in microscopic using the energy level diagram. Therefore (38) is rewritten as (40), with $g_m$ dependent on $i$ and $j$ through $V_{gs} - V_t$.

$$J_{ij} = g_{m1_{ij}} g_{m2_{ij}} R^2 e V_{peak} \tag{40}$$

To see the strength of the interaction, consider the nearest neighbor, as shown by going from Figure 10ai to Figure 10aii at $E = 5$ (see dotted line). This is responsible for the nearest-neighbor interaction $J_{65,}$ shown by the curved arrow in Figure 10ai.

Meanwhile the interaction on any electron (say starting from $E = 6$ in Figure 10ai), going through the interaction path $M1_G - M1_D - M2_G - M2_S - M1_S - M1_G$, induces another electron, see (see dotted line) at $E = 5$. It should be noted that because of this inducing, the number of electrons on $M_1$ and $M_2$ remains the same, even though on individual transistor, they change (e.g. in Figure 10ai $M_1 = 3$ and $M_2 = 3$ and in Figure 10aii $M_1 = 4$, $M_2 = 2$ so that the sum stays constant at 6).

Let us now discuss a next-nearest neighbor interaction This is shown in Figure 10b, specifically going from Figure 10bi to Figure 10bii at $E = 2$ (see dotted line)). This is responsible for the interaction $J_{62}$ (solid curve at Figure 10bi). The interaction on any electron goes through the same interaction path as in Figure 10a, inducing another electron (see dotted line).

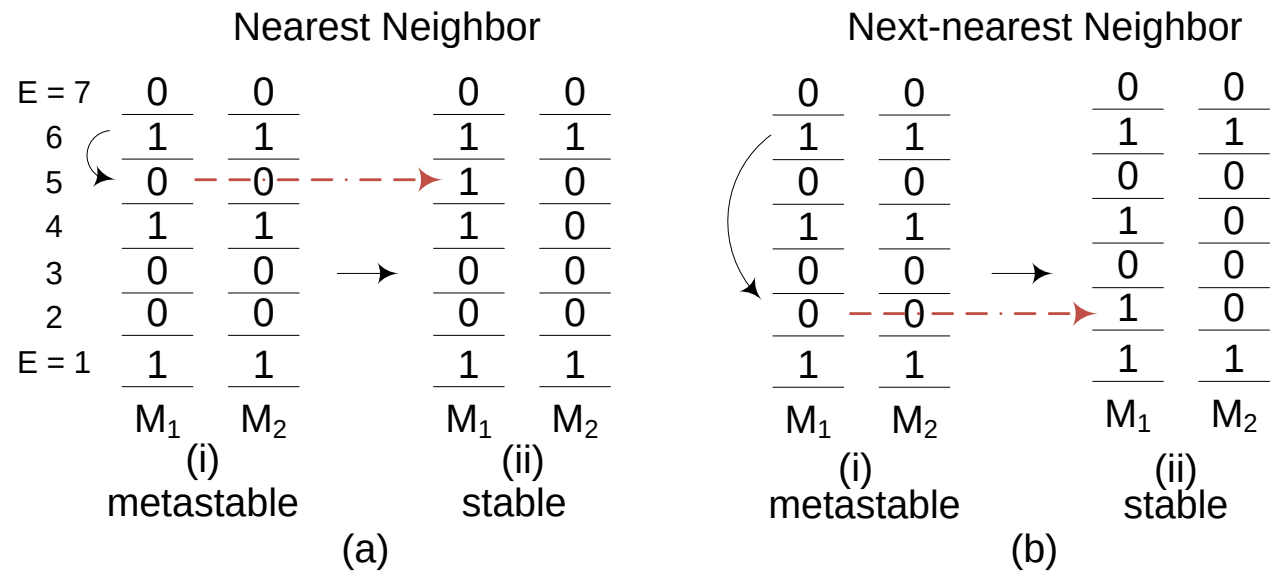

Figure 10. A CHANGE IN OCCUPANCY CORRESPONDING TO A NEAREST (B) NEIGHBOR INTERACTION. THE HORIZONTAL ARROWS SHOW HOW THE OCCUPANCY DISTRIBUTION ON M1 AND M2 CHANGE AFTER THE INTERACTION. INTERACTING ENERGY LEVELS ARE DEPICTED BY THE CURVED ARROWS.

Now to show interaction strength in Figure 10a is stronger than Figure 10b, we work out an example. Assume that the energy levels $E$ are equally spaced, and dimensionless. The scenarios presented in Figure 10ai and Figure 10bi both are at metastable state[21]. Following (39) for both transistors $V_{gs} - V_t \propto \frac{11}{3}$

The $J_{65}$ for the nearest-neighbor interaction is proportional to $(V_{gs1} - V_t)(V_{gs2} - V_t)$ (from (40)), since $g_m \propto V_{gs} - V_t$, and hence depends explicitly on the energy index.[22] The corresponding $V_{gs1} - V_t = 4$ and $V_{gs2} - V_T = \frac{7}{2}$. The resulting $J$ is 14.

In the next-nearest neighbor interaction, the $V_{gs1} - V_t$ and $V_{gs2} - V_t$ are $\frac{13}{4}$ and $\frac{7}{2}$ respectively. The resulting $J$ is $\frac{91}{8}$, which is less than $J$ of 14 for nearest neighbor. This shows that next-nearest neighbor interactions matter less than nearest-neighbor interactions. This argument generalizes to other energy levels.[23]

[21] There are other configurations corresponding to metastable state i.e. as long as the $V_{gs} - V_T$.for both $M_1$ and $M_2$ are equal e.g. configuration $M_1(E_1, E_2, \cdots) = (1001010)$ and configuration $M_2(E_1, E_2, \cdots) = (1010001)$. Following (39) they both give $V_{gs} - V_t \propto \frac{11}{3}$. However for simplicity of illustration, we choose as the example where configurations are the same.

[22] The transitions in Figure 10 (a) and (b) come close enough to represent $J$ as the sum of $V_{gs} - V_t$ to present the argument.

[23] In general the situation is: most of occupied energy level starts around a metastable state (Figure 4), as that is where noise spike is (for $T = 0$, at mid-level, at $T > 0$ a few $kT$ around, but still far from stable) e.g. such as (a) $i = 3 \to 4$ vs. (b) $i = 3 \to 7$ (both of which start at metastable, are encountered), while situations with counter-example such as: (a), going from $i = 6 \to 7$ (close to stable state/nearest neighbour), $J$ is smaller than (b), going from $i = 3 \to 5$ (towards metastable state/next-nearest neighbour), is not likely to happen (the counter example $i = 6 \to 7$, which starts at stable state is not encountered), and cases are ignored.

### C. Partition Function and Free Energy

As discussed in section III.B.4.c(1), the existence of many configurations (many degrees of freedom) at a finite temperature (main difference between present section and section III.B) can physically be traced back to fluctuation[24]. To capture quantitatively all these configurations, one resorts to the partition function, which contains the probability (Boltzmann factor $e^{-\beta E_\nu}$) for a particular energy $E_\nu$, given the configuration $\nu$ at thermal equilibrium. The ensemble average of $\Delta$, denoted as $\langle\Delta\rangle$, is the observed average value of $\Delta$. $\langle\Delta\rangle$ is the average differential current in the Schmitt Trigger in the absence of noise. As is evident, for any configuration of occupancy, $\Delta$ indicates the charge (in the range of $kT$, centered around $E_f$) that participates in current flow [27], and hence represents the current i.e. consistent with the definition of $\Delta$ given in (11) (macroscopic, given in terms of current) of III.B.

The partition function $Q$ (a function of temperature), is the sum of all configurations of the system weighed by $\beta = \frac{1}{kT}$ via exponentiation[25] i.e. $Q = \sum_\nu \exp(-\beta E_\nu)$. Using $Q$ and $\langle\Delta\rangle$ as defined in (41), the ensemble average is $\langle\Delta\rangle = Q^{-1} \sum_\nu \sum_j q_j \, e^{-\beta E_\nu}$. Substituting $E_\nu$ from (30) yields (41).

$$Q(N, \mathcal{E}, \beta) = \sum_{q1} \cdots \sum_{q_N} \exp\left(\beta\left(\frac{1}{2}\sum_{\langle ij\rangle} J_{ij}\, q_i q_j + \mathcal{E}\left(\sum_{i(M_1)} q_i - \sum_{i(M_2)} q_i\right)\right)\right) \tag{41}$$

In (41), $N$ is the number of charges in the ensemble and $\mathcal{E}$ is the external electric field applied to the charges arising from voltage difference between $C_{gs1}$ and $C_{gs2}$[26]. The role of $\mathcal{E}$ in (41) is analogous to the external magnetic field $H$ in the Ising model of magnetization. $\mathcal{E}$ is the conjugate variable of $N$.

Let us use the mean-field approximation to simplify $Q$ in order to investigate closed-form solutions. This approximation assumes that the number of charges and the number of neighbors for a given charge is large. This does not violate the restriction of interactions to nearest neighbors, since all the neighbors of a given charge are "nearest" to each other. Finally, since we are interested in fluctuations around the metastable point, assume $\mathcal{E}$ is negligible. Then, $q_i$ can be written as (42).

$$\begin{aligned} q_i &= q_i - \langle q_i\rangle + \langle q_i\rangle \\ &= q_i - \Delta + \Delta \end{aligned} \tag{42}$$

Using the identity (42) $q_i$ is substituted into the Hamiltonian with $q_i q_j$ as given in (43).

$$\begin{aligned} q_i q_j &= (q_i - \Delta + \Delta)(q_j - \Delta + \Delta) \\ &= (q_i - \Delta)(q_j - \Delta) + \Delta(q_i + q_j) - \Delta^2 \end{aligned} \tag{43}$$

Under mean field approximation the term $(q_i - \Delta)(q_j - \Delta)$ is negligible. [20] The Hamiltonian can be approximated via (44), with $J$ defined such that $\sum_{\langle ij\rangle} J_{ij} = NJ$. One can simplify $J$ in this way because the interaction strength is significant only for nearest-neighbor interactions. Section III.B.4.c discusses this in further detail. These simplifications yield (45), which expresses (41) as a function of $\Delta$ in the mean-field approximation.

$$H \approx -\frac{1}{2} NJ\Delta^2 + J\,\Delta \sum_i q_i$$

[24] The first cause is inversion is flipped (due to $kT$). The other is scattering, also due to $kT$ that occurs for electrons in the ensemble when the electrons travels through the resistors during the virtual transfer. The first cause also explains the '10'as opposed to '11' (the expected) adjacent energy levels, or more energy favorable (similar to up/up), and is 'adjacent', as inversion brings about next (adjacent) charge (at adjacent level, due to Pauli Exclusion Principle), whereas scattering can be any level.

[25] The exponential dependence (see [20], pg. 63) comes from the Boltzmann distribution as a consequence of the ensemble. We are using Ising model formulation, which uses the canonical ensemble partition function Q and exponential dependence (see [19], pg. 395-397). This can describe phase transition (see [19], pg. 395). The ensemble is canonical because the tail current conserves the number of charges on the capacitor. If one invokes the grand canonical partition function $\Xi$, the pair of conjugate variables $(\mu, N)$ correspond to the pair $(H, M)$ in the Ising model, and $(\mathcal{E}, N)$ in Method 2. $\mu$ is the chemical potential, $H$ is the external magnetic field, and $M$ is the magnetization. The description using $\Xi$ is equivalent to the Ising model. Just as $H$ flips spins in the Ising model, so $\mathcal{E}$ changes the distribution of charges in Method 2.

[26] Like discussion in pg. 126 [20], a voltage difference will tilt the free energy plot shown in Figure 11. The metastable state corresponds to both $\mathcal{E} = 0$ and $\Delta = 0$. This is like pg. 125 [20], where metastable state corresponds to $H$ and $M$ being zero. The fluctuation (equal to the partial derivative of $\Delta$ with respect to voltage difference) is not zero at $\Delta = 0$ in Method 2 just as the fluctuation (equal to the partial derivative of $M$ with respect to $H$) is not zero in the Ising model (pg. 128 of [20]). See footnote 28 below. Finally, since we include $\Delta$, the energy difference due to microscopic distribution of charges on $C_{gs1}$ and $C_{gs2}$ is considered in determining $\mathcal{E}$ in addition to the voltage difference.

$$Q = e^{-\beta N J \Delta^2} \left( \sum_{q=\pm 1} e^{\beta J \Delta q} \right)^N$$
$$= e^{-\frac{\beta N J \Delta^2}{2}} (2 \cosh(\beta J \Delta))^N \tag{44}$$

(45)

Equation (46) gives the free energy per unit charge $G(N, T)$.

$$G = -\frac{1}{\beta} \ln Q(N, T)$$
$$= \frac{1}{2} J \Delta^2 - \frac{1}{\beta} \ln(2 \cosh(\beta\, J\, \Delta)) \tag{46}$$

Using the Taylor series defined in (47) on the second term in (46) about $\Delta = 0$ yields (48).

$$\ln(2 \cosh(\beta J \Delta)) = \ln 2 + \frac{1}{2} (\beta J \Delta)^2 - \frac{1}{12} (\beta J \Delta)^4 + O(\beta J \Delta)^6 \tag{47}$$

$$G = \frac{kT}{2} \left( \left( 1 - \frac{(g_m\, R)^2\, e\, \Delta V_{peak}}{kT} \right) \Delta^2 + \frac{\Delta^4}{6} + const. \right) \tag{48}$$

Figure 11 indicates that the plot shows metastability when $(g_m R)^2 e V_{peak} < kT$. This leads naturally to the definition of a critical temperature $T_c$ as defined in (49). Using $T_c$ in (24) shows that $J = kT_c$. When the temperature decreases, the energy difference between metastable state and stable states decreases until a point where the metastability of the system vanishes. For the case of $T < T_c$ where the metastability of the system exists, the energy difference is (50).

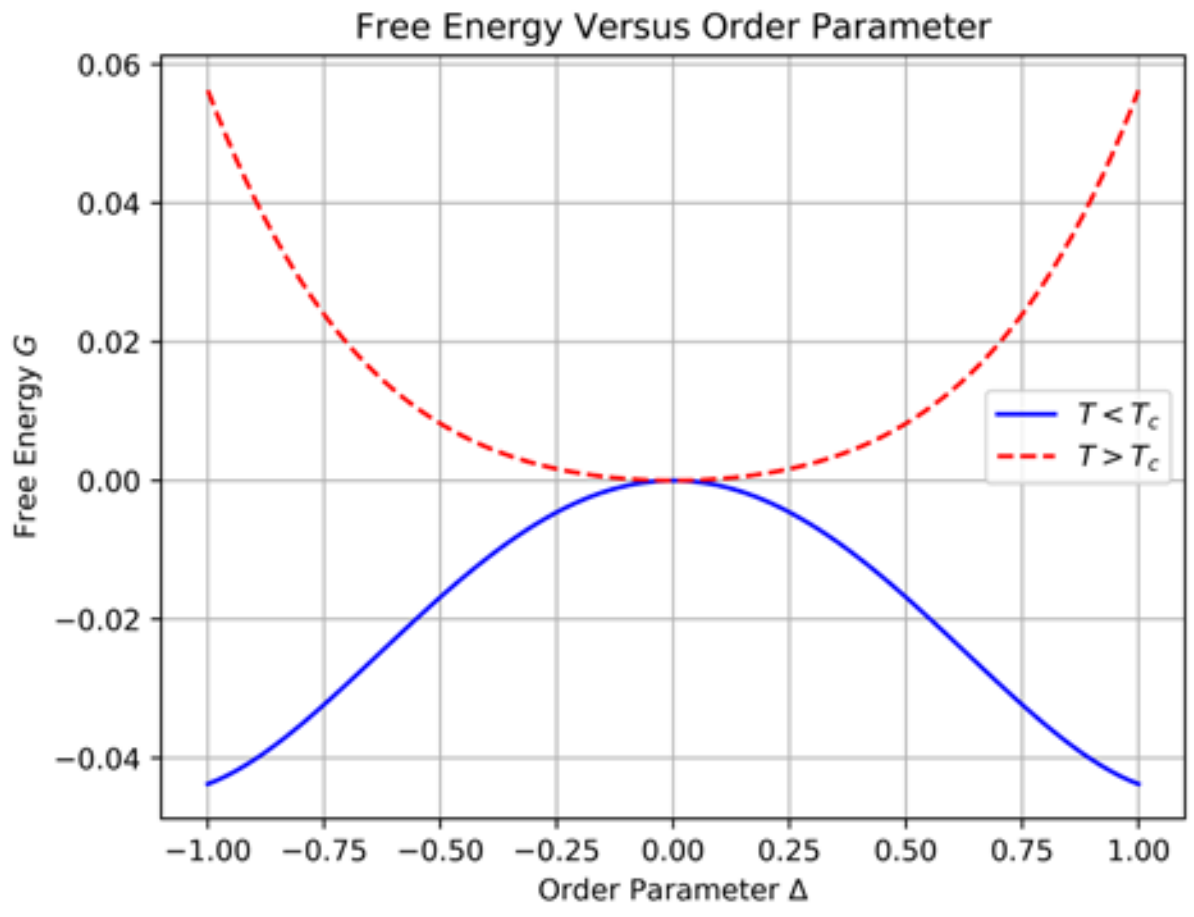

Figure 11. THE FREE ENERGY VS $\Delta$ FOR DIFFERENT RELATIONS BETWEEN $T$ AND $T_c$.

$$T_c = \frac{(g_m R)^2 e V_{peak}}{k} \tag{49}$$

$$\Delta G = \frac{3}{2} \left( 1 - \frac{T_c}{T} \right)^2 \tag{50}$$

When $T < T_c$, the free energy obtained is an even function with two minima. The form of (48) is similar to the potential energy obtained in section II. However in section II, it is made explicit that it comes from the circuit topology of the cross coupled pair with the tail current source that we obtain the terms $\Delta^2$ and $\Delta^4$ in the end. Section II models the system as a single particle that follows the principle of stationary action. Method 2 differs from section II by treating the system as made of many particles with many energy configurations characterized by the thermal distribution. Considering the many configurations of the system allows calculation of noise directly from the circuit dynamics without having to map to an equivalent thermodynamic system. This shows the microscopic origin of noise in the circuit. Summing up the energy of all configurations yields the free energy in terms of temperature. Both approaches show that the electronic bistable system has two stable states on either side of a metastable state. However, (48) depends on $kT$ and (15) does not. This difference highlights:

- (48) is free energy (includes $T$ effect) whereas (15) is potential energy (no $T$ effect)
- Noise can be calculated directly from (48) (shown in next section), while for (15) it is a roundabout way: identify (15) as a system with metastable point→add thermal noise to this metastable system→map equation of motion on this noisy metastable system to equation of state of an equivalent thermodynamic metastable system in→as this ther modynamic system, like any thermodynamic system, has free energy, like (48), again noise can be calculate, but since unlike (48), (15) only maps into equation of state, not free energy→use equation of state to calculate noise indirectly
- Put another way following [28], gain in energy from aligning charge outweighs (not as bad as) loss in entropy (charge align) [27]

D. The Normalized Current Fluctuation

From the partition function $Q$ given in (45), we determine the current fluctuation. The fluctuation of the normalized current $\Delta$, is determined and is given by second partial derivative of the logarithm of the partition function, with respect to the conjugate variable $\mathcal{E}$ [28], as shown in (51)([20] pg. 71). $\mathcal{E}$ is an external electric field applied to the bistable circuit.

$$\frac{\partial^2}{\partial \mathcal{E}^2}(\ln Q) = \beta^2 N(\langle \Delta^2 \rangle - \langle \Delta \rangle^2) = \beta N \frac{\partial \langle \Delta \rangle}{\partial \mathcal{E}} \tag{51}$$

The root-mean square fluctuation in normalized current $\sigma_\Delta$ is defined as the standard deviation of the normalized current $\Delta$ as shown in (52).

$$\sigma_\Delta = \sqrt{\langle \Delta^2 \rangle - \langle \Delta \rangle^2} \tag{52}$$

By substituting (51) into (52), $\sigma_\Delta$ can be represented by (53).

$$\sigma_\Delta = \sqrt{\frac{\partial \langle \Delta \rangle}{\partial (\beta \mathcal{E})}} \tag{53}$$

One can determine $\Delta$ from (53) by optimizing the free energy with respect to $\Delta$, as in (54). The result of this is (55).

$$\frac{\partial G}{\partial \Delta} = J\Delta - J\tanh(\beta J \Delta) = 0 \tag{54}$$

$$\Delta = \tanh(\beta J \Delta) \tag{55}$$

Since $\tanh x = x - \frac{x^3}{3} + O(x^5)$ about $x = 0$, equation (55) can be approximated as (56).

$$\Delta = \frac{J}{kT}\Delta - \frac{1}{3}\left(\frac{J}{kT}\right)^3 \Delta^3 + O((\beta J \Delta)^5) \tag{56}$$

When $kT > J$, the only solution is $\Delta = 0$. The system is in a disordered state and the mean of the normalized current is zero. When $kT < J$, spontaneous symmetry breaking occurs. [20] There are two non-trivial solutions where the system spontaneously picks one of the stable states. Solving the cubic equation (56) gives (57).

$$\Delta = \pm \sqrt{3\left(1 - \frac{kT}{J}\right)} \tag{57}$$

---

[27] On the other hand, for plot in Figure 2 can be interpreted with loop gain $g_m R$ changes, so that eventually, the bistable (solid line) becomes less bistable (in Figure 2, as there is no thermal energy, due to no regeneration/no positive feedback)

[28] Even though $\mathcal{E} = 0$, $\frac{\partial \ln Q}{\partial \mathcal{E}}$ is not 0.

By taking partial derivative of (56) with respect to $\beta\mathcal{E}$, one obtains equation (58).

$$\frac{\partial\Delta}{\partial(\beta\mathcal{E})} = \text{sech}(\beta J\Delta)\left(1 + J\frac{\partial\Delta}{\partial\mathcal{E}}\right)$$
$$= (1 - \tanh^2(\beta J\Delta))\left(1 + \beta J\frac{\partial\Delta}{\partial(\beta\mathcal{E})}\right)$$

(58)

Then, by substituting (56) into (58), one obtains (59).

$$\frac{\partial\Delta}{\partial(\beta E)} = \frac{1 - \Delta^2}{1 - \beta J(1 - \Delta^2)}$$

(59)

By substituting (57) into (59), the partial derivative in (59) can now be expressed in terms of $\beta J$, as shown in (60). For $\beta J \approx 1$ and interaction energy $J$ as given in (24), equation (59) simplifies to (61). The root mean square fluctuation of normalized current is therefore given by (62).

$$\left(\frac{\partial\Delta}{\partial(\beta\mathcal{E})}\right)_\beta = \frac{\left(1 - 3\left(1 - \frac{1}{\beta J}\right)\right)}{1 - \beta J\left(1 - 3\left(1 - \frac{1}{\beta J}\right)\right)}$$

(60)

$$\left(\frac{\partial\Delta}{\partial(\beta\mathcal{E})}\right)_\beta \approx \frac{5}{2(\beta J - 1)} = \frac{5}{2\left(\frac{T_c}{T} - 1\right)}$$

(61)

$$\sigma_\Delta \propto \frac{1}{\sqrt{\frac{T_c}{T} - 1}}$$

(62)

Equation (62) is the fluctuation model for a bistable circuit, such as Schmitt trigger, near the free energy extremum. The extremum corresponds to the metastable point. Referring to Appendix C, the relaxation process has a fast time scale, governed by the dynamic of the Schmitt trigger, and a slow time scale, dictated by the timing capacitor. As the noise spikes occurs during the regeneration process of the fast time scale, we can approximate the fluctuation of the Schmitt trigger to be the same as the fluctuation of the relaxation oscillator.

In order to compare the temperature dependency in Method 1, whose fluctuation formula is given in eq. 49 of [1], from (62) we obtain the relative fluctuation observed per unit time interval (equivalently per frequency interval) and we have:

$$\frac{\frac{\sigma_\Delta}{\sqrt{\Delta f}}}{\Delta} \propto \frac{1}{\sqrt{\frac{T_c}{T} - 1}}$$

(63)

Thus, in summary, the expression for root-mean square fluctuation in normalized current obtained in (63) is consistent with the fluctuation expression in [1]. Both the fluctuation models are obtained with the assumption that the system is in thermal equilibrium. The fluctuation model in [1] compares the equation of states between the van der Waals gas and the relaxation oscillator and maps the corresponding variables to find the fluctuation of current for the relaxation oscillator. However, the equation of states, like van der Waals gas, works with the average pressure, volume and density will miss the microscopic aspect of physics. By investigating the microscopic physic of the circuit, we are able to obtain the thermodynamic quantities such as the free energy of the system, and obtain fluctuation from first principle in physics (3rd row of Table 2). However, when the temperature goes down enough, the underlying physics in the above free energy derivation, i.e. classical, may become invalid as quantum noise might manifest. This is not taken account in the derivation of the fluctuation model in this section. Because of the difficulty of finding eigenenergy in an interaction system, (eigenfunction is needed in the generalization of the free energy calculation method to quantum case), such as the present relaxation oscillator, finding free energy is also difficult, as stated in Table 2. Therefore, section IV introduces Method 3.

## IV. METHOD 3: LANGEVIN EQUATION BASED

Following Table 2 row 3, Method 3 involves the Langevin approach. The salient features of Method 3 are that it is good for system which can be represented by one degree of freedom, similar to the macroscopic approach taken in Section II. However, the microscopic degrees of freedom as discussed in Section III are decoupled from the macroscopic degree of freedom. The microscopic degrees of freedom are represented by immersion in a heat bath. Compared to Method 2, instead of starting from statistical representation of the system at thermal equilibrium, i.e. its free energy (and only the bistable part), and then obtaining statistical properties of the system, such as average and fluctuation of current, Method 3 starts with an equation of motion for the macroscopic degree of freedom in the statistical representation i.e. $\rho(t)$ or density matrix of the excess carrier ensemble. From $\rho(t)$ throughout the period, it then obtains the time properties, such as average and fluctuation. In Langevin approach, the noise is assumed additive and thus the statistical representation is split into a deterministic part and a noise source. Hence one needs to find i) equation of motion of deterministic part ii) find the time average of the system by integrating the noise part contribution along this equation of motion throughout the cycle/period. The equation of motion corresponding to the deterministic part of the relaxation oscillator throughout the whole cycle needs to be obtained.

The system modeled by the Langevin equation is linear and has one degree of freedom, making it easy to solve. The trade-off is that the Langevin equation does not model the non-linear aspects of the relaxation oscillator as discussed in Method 1, and so it is valid only around the region which is intended i.e. the metastable point. The lack of non-linearity means that the resulting equation does not model phase change, or shows the critical temperature $T_c$ as derived in Method 2. Furthermore, to model as a Langevin equation, this requires that in obtaining the deterministic equation of motion, the sign of one of the current equations is flipped. This ensures that the system behaves like a damped harmonic oscillator about $\Delta = 0$.

Separate the oscillator cycle into a fast regeneration phase, and a slow phase where the capacitor $C$ is charging. This separation is shown in Appendix C. Following Appendix C, the large part of noise comes from the fast phase. We now develop the deterministic equation of motion for the fast phase. To determine the equation of motion describing the system, it is simpler use the floating capacitor oscillator shown in Figure 12 instead of the ground capacitor oscillator in Figure 1. Figure 1 and Figure 12 have equivalent circuit topologies. [11]

At regeneration point we ignore the capacitor, $C$ since the voltage across the capacitor is zero. Next, we linearize the circuit since we are only interested in the dynamics of a small region around the metastable state, where the fluctuation is the biggest.

### A. Langevin Equation Setup and Noise Sources

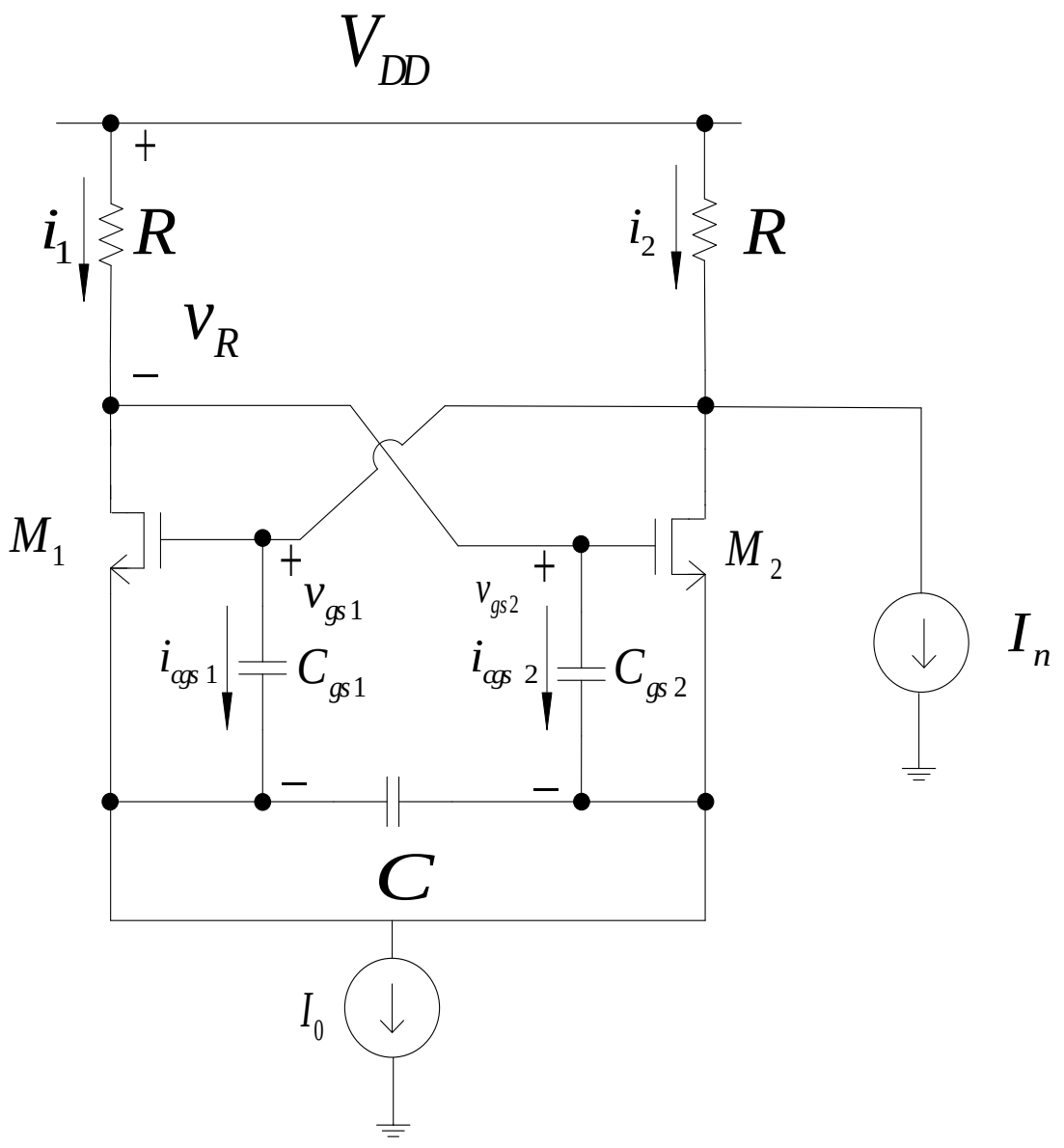

Figure 12. CIRCUIT DIAGRAM OF FLOATING CAPACITOR RELAXATION OSCILLATOR

Like Method 1, this section uses macroscopic tools such as KCL and KVL to develop the deterministic equation of motion. However, we now develop an equation of motion for voltage across one resistor $v_R$ rather than the differential current. This results in the equation of motion being a second-order ODE, instead of a first-order ODE as in Method 1. This is equivalent to combining the equations of motion for the common and differential current modes in Method 1.

By using Kirchhoff's Voltage Law (KVL) around the loop of transistor $M_1, M_2$, equation (64) is obtained, where $i_1$ and $i_2$ are the small-signal currents flowing through the transistors $M_1$ and $M_2$. $v_{gs1}$ and $v_{gs2}$ are the gate-to-source voltages of the transistors. Assume $M_1$ and $M_2$ have the same $g_m$.

$$(i_1 - i_{cgs1})R + v_{gs2} - (i_2 - i_{cgs2})R - v_{gs1} = 0 \tag{64}$$

By setting $v_{gs1} = \frac{i_1}{g_m}$ and $v_{gs2} = \frac{i_2}{g_m}$, one obtains equation (65).

$$(i_{cgs1} - i_{cgs2})R = (i_1 - i_2)\left(R - \frac{1}{g_m}\right) \tag{65}$$

Next we include the parasitic capacitance $C_{gs}$ and its displacement current in our analysis and rewrite equations (64) as (65). From capacitor considerations, $i_{cgs1} = C_{gs1}\frac{dv_{gs1}}{dt}$. Since $v_{gs1} = \frac{i_1}{g_m}$, the relation between $i_1$ and $i_{cgs1}$ is given by (66). $i_{cgs2}$ has a similar derivation.

$$i_{cgs1} = \frac{C_{gs1}}{g_m}\frac{di_1}{dt} \tag{66}$$

Substituting (66) into (65) gives (67). Setting $C_{gs1} = C_{gs2}$ gives (68).

$$R\left(C_{gs1}\frac{di_1}{dt} - C_{gs2}\frac{di_2}{dt}\right) = (i_1 - i_2)(g_m R - 1) \tag{67}$$

$$RC_{gs}\frac{d(i_1 - i_2)}{dt} = (i_1 - i_2)(g_m R - 1) \tag{68}$$

Following [1], we normalize the difference in current and set it to a variable $\Delta = \frac{i_1 - i_2}{I_0}$. This is the same as $\Delta$ in Method 1. Substituting $\Delta$ we have equation (69), which also has roughly the same factor in front of $\Delta$, as (14), and equation (69) lacks a cubic nonlinear term, as expected.

$$\frac{d\Delta}{dt} = \frac{g_m R - 1}{RC_{gs}}\Delta \tag{69}$$

Due to the tail current source, $i_1 + i_2 = I_0$, this means $i_1 = \frac{I_0}{2}(1 + \Delta)$. Using this condition, (69) gives the derivative of $i_1$ as a function of $\Delta$.

$$\frac{di_1}{dt} = I_0\frac{\Delta}{2}\frac{g_m R - 1}{RC_{gs}} \tag{70}$$

Equation (71) gives the derivative of $i_2$ as a function of $\Delta$.

$$\frac{di_2}{dt} = -I_0\frac{\Delta}{2}\frac{g_m R - 1}{RC_{gs}} \tag{71}$$

The coupled first-order ODEs result in a second-order ODE when solving for $i_1$. The solution to the equation is therefore (72).

$$\frac{d^2 i_1}{dt^2} = \frac{1}{2}\frac{g_m R - 1}{RC_{gs}}\frac{di_1}{dt} + \frac{1}{4}\left(\frac{g_m R - 1}{RC_{gs}}\right)^2 (2i_1 - I_0) \tag{72}$$

In order to model the system as a damped harmonic oscillator about $\Delta = 0$, we flip the sign of the second derivative term in (72). Substituting $v_R = -i_1 R$ into (72) gives (73). This equation describes the time evolution of the voltage across the resistor.

$$\frac{d^2 v_R}{dt^2} + \frac{1}{2}\frac{g_m R - 1}{g_m RC_{gs1}}\frac{dv_R}{dt} + \frac{1}{4}\frac{(g_m R - 1)^2}{(g_m R)^2 C_{gs1}C_{gs2}}v_R = 0 \tag{73}$$

Next define a damping coefficient $\gamma$ according to (74). Define an angular frequency $\omega_0$ according to (75).

$$\gamma \equiv \frac{g_m R - 1}{R C_{gs1}} \tag{74}$$

$$\omega_0 \equiv \frac{g_m R - 1}{g_m R \sqrt{C_{gs1} C_{gs2}}} \tag{75}$$

By substituting (74) and (75) into (73), the simplified second order differential equation is given by (76). This is the equation of motion of voltage across the resistor.

$$\frac{d^2 v_R}{dt^2} + \frac{1}{2}\gamma \frac{d v_R}{dt} + \frac{1}{4}\omega_0^2 \, v_R = 0 \tag{76}$$

The transform is (77).

$$v_R(\omega) = \frac{V_{peak}\gamma}{\gamma^2 + (\omega - \omega_0)^2} \tag{77}$$

Next, we consider the noise. From the Langevin equation, the current noise fluctuation across the resistor is defined as $I_n(t)$. As $I_n(t)$ corresponds to colored Gaussian noise, then its noise spectrum is defined according to (78), (79), and (80). [29]

$$\langle I_n(t) I_n(t') \rangle \equiv \alpha(t - t') \tag{78}$$

$$\alpha(\omega) = \int_{\infty}^{\infty} \exp(-i\omega t)\, \alpha(t) \mathrm{d}\omega \tag{79}$$

$$\alpha(\omega) = \frac{\hbar\omega}{R} \coth\left(\frac{\beta\hbar\omega}{2}\right) \tag{80}$$

$\alpha(\omega)$, current spectral density, is the transform of $\alpha(t)$, the auto-correlation function of noise current:

$$\langle \delta E^2 \rangle = \int_0^{T_0} \int_0^{T_0} \langle V\,(t) V\,(t') I_n(t) I_n(t') \rangle \mathrm{d}t' \,\mathrm{d}t \tag{81}$$

Equation (81) essentially approximates the noise of a relaxation oscillator by means of a linearized damped harmonic oscillator.

The Fourier transform of (81) gives (82), where $V_{peak}$ is the initial voltage of the relaxation oscillator at metastable state.

$$\begin{aligned}\langle \delta E^2 \rangle &= \frac{1}{2\pi} \int_{-\infty}^{\infty} |v(\omega)|^2 \alpha(\omega)\, \mathrm{d}\omega \\ &= \frac{1}{2\pi} \int_{-\infty}^{\infty} \left(\frac{V_{peak}\gamma}{\gamma^2 + (\omega - \omega_0)^2}\right)^2 \frac{\hbar\omega}{R} \coth\left(\frac{\beta\hbar\omega}{2}\right) \mathrm{d}\omega\end{aligned} \tag{82}$$

In the high temperature regime, $\beta$ approaches zero. Expanding $\coth\left(\frac{\beta\hbar\omega}{2}\right)$ about $\beta = 0$ yields (83). As a result, the $\hbar\omega$ term from the expansion cancels with $\hbar\omega$ in the integrand in (82). Equation (82) therefore simplifies to (84), and the result of the integral is (85).

$$\coth\left(\frac{\beta\hbar\omega}{2}\right) = \frac{2}{\beta\hbar\omega} + O(\beta\hbar\omega) \tag{83}$$

$$\langle \delta E^2 \rangle = \frac{V_{peak}^2}{\beta R \pi\, \gamma^2} \int_{-\infty}^{\infty} \frac{1}{\left(1 + \left(\frac{\omega - \omega_0}{\gamma}\right)^2\right)^2} \,\mathrm{d}\omega \tag{84}$$

$$\langle \delta E^2 \rangle = \frac{kTV_{peak}^2}{2R}\left(\frac{1}{\gamma}\right)$$
(85)

In the low-temperature limit, $\beta$ approaches infinity, and so coth $\left(\frac{\beta\hbar\omega}{2}\right)$ approaches 1. Substituting this result into (82) yields (86). Equation (86) simplifies to (87).

$$\langle \delta E^2 \rangle = \frac{V_{peak}^2 \hbar}{2\pi R\gamma^2}\int_{-\infty}^{\infty}\frac{\omega}{\left(1+\left(\frac{\omega-\omega_0}{\gamma}\right)^2\right)^2}\mathrm{d}\omega$$
(86)

$$\langle \delta E^2 \rangle = \frac{\hbar\omega_0 V_{peak}^2}{4R}\left(\frac{1}{\gamma}\right)$$
(87)

Next, we obtain the current fluctuation from (87). In the thermodynamic system, the energy fluctuation is given by (88), where $\mu$ is the chemical potential, $N$ is the number of particles (charges), $U$ is the internal energy and $C_v$ is the volume specific heat. During the phase change, the second term of (88) dominates, and the fluctuation of energy can be expressed as (89) [15]. [29]

$$\langle \delta E^2 \rangle = kT^2 C_v + kT\left(\frac{\partial U}{\partial N}\right)_{T,V}\left(\frac{\partial U}{\partial \mu}\right)_{T,V}$$
(88)

$$\langle \delta E^2 \rangle = kT\left(\frac{\partial U}{\partial N}\right)_{T,V}\left(\frac{\partial U}{\partial \mu}\right)_{T,V} + O(T^2)$$
(89)

$$\langle \delta E^2 \rangle = \left(\frac{\partial U}{\partial N}\right)_{T,V}^2 \langle \delta N^2 \rangle$$
(90)

$$\left(\frac{\partial U}{\partial N}\right)_{T,V} = \mu - T\left(\frac{\partial \mu}{\partial T}\right)_{N,V}$$
(91)

The chemical potential of the semiconductor device is the Fermi energy. Since the Fermi level energy does not change significantly within temperature of interest (77K to 273K), the second term of (91) is negligible. Substituting (91) into (88) gives equation (92).

$$\langle \delta E^2 \rangle = \mu^2 \langle \delta N^2 \rangle$$
(92)

For our system (relaxation oscillator) specifically, the potential energy needed to add a charge on the top plate of capacitor is on the order of $eV_{peak}$. Thus, $\mu \propto eV_{peak}$ and the fluctuation in particle number $\langle \delta N^2 \rangle^{\frac{1}{2}}$ is given by (93).

$$\langle \delta N^2 \rangle^{\frac{1}{2}} \propto \frac{1}{eV_{peak}}\langle \delta E^2 \rangle^{\frac{1}{2}}$$
(93)

B. Consistency of Noise Model at High Temperature Regime with Method 1 and 2

In the high-temperature regime, $\beta$ is close to 0. The particle number fluctuation comes from substituting (85) into (93), and with proper scaling, giving:

$$\frac{\left\langle \frac{\delta I^2}{\Delta f}\right\rangle^{\frac{1}{2}}}{I_0} \propto \frac{1}{V_{peak}}\sqrt{\frac{kT}{g_m}\left(\frac{1}{g_m R(g_m R - 1)}\right)}$$
(94)

[29] Note that the following equation is, strictly speaking, developed under thermal equilibrium. Again, using the ergodic theorem, $\langle \delta E^2 \rangle$ is now interpreted as statistical variance (obtained from time variance), and so is interpreted as the variance at thermal equilibrium

The fluctuation of the normalized current obtained in (94) has similar form as equation 49 in [1], obtained from mapping variables with the van der Waals gas (method 1), and (62) from Section III.D (method 2), obtained from free energy.

It is also found that when $g_m R \rightarrow 1$, these equations are consistent. Specifically, the difference between (94) and equation 49 in [1] (and (63)) is due to the fact that equation (94) is derived by linearizing around metastable point.

Equation (94) approaches (63) in the limit of $T_c$ being large, showing that Method 3 is consistent with Method 2. The limit of $T_c$ being large is equivalent to the limit of $g_m R$ being large.

In summary, comparison of present method with methods 1 and 2 agrees at high temperature or classical regime. This gives confidence to the present method, in particular when we use it in the low temperature or quantum regime.

C. Noise Model at Low Temperature Regime and New Noise Contribution from Quantum Effects

1. **Quantum noise: $\hbar\omega_0$ Interpretation**

While (94) captures both the effects of thermal noise and the dynamic of the circuit, fluctuation, as predicted by (94), will go to zero when temperature goes to zero, which is counter-intuitive. Thus, at low temperature, we again use (82), but this time substitute into low temperature approximation, (87):

$$\frac{\left\langle \frac{\delta I^2}{\Delta f} \right\rangle^{\frac{1}{2}}}{I_0} \propto \frac{1}{V_{peak}} \sqrt{\frac{\hbar\omega_0}{g_m}\left(\frac{1}{g_m R(g_m R - 1)}\right)} \tag{95}$$

Unlike (94) this does not have any temperature dependency, which is a characteristic of low temperature model. Its quantum nature is apparent from the factor $\hbar\omega_0$. The exact calculation of this factor, however, deserves further attention. If we use $\omega_0$ as in (76), then the time scale (on the time scale of $(\frac{2\pi}{\omega_0})$ is much larger than the regeneration time (small: see discussion after equation 8 in [11]), which is when noise spike appears. Fluctuation so calculated is inaccurate.

2. **Quantum noise: Effective $\hbar\omega_0$ Interpretation**

To correct the problem of $\omega_0$ being too low, we refine (95) to better estimate $\omega_0$, now denoted as effective $\omega_0$. This is done by recognizing that the noise spike is sharp since it occurs during the regeneration time $t_{reg}$. To capture the high frequency component in the fast regeneration, we heuristically construct a sinusoidal signal with effective $\omega_0$ and amplitude $V_{peak}$. This signal has the same slope at zero crossing as the slope of $\frac{V_{peak}}{t_{reg}}$. The solution to this is to let effective $\omega_0$ be:

$$\omega_{0eff} = \frac{2\pi}{t_{reg}} \tag{96}$$

The sinusoidal signal, which captures the high-frequency aspect of regeneration in the relaxation oscillator, can be interpreted as the resonance frequency of (76) and (77), but with effective $\omega_0$ instead. The noise is then obtained by substituting this into (95) and:

$$\frac{\left\langle \frac{\delta I^2}{\Delta f} \right\rangle^{\frac{1}{2}}}{I_0} \propto \frac{1}{V_{peak}} \sqrt{\left(\frac{2\pi\hbar}{t_{reg}}\right)\frac{1}{g_m}\left(\frac{1}{g_m R(g_m R - 1)}\right)} \tag{97}$$

This is interpreted as the fluctuation arising from the 'equivalent' regeneration in the relaxation oscillator. Next, we define the crossover temperature $T_{cross}$ as the temperature when the high-temperature limit of the noise is equal to the low-temperature limit.[30] The low-temperature noise is (97). The high-temperature noise is (94). Assuming that the proportionality constants and $g_m R$ do not change with temperature, the crossover temperature is:

[30] The proportionality constant can shift $T_{cross}$ but there will still be two regions of temperature dependency

$$T_{cross} = \frac{2\pi\hbar}{kt_{reg}} \tag{98}$$

### 3. Consideration of Quantum Fluctuation Arising from Time Evolution

This section explains macroscopically (Langevin equation/1 degree of freedom) the high/low temperature behavior. The noise behavior was explained microscopically, but only in thermal equilibrium. In equilibrium, the time evolution of the quantum state of the excess carriers was ignored. As $t_{reg}$ decreases, fluctuation due to time evolution is going to play a larger role. $t_{reg}$ comes from simulating the circuit using Eldo. Device-level simulations are also available in Sentaurus. The predicted $t_{reg}$ of $0.4\, ps$ is on the order of the mean free time $\tau_{mean}$ in silicon. This means we are approaching a situation where there is not enough time for scattering processes to establish thermal equilibrium.[31] In order to consider this, we use the density matrix $\rho(t)$. We need to look at its behavior within $t_{reg}$ when it is not necessarily in equilibrium, and observe its fluctuation. A microscopic discussion/many degree of freedom, on the quantum aspect, is presented in this section via an example. We will do a qualitative description using this example. Fortunately, within $t_{reg}$, we can simplify $\rho(t)$(which was mentioned previously as hard to obtain), and make useful observations.

#### *a. Discussion of Effective $\hbar\omega_0$ Using Time Evolution of a Single Electron in a Pure State*

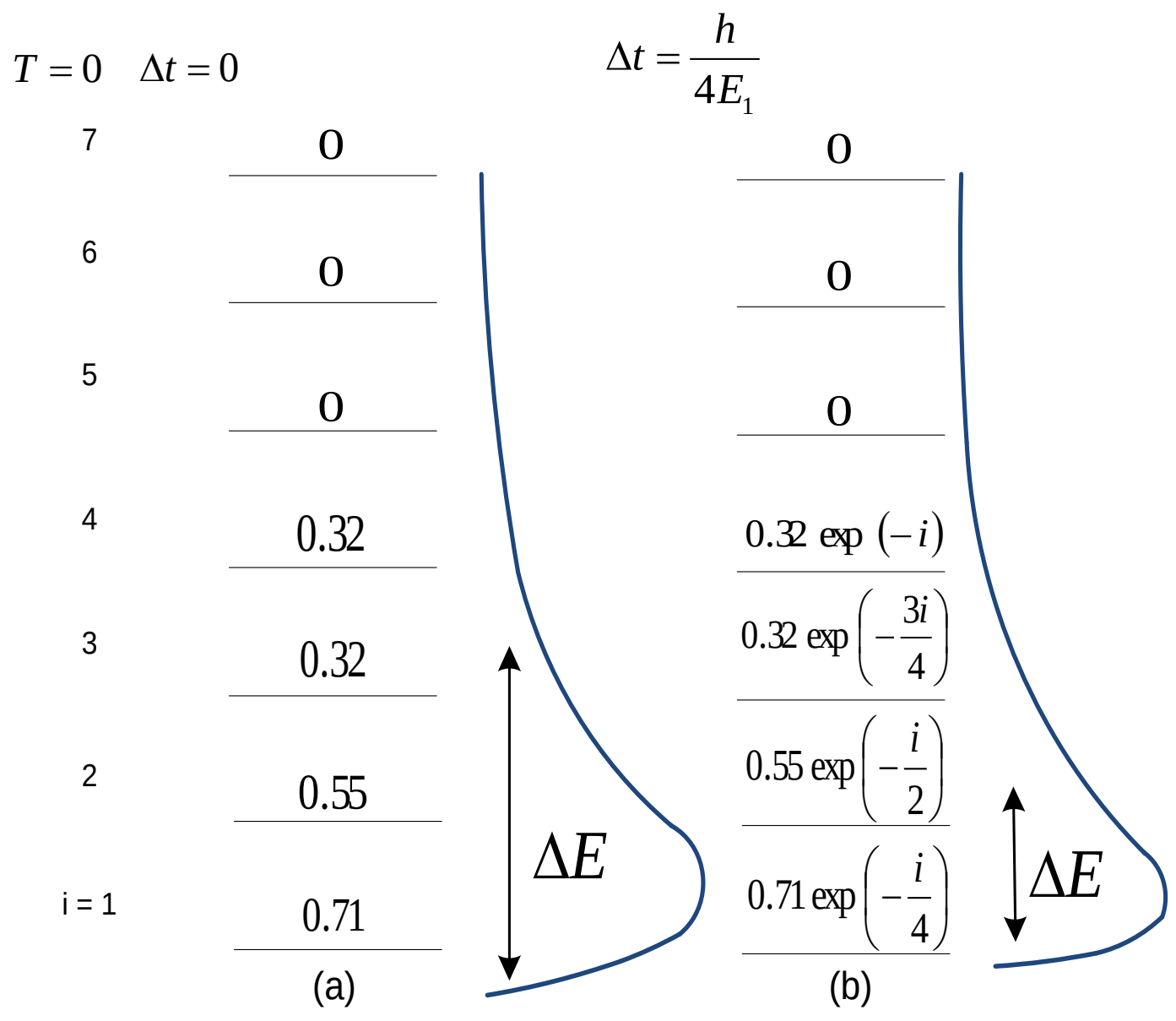

Figure 13. THE TIME EVOLUTION OF A SINGLE ELECTRON $q_1$.

To calculate the fluctuation at the metastable state (both $M_1$, $M_2$ are on and at equal strength), we take the Hamiltonian of the ensemble of excess carriers in this condition (thus this Hamiltonian is different from (25)). This Hamiltonian and the corresponding eigenstates are assumed to remain constant within $t_{reg}$. $|\alpha'\rangle$ denotes an energy eigenstate of this Hamiltonian. Meanwhile the ensemble is represented by the density matrix $\rho$ (background on $\rho$ given in Appendix E). We further assume that the ensemble is non-equilibrium, and that $w_i$ (see Appendix E, (125)) also remains constant during $t_{reg}$, and only has significant change afterwards, when eventually it settles to equilibrium, before the end of the metastable state. Let us start with the eigenenergies of the system (let say on top plate of capacitor $C_{gs}$ of $M_1$. We assume it is in a superposition state i.e. it is given as the superposition of energy eigenstates of the Hamiltonian (e.g. shown as $E_i = \{1, 2, \cdots, 7\}$ in Figure 13 (compared to the electron with energy level $i = 1$ of $M_1$ in Figure 4b; here the electron's $c_n$ (time evolution) is explicitly shown; as will be explained below, here there is $\Delta E$ (due to time evolution)).

[31] As temperature drops $\tau_{mean}$ increases, (which can be further checked by device-level simulations, such as Sentaurus) and beyond $T_{cross}$, we may experience such a situation.

Now as an example, let us consider a single $q_1$, having state $|\alpha\rangle$ (see Appendix D, pt1). Further example of one electron, the expansion coefficients $c_n$ are as shown at $t = 0$. $c = (0.71,\ 0.55, \cdots, 0)$. It then oscillates via $\exp\left(-i\frac{E_\alpha t}{\hbar}\right)$ where $E_\alpha$ is the eigenvalue of $|\alpha\rangle$. $E_1, E_2$ and $E_3$ correspond to $i = 1, 2, 3$ and $c_1, c_2, c_3$. As time progresses, time evolution makes $c_n$ change e.g. at $t = \frac{\hbar}{4E_1}$ (2nd column of Figure 13), the expansion coefficients become $c = \left(0.71e^{-\frac{it}{4}}, 0.55\ e^{-\frac{it}{2}}, \cdots, 0\right)$. The rate of change of the expansion coefficient depends on the energy of the associated eigenstate.

Generalizing the example to $n$ eigenstates, the state $|\alpha\rangle$ has the form $|\alpha\rangle = \sum c_{\alpha'}|\alpha'\rangle$. After some time $t$, the state is given by equation (99).

$$|\alpha, t\rangle = \sum_{\alpha'=1}^{N} c_{\alpha'} \exp\left(-i\frac{E_{\alpha'}t}{\hbar}\right)|\alpha'\rangle \tag{99}$$

Next we calculate the autocorrelation function, given by $C(t) = \langle\alpha, t = 0|\alpha, t\rangle$. When $|\alpha\rangle$ is an energy eigenstate, the autocorrelation function is given by (100). With a continuum of eigenvalues, equation (100) becomes (101). In (101), $\rho(E)$ is the density of energy eigenstates, and $c_\alpha$ maps to $g(E)$. In the continuous case, $g(E)^2\rho(E)$ has a peak around $E_0$, with width $\Delta E$.

$$C(t) = \sum 2|c_{\alpha'}| \exp\left(-i\frac{E_{\alpha'}t}{\hbar}\right) \tag{100}$$

$$C(t) = \exp\left(-i\frac{Et}{\hbar}\right)\int |g(E)|^2\rho(E)\exp\left(-\frac{iEt}{\hbar}\right)\,\mathrm{d}E \tag{101}$$

As $t$ becomes large, the integrand oscillates rapidly unless the energy interval $\Delta E$ is smaller compared with $\frac{\hbar}{t}$. The characteristic time at which the modulus of $C$ starts becoming appreciably different from 1 is given by (102).

$$t = \frac{\hbar}{\Delta E} \tag{102}$$

Equation (102) describes the nature in which the state ket of $q_1$ starts to lose its character (less correlated) from initial ket, as discussed above. Meanwhile looking at the expectation value of some observable $B$, when taken with respect to the basis of a superposition of energy eigenstates (or a non-stationary state, see Appendix D, pt2), then equation (103) gives the expectation value of this observable. As a consequence, $\langle B\rangle$ oscillates with a frequency $\omega_0$ that is related to $E_{\alpha'} - E_{\alpha''}$.

$$\langle B\rangle = \sum\sum c_{\alpha'}\ c_{\alpha''}\langle\alpha'|B|\alpha''\rangle \exp\left(-i\frac{(E_{\alpha'} - E_{\alpha''})t}{\hbar}\right) \tag{103}$$

In our case we take $B$ to be the momentum operator of $q_1$, which gives us the current of $q_1$. Now turning to case of the current coming from ensemble of excess carriers i.e. many electrons $(q_1, q_2, \cdots)$, this is explained in Section IV.C.3.b below, which calculates the current of the ensemble, which again needs the density matrix $\rho$. As previously mentioned, $\rho$ is complicated, and so we apply simplification. To illustrate we begin with some example.

***b. Discussion of Effective $\hbar\omega_0$ via Time Evolution of Two Electrons in a Mixed State***

| | $T=0 \quad \Delta t=0$ | | $\Delta t=\frac{h}{4E_1}$ | | $\Delta t=\frac{h}{E_1}$ | |
|---|---|---|---|---|---|---|
| 7 | 0 | 0 | 0 | 0 | 0 | 0 |
| 6 | 0 | 0 | 0 | 0 | 0 | 0 |
| 5 | 0 | 0 | 0 | 0 | 0 | 0 |
| 4 | $\frac{0.32}{\sqrt{2}}$ | $\frac{0.32}{\sqrt{2}}$ | $\frac{0.32}{\sqrt{2}}\exp(-i)$ | $\frac{0.32}{\sqrt{2}}\exp(-i)$ | $\frac{0.32}{\sqrt{2}}\exp(-4i)$ | $\frac{0.32}{\sqrt{2}}\exp(-4i)$ |
| 3 | $\frac{0.32}{\sqrt{2}}$ | $\frac{0.45}{\sqrt{2}}$ | $\frac{0.32}{\sqrt{2}}\exp\left(-\frac{3i}{4}\right)$ | $\frac{0.45}{\sqrt{2}}\exp\left(-\frac{3i}{4}\right)$ | $\frac{0.32}{\sqrt{2}}\exp(-3i)$ | $\frac{0.45}{\sqrt{2}}\exp(-3i)$ |
| 2 | $\frac{0.55}{\sqrt{2}}$ | $\frac{0.71}{\sqrt{2}}$ | $\frac{0.55}{\sqrt{2}}\exp\left(-\frac{i}{2}\right)$ | $\frac{0.71}{\sqrt{2}}\exp\left(-\frac{i}{2}\right)$ | $\frac{0.55}{\sqrt{2}}\exp(-2i)$ | $\frac{0.71}{\sqrt{2}}\exp(-2i)$ |
| i = 1 | $\frac{0.71}{\sqrt{2}}$ | $\frac{0.45}{\sqrt{2}}$ | $\frac{0.71}{\sqrt{2}}\exp\left(-\frac{i}{4}\right)$ | $\frac{0.45}{\sqrt{2}}\exp\left(-\frac{i}{4}\right)$ | $\frac{0.71}{\sqrt{2}}\exp(-i)$ | $\frac{0.45}{\sqrt{2}}\exp(-i)$ |
| | (a) | | (b) | | (c) | |

Figure 14. THE TIME EVOLUTION OF AN ENSEMBLE OF TWO ELECTRONS $q_1$ AND $q_2$ WITH $w_1 = w_2 = \frac{1}{2}$

Next we describe the ensemble using $\rho$. [30] The background on density matrix $\rho$ is reviewed in Appendix E. Applying that to our present situation, as a simple illustration we look at a description of an ensemble with two electrons, $q_1, q_2$ on the top plate of $C_{gs}$, as described in Figure 14, where the electrons are in states $|\alpha_1\rangle$ and $|\alpha_2\rangle$, respectively.

At metastable state this is as shown in Figure 14a. Each e describes a pure state. Since the expansion coefficients are different for 2e, there are two different pure states. Thus the two-electron ensemble is in a mixed state. The weight of each of them is $w_i$ $(w_1, w_2)$ and as stated above since the ensemble is at non-equilibrium, this weight does not follow a Boltzmann distribution.

Assume $w_1 = w_2 = \frac{1}{2}$, and as stated above, remains constant during $t_{reg}$. The time evolution of the state kets $|\alpha_1\rangle$ and $|\alpha_2\rangle$ are just like the single electron case. Thus Figure 14b,c shows the two kets dependency on time, which is like Figure 13a,b,c.

Since from above we assume the ensemble distribution remains constant during $t_{reg}$, then $w_1 = w_2 = \frac{1}{2}$ (In general $w_i$, usually changes with time and the resulting density matrix follows the von Neumann equation). Later on it settles (via e.g. phonon scattering) into thermal equilibrium (resulting $\rho_{kk}$ has Boltzmann factor), and the ensemble of electrons settles into a Fermi-Dirac distribution. For example, referring to Figure 7 ($T > 0$, 1st-3rd picture, $M_1$), the occupancy $1000110 \rightarrow 0011100 \rightarrow 0101010$ describes a Fermi-Dirac distribution. After the density matrix settles beyond $t_{reg}$, the fluctuation depends on $kT$, instead of $\hbar\omega_0$. The fluctuation is thermal, rather than quantum, and no longer depends on the time evolution.

Now back to the actual ensemble of excess carrier, whose fluctuation is calculated by using density matrix $\rho$ of the ensemble. This fluctuation would allow us to measure fluctuation of the current $j$.

$j$ is proportional to the gradient or momentum operator $p$. Thus the current $j$ is evaluated using the momentum operator $p$. Thus $\langle p\rangle$, like $\langle B\rangle$, oscillates with effective frequency $\omega_0$, as given in (103). Next take a look at fluctuation [16] of $j$ via fluctuation of $p$, as shown in (104).

$$\Delta p = \langle (p - \langle p\rangle)^2\rangle^{\frac{1}{2}} = \sqrt{\langle p^2\rangle - \langle p\rangle^2} \tag{104}$$

Since $\langle p\rangle$ oscillates with frequency $\omega_0$, $\langle p^2\rangle$ also oscillates and therefore $\Delta p$ also oscillates, like $\Delta E$. In a time frame less than $\frac{\hbar}{\Delta E}$, the $c_{\alpha'}c_{\alpha''}$, terms in $\sum\sum c_{\alpha'}c_{\alpha''}$ corresponding to $\Delta E$ do not cancel enough and therefore contribute to $\langle B\rangle$ (for that matter to $\langle p\rangle, \langle p^2\rangle$ as well). [32] Thus this group of terms is responsible for calculation of $\Delta p$. Thus $\Delta p$, or ($\Delta j$, $\Delta i$) is a function of $\frac{\hbar}{t}$ or effective $\hbar\omega_0$.

As an example, pictorially, as $\Delta E$ is spread around eigenenergy $E_\alpha(i = 1)$ in Figure 13a (with $\Delta E$ ranging from $i = 1$ to $i = 4$), then the set $\{c_1^*c_1, c_2^*c_2, \cdots, c_6^*c_6\}$ ($c_5^*c_5$ and $c_6^*c_6$ are both zero, and so do not contribute to the sum in this case) are involved in the calculation of $\sum\sum c_{\alpha'}c_{\alpha''}$, or contributing in (103), and hence $\Delta i$. As time progresses $\Delta E$ shrinks (say at $t = \frac{\hbar}{4E_1}$ due to the phase factor $\exp\left(-i\frac{E_{\alpha'}t}{\hbar}\right)$). Note in $t = \frac{\hbar}{4E_1}$, the magnitude remains $c = (0.71,\ 0.55, \cdots, 0)$, but the phase changes, and so the contributing group shrinks. (This is the time uncertainty principle). Equating this to $kT$ at $T = T_{cross}$ also gives $T_{cross} = \frac{2\pi\hbar}{kt_{reg}}$. [33]

In summary, even if quantum noise does arise from time evolution, $T_{cross}$ remains as given by (98)

## V. Experimental Results

Three relaxation oscillators based on the designs $RO_1$ and $RO_2$ were fabricated using $0.13\ \mu m$ CMOS technology, with microphotograph shown in Figure 15. Table 3 gives their design parameters. (I is 100uA and $V_{peak}$ is around 100mV for RO1, 2). W/L ratio is large so that effect of 1/f noise (not our focus) is mitigated. The $\left(\frac{W}{L}\right)_{1-2}$ of $RO_2$ is twice the $\left(\frac{W}{L}\right)_{1-2}$ of $RO_1$, doubling $C_{gs}$. Next, the current fluctuation of the relaxation oscillator is measured when the temperature varies. The cryogenic experiment setup is shown as Figure 16.

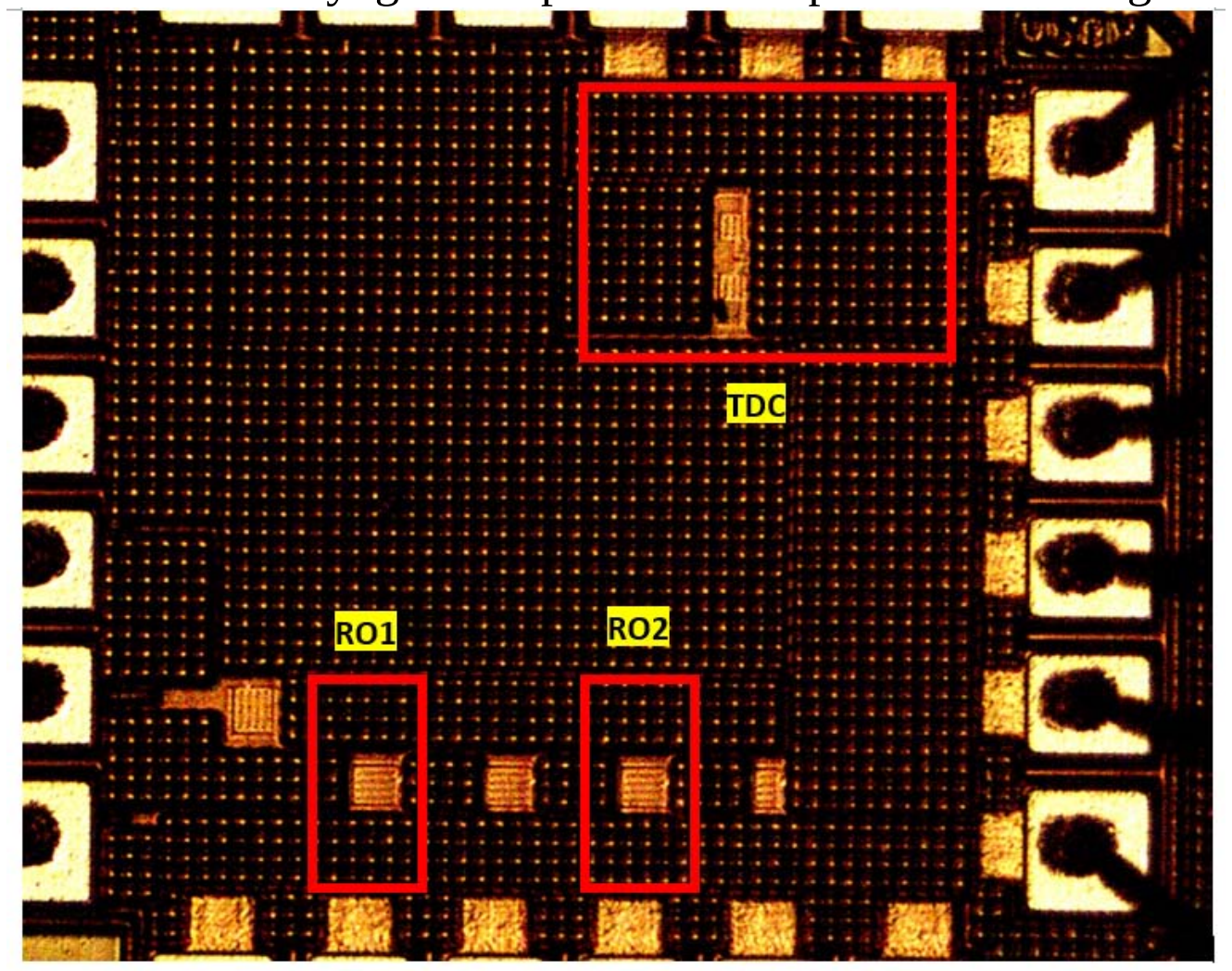

Figure 15. Microphotgraph of the Relaxation Oscillators $RO_1$ and $RO_2$

TABLE 3 Design Parameters for $RO_1$ and $RO_2$

| Parameter | $RO_1$ Value | $RO_2$ Value |
|---|---|---|
| $W(\mu m)$ | 7 | 14 |
| $L(\mu m)$ | 0.13 | 0.13 |
| $R(k\Omega)$ | 5 | 5 |
| $C(pF)$ | 0.4 | 0.4 |
| $g_m R$ | 1.1 | 1.56 |

The jitter is given as rms noise voltage in timing waveform divided by slope of waveform, at regeneration point. We assume as value of noise spike increases, rms noise voltage also increases, and thus jitter increases (assuming slope of waveform or slew rate (SR) stays constant (see Appendix F)). Thus, jitter value is a reflection of value of noise spike $\frac{\left\langle\frac{\delta I^2}{\Delta f}\right\rangle^{\frac{1}{2}}}{I_0}$. Therefore, the phase noise $\frac{\left\langle\frac{\delta I^2}{\Delta f}\right\rangle^{\frac{1}{2}}}{I_0}$ at a fixed frequency offset is used to represent $\frac{\langle\delta I^2\rangle^{\frac{1}{2}}}{I_0}$ and thus the jitter and over the temperature range. For $RO_1$, the ambient temperature of the relaxation oscillators ranges from $77\ K$ to $300\ K$. For every temperature step, the phase noise at $1\ MHz$ offset (so that effect of $\frac{1}{f}$ noise is mitigated) of the $RO_1$ oscillator frequency is captured using Keysight N9010B EXA Signal Analyzer. At $77\ K$, the phase noise (log plot) of $RO_1$ is captured and shown in Figure 17.

[32] As a side note, the present situation is non-equilibrium; but the fluctuation is due to insufficient cancellation of $\Delta E$ (and hence $\Delta p$), due to different rate of time evolution. Meanwhile back in (61), the situation is in thermal equilibrium which relates $\Delta N$ to $\Delta E$(derived under thermal equilibrium), arising from noise source in (61) (also arises from thermal equilibrium).

[33] From [17], for 130nm CMOS down to $77K$, the transient time simulation remains within same order of magnitude, using $300K$ device model.

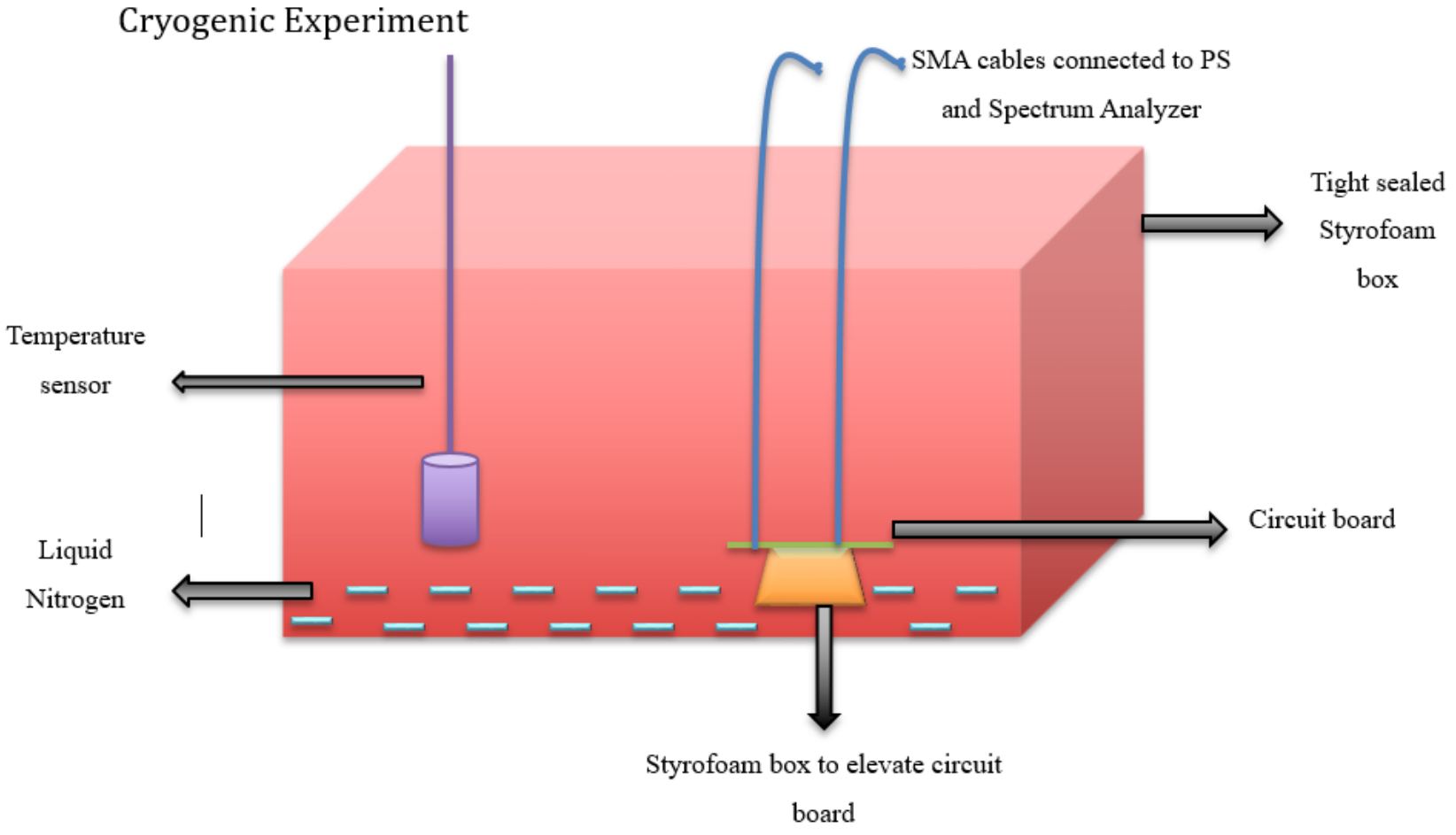

Figure 16. A BLOCK DIAGRAM OF THE APPARATUS USED TO COOL THE RELAXATION OSCILLATORS TO $77\ K$

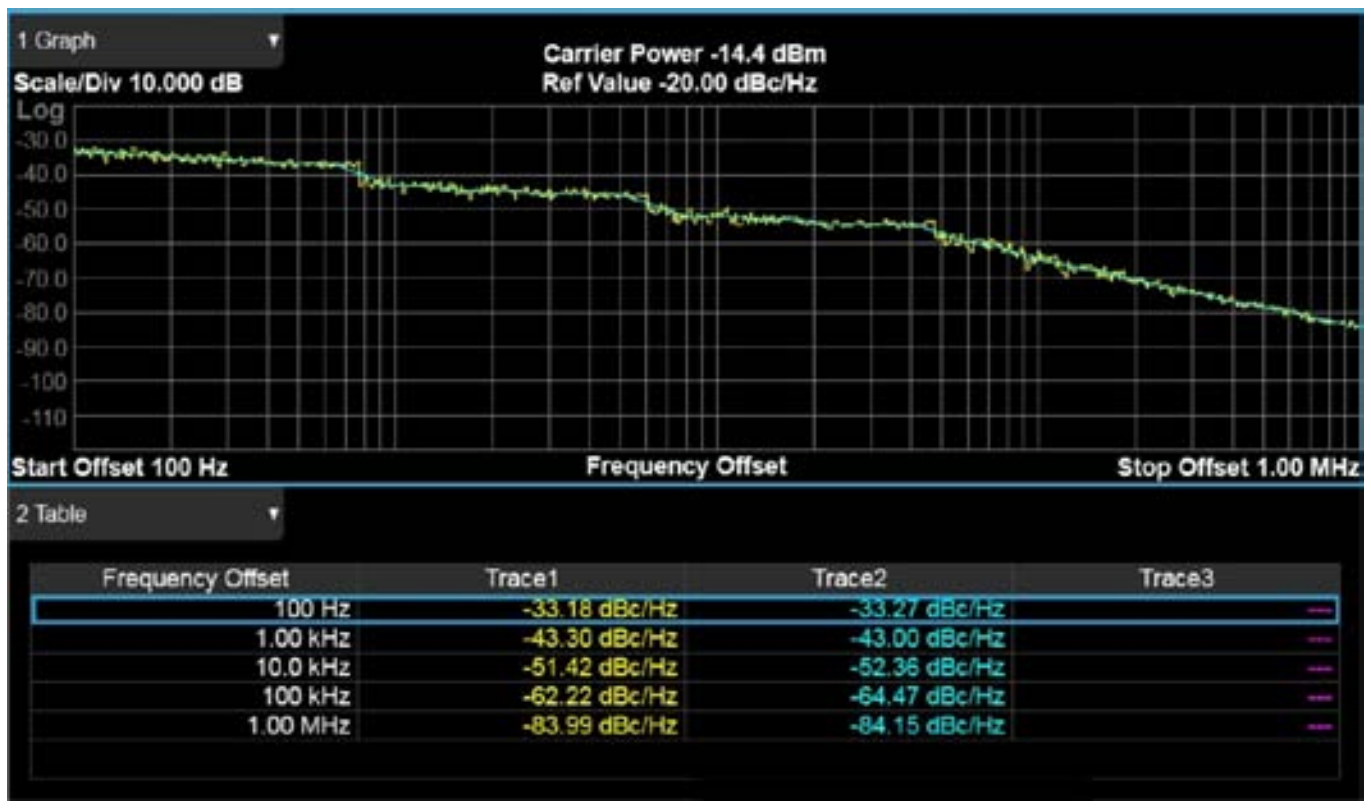

Figure 17. THE MEASURED PHASE NOISE AS A FUNCTION OF FREQUENCY OFFSET OF $RO_1$, CHIP 1 AT $77\ K$

The phase noise (at offset frequency of 1MHz from carrier frequency) of $RO_1$ (chip 1) versus temperature is shown in Figure 18. The trend is consistent with noise behaviors' dependency on temperature in two different ranges. At high temperature, $PN$ varies monotonically with temperature, as in (63). At low temperature, $PN$ stays roughly constant with temperature as in (97).[34] The trend, as given in (63) is plotted as the red dotted line, and the trend as given in (97) is plotted as the blue solid line. These trends agree rather well with experiment results. The crossover temperature $T_{cross}$, depends on value of effective $\hbar\omega_0$ and will be discussed later on. The oscillation frequency is measured as $41\ MHz$.

The phase noise (at offset frequency of $1\ MHz$ from carrier frequency) of $RO_1$ (chip 1 and chip 2) versus temperature are shown in Figure 19. With same design parameters, the noise difference is due to process variation (they oscillate at slightly different frequencies. $RO_1$ oscillates at $41\ MHz$, and $RO_1$ on the second chip oscillates at $40\ MHz$).

[34] a) The proportionality constant in the model is selected so that it fits (least square fit) the experimental results

b)The dominant $T$ dependency comes from the $kT$ factor. Any device temperature dependency from $g_m$ or $R$ is estimated using simulated (Eldo) results (see Appendix F; see also [33]). These dependencies are not significant compared to that in (82) than the relationship presented here, and are therefore neglected for simplification.

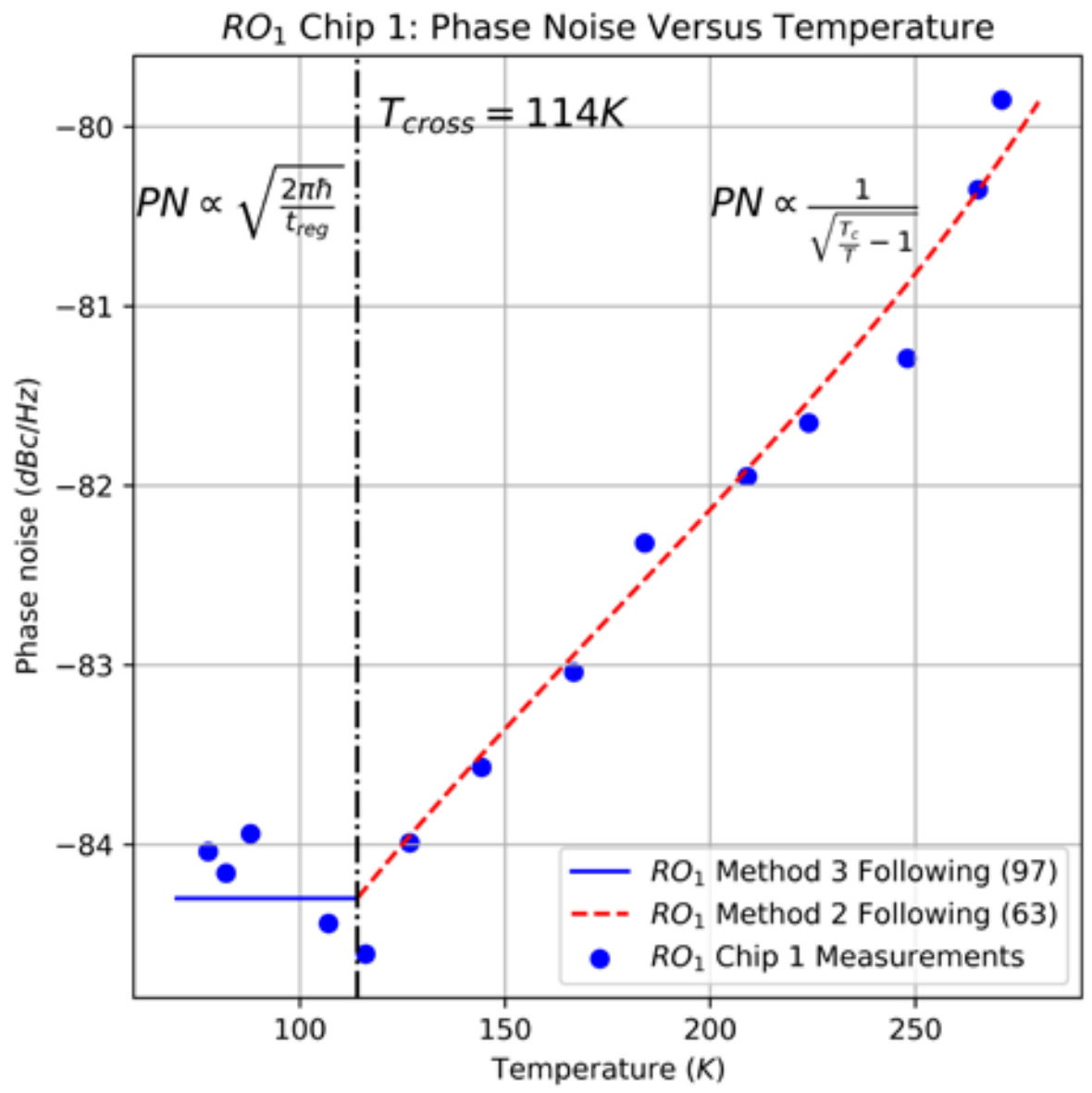

Figure 18. COMPARISON BETWEEN EXPERIMENT DATA OF $RO_1$ CHIP 2 WITH THE TWO REGIONS PHASE NOISE MODEL.

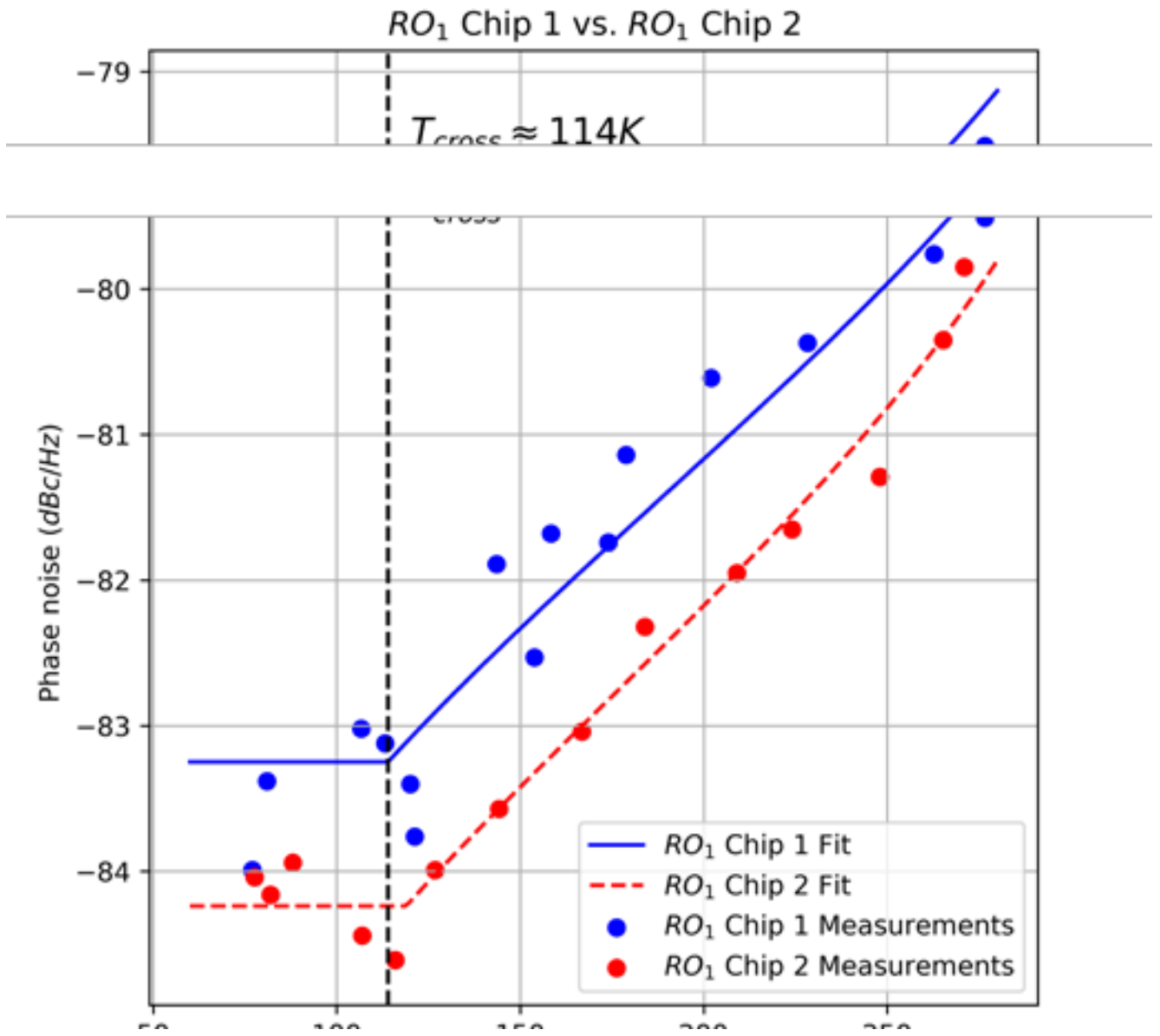

Figure 19. COMPARISON BETWEEN EXPERIMENT DATA AND MODEL FIT OF $RO_1$ CHIP 1 AND $RO_1$ CHIP 2.

The phase noise (at offset frequency of $1\,MHz$ from carrier frequency) of $RO_1$ (Chip 1 and Chip 2) and $RO_2$ (chip 1) versus temperature are shown in Figure 20. Here it is seen that the trend for $RO_2$ Chip 1 is the same as $RO_1$ Chip1 and $RO_1$ Chip 2, but the crossover temperature is lower. This is consistent with the lowering of effective $\hbar\omega_0$, since larger $C_{gs}$ results in a longer $t_{reg}$ and thus lower crossover temperature. In this case, the crossover temperature falls from $120\,K$ to $100\,K$.[35] Eldo simulation estimates $t_{reg} = 0.4\,ps$ for $RO_1$ and $t_{reg} = 0.5\,ps$ for $RO_2$, with swing around $100\,mV$.[36] From Figure 20 this estimates an effective $\hbar\omega_0$. Equating this energy to $kT$ in (94) gives an estimate of the crossover temperature. For $RO_1$, the crossover temperature is $114\ K$ and for $RO_2$ this is given as $96\ K$.[37]

[35] Measurement is extrapolated due to buffer design.
[36] This is around the overdrive voltage

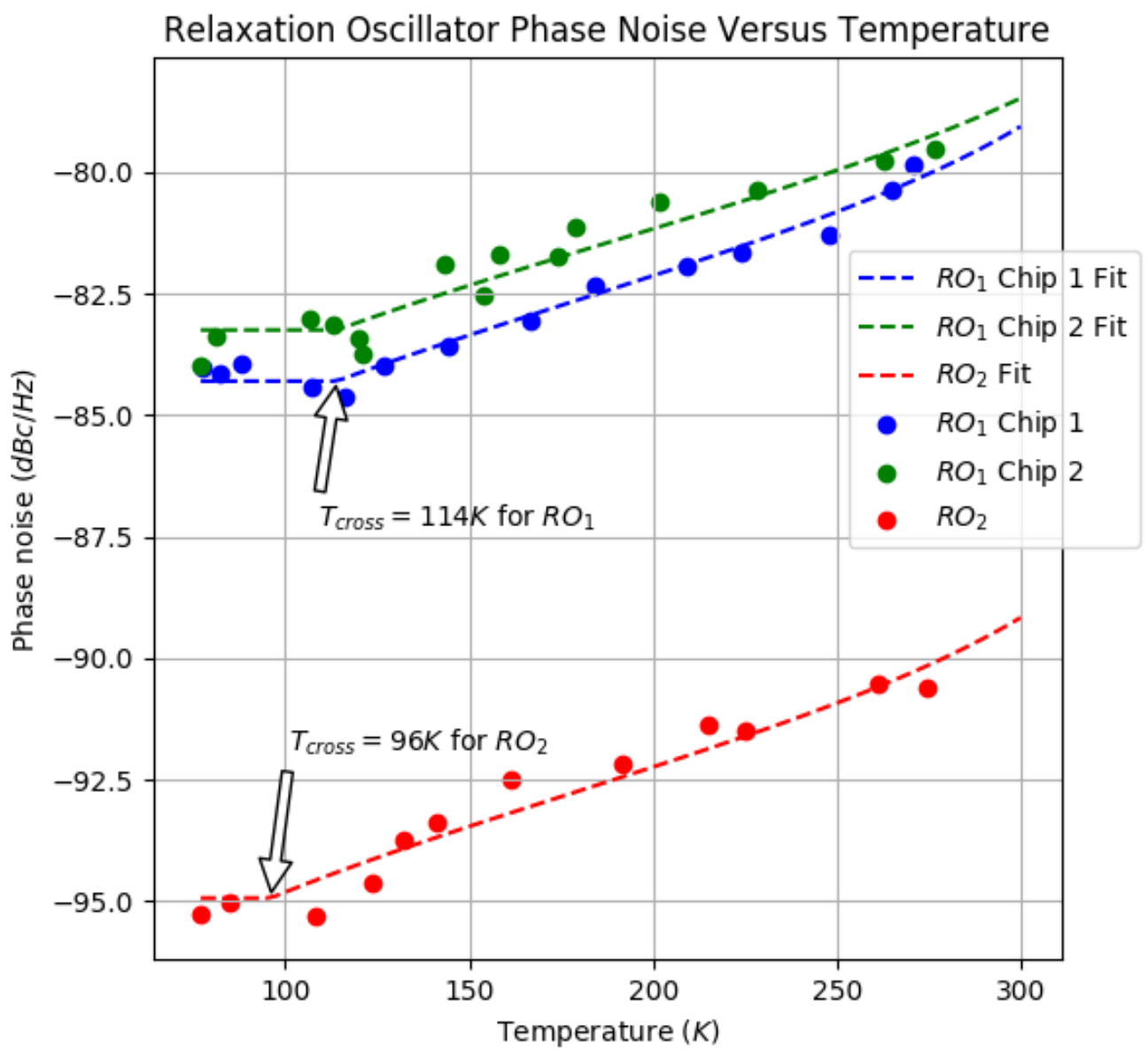

Figure 20. COMPARISON BETWEEN EXPERIMENT DATA OF ALL THREE RELAXATION OSCILLATORS.

Finally $RO_1$ from two experiment runs are compared in Figure 19. The oscillation frequency of $RO_1$ Chip 1 is $55\ MHz$ and the oscillation frequency of $RO_1$ Chip 2 is 40 $MHz$.

## VI. CONCLUSION

The noise spike model of relaxation oscillator with design parameter loop gain close to 1 is investigated as temperature varies (range down to 77 $K$), for application in a Time-to-Digital Converter (TDC).

In Section II Method 1 is performed with the underlying assumption that the system can be represented by one degree of freedom. The physical interpretation of the potential energy is given to be of the difference in electrostatic potential energy between the gate-to-source capacitors in transistors, $C_{gs1}$ and $C_{gs2}$. The system consists of the electrons on the top plates of $C_{gs1}$ and $C_{gs2}$, whereas these electrons travel through the resistors and experience the interaction energy from the cross-coupled topology of the transistors. It is shown that the metastable state has the highest potential energy, and through regeneration, will have the energy lowered to the lowest potential energy and settle to a stable state. It is found that the potential energy with this nature is due to the symmetrical cross-coupling of the electronic circuit, thus relating the potential energy with the design parameters of the electronic circuit. The limitation of Method 1 is that it works only with system that has a corresponding thermodynamic system to map to. It does introduce energy concept and sets the stage for method 2, a more general method, using the free energy method, which is general for any thermodynamic system.

Specifically Method 1 does not explicitly investigate the microscopic aspect (e.g. charges distribution over energy levels in phase space/eigenenergies). This limitation is overcome in Section III (Method 2) by investigating the microscopic behavior (e.g. different configuration in which these internal degrees of freedom manifest). The free energy obtained shows the dependency of temperature in the equation. From the free energy, the noise model is obtained and it is consistent with the existing noise model in [9], but with more insights on the thermodynamic quantities. However, when the temperature goes down enough, Method 2 (i.e. classical) becomes invalid as the quantum noise becomes important.

An alternative, Method 3, based on Langevin approach, is presented in IV. The system is represented by one degree of freedom but with the internal degree of freedom decoupled and assumed to be represented by the immersion of the system in a heat bath with many degrees of freedom. The manifestation of the quantum noise, $h\omega_0$, is obtained in the noise source in the low temperature limit, with $\omega_0$ characterizes the regeneration time of relaxation oscillator.

[37] Even though the proportionality constant can change $T_{cross}$ on $RO_1$ and $RO_2$, the shift from $RO_1$ to $RO_2$, and most importantly, the trend of the shift remains.

The experimental results, presented in Section V show the phase noise of the relaxation oscillator has two regions and that there exists a crossover temperature, above which the phase noise is a monotonic increasing function of $T$ and below which the noise is relatively constant.

## ACKNOWLEDGMENTS

The authors acknowledge help from Profs. Heunis, Leggett, Majedi, Mann, McCourt, Nielsen, Roy, and Wickham.

## APPENDIX A PHASE PORTRAIT OF THE SCHMITT TRIGGER, SHOWING STABLE AND METASTABLE STATES

The crossover temperature between these both regions is investigated. With the design parameter of the relaxation oscillator varies, the crossover temperature shifts as predicted by the noise model. This is shown experimentally using 0.13um CMOS technology over a temperature range of 300K to 77K. In this appendix, to model the Schmitt trigger as a 2 level system, both $C_{gs1}$ and $C_{gs2}$ are explicitly shown, as in Figure 1. At the metastable state, $i_{d1} = i_{d2} = I_0$, $R_1 = R_2 = R$, and $k_{n1} = k_{n2} = k_n$. Assume long-channel approximation (ignore channel length modulation). The circuit is now described by two coupled first order differential equations that include $C_{gs1}$ and $C_{gs2}$. By applying KCL at the drain nodes of $M_1$ and $M_2$, equations (105), (106) and (107) are obtained:

$$i_{cgs1} = i_{R2} - i_{d2} \tag{105}$$

$$i_{cgs2} = i_{R1} - i_{d1} \tag{106}$$

$$i_{R1} + i_{R2} = I_0 \tag{107}$$

Next applying KVL around the loop of resistors, $R_1$, $R_2$ and transistors, $M_1$, $M_2$, one obtains (108).

$$i_{R1} - i_{R2} = \frac{v_{gs2} - v_{gs1}}{R} \tag{108}$$

By symmetrizing around the common mode signal, $\frac{I_0}{2}$, $i_{R1}$ and $i_{R2}$ are obtained. The results are (109) and (110).

$$i_{R1} = \frac{I_0}{2} + \frac{v_{gs2} - v_{gs1}}{2R} \tag{109}$$

$$i_{R2} = \frac{I_0}{2} - \frac{v_{gs2} - v_{gs1}}{2R} \tag{110}$$

The capacitor currents, $i_{cgs1}$ and $i_{cgs2}$ are related to voltages $v_{gs1}$ and $v_{gs2}$ as shown below in (111) and (112).

$$i_{cgs1} = C_{gs1} \frac{dv_{gs1}}{dt} \tag{111}$$

$$i_{cgs2} = C_{gs2} \frac{d_{vgs2}}{dt} \tag{112}$$

Transistor device equations using long channel approximation (i.e. square law) give (113) and (114).

$$v_{gs1} = \sqrt{\frac{2\, i_{d1}}{k_n}} \tag{113}$$

$$v_{gs2} = \sqrt{\frac{2\, i_{d2}}{k_n}} \tag{114}$$

The current flowing through the transistors, $i_{d1}$ and $i_{d2}$ can be related to $v_{gs1}$ and $v_{gs2}$ through equations (115) and (116).

$$\frac{di_{d1}}{dt} = k_n\left(\sqrt{\frac{2i_{d1}}{k_n}} - v_t\right)\frac{dv_{gs1}}{dt}$$

(115)

$$\frac{di_{d2}}{dt} = k_n\left(\sqrt{\frac{2i_{d2}}{k_n}} - v_t\right)\frac{dv_{gs2}}{dt}$$

(116)

By substituting (109)-(116) into (107) and (108), two coupled differential equations are obtained as shown in (117), (118).

$$C_{gs1}\frac{di_{d1}}{dt} = \frac{1}{k_n\left(\sqrt{\frac{2i_{d1}}{k_n}} - v_t\right)}\left(\frac{I_0}{2} + \frac{1}{2R}\left(\sqrt{\frac{2id_2}{k_n}} - \sqrt{\frac{2i_{d1}}{k_n}}\right) - i_{d2}\right)$$

(117)

$$C_{gs2}\frac{di_{d2}}{dt} = \frac{1}{k_n\left(\sqrt{\frac{2i_{d2}}{k_n}} - v_t\right)}\left(\frac{I_0}{2} - \frac{1}{2R}\left(\sqrt{\frac{2id_2}{k_n}} - \sqrt{\frac{2i_{d1}}{k_n}}\right) - i_{d1}\right)$$

(118)

Then, the phase portrait of an example of bistable system described in (117), (118) is plotted in Figure 21. It shows that the Schmitt trigger has one metastable state at $\left(\frac{i_{d1}}{I_0}, \frac{i_{d2}}{I_0}\right) = (0.5, 0.5)$ and two stable states at $\left(\frac{i_{d1}}{I_0}, \frac{i_{d2}}{I_0}\right) = (0, 1)$ and $\left(\frac{i_{d1}}{I_0}, \frac{i_{d2}}{I_0}\right) = (1, 0)$. Hence it behaves as a two-level system. It is also shown that the dynamic of the system is symmetrical at $i_{d1} = i_{d2}$.

Figure 22 shows the dynamic of the system going down from either side of the metastable state. As shown in Figure 22, the system trajectory either moves toward the stable state $\left(\frac{i_{d1}}{I_0}, \frac{i_{d2}}{I_0}\right) = (0, 1)$, or moves toward the stable state $\left(\frac{i_{d1}}{I_0}, \frac{i_{d2}}{I_0}\right) = (0, 1)$. The time evolution of the normalized currents, $\left(\frac{i_{d1}}{I_0}, \frac{i_{d2}}{I_0}\right)$ or the case of Figure 22a is shown in Figure 23.

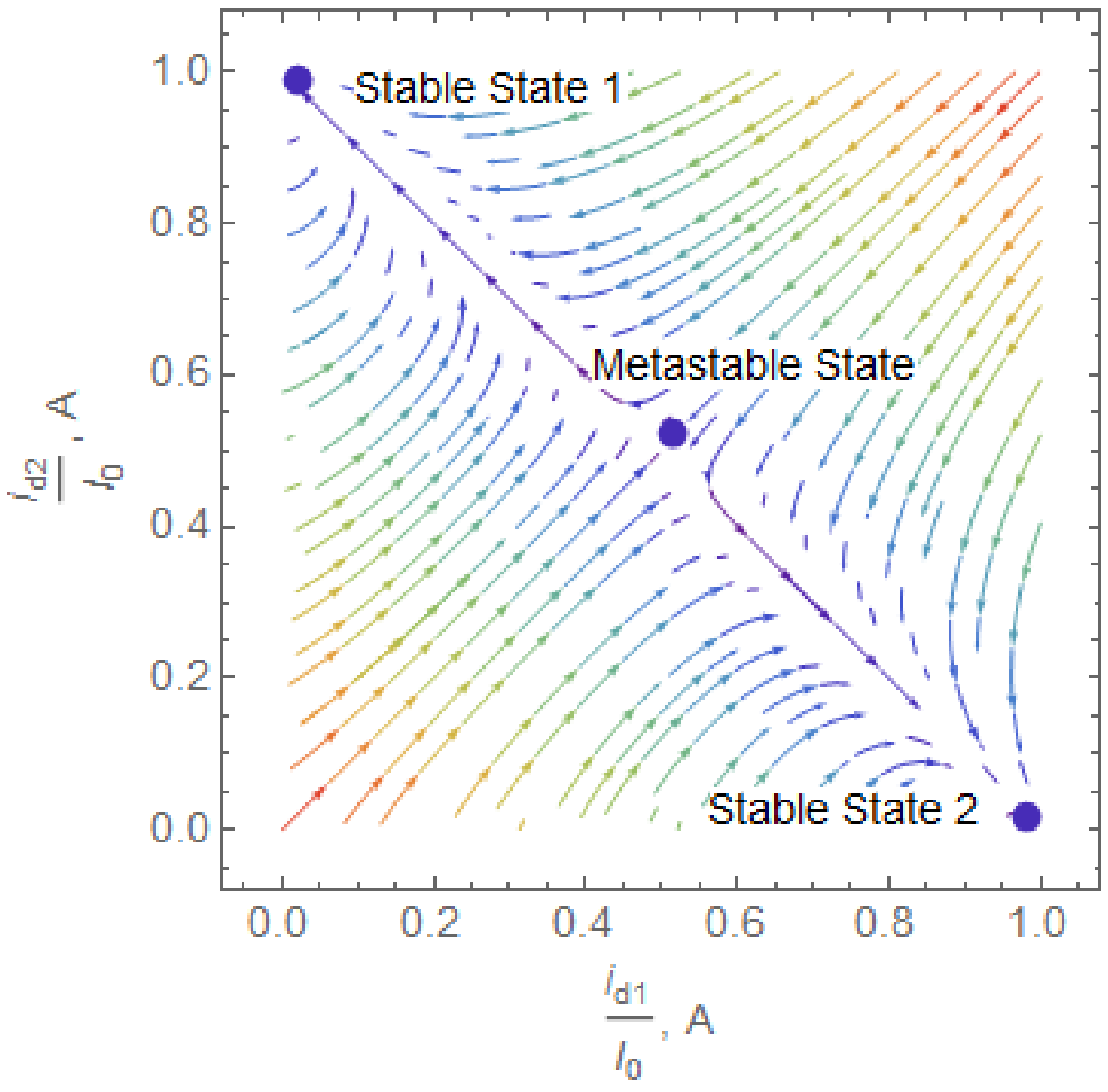

Figure 21. PHASE PORTRAIT OF (117),(118) USING NORMALIZED CURRENTS $\frac{i_{d2}}{I_0}$ AND $\frac{i_{d1}}{I_0}$

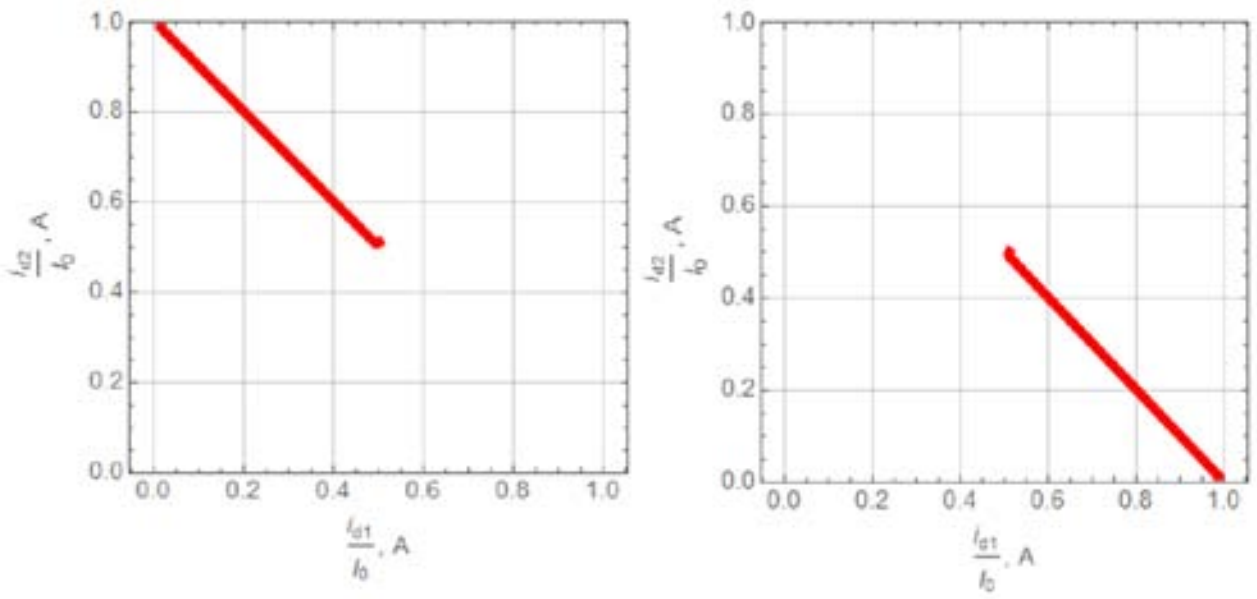

Figure 22. PHASE CURVE OF NORMALIZED CURRENTS, $\frac{i_{d2}}{I_0}$ VERSUS $\frac{i_{d1}}{I_0}$

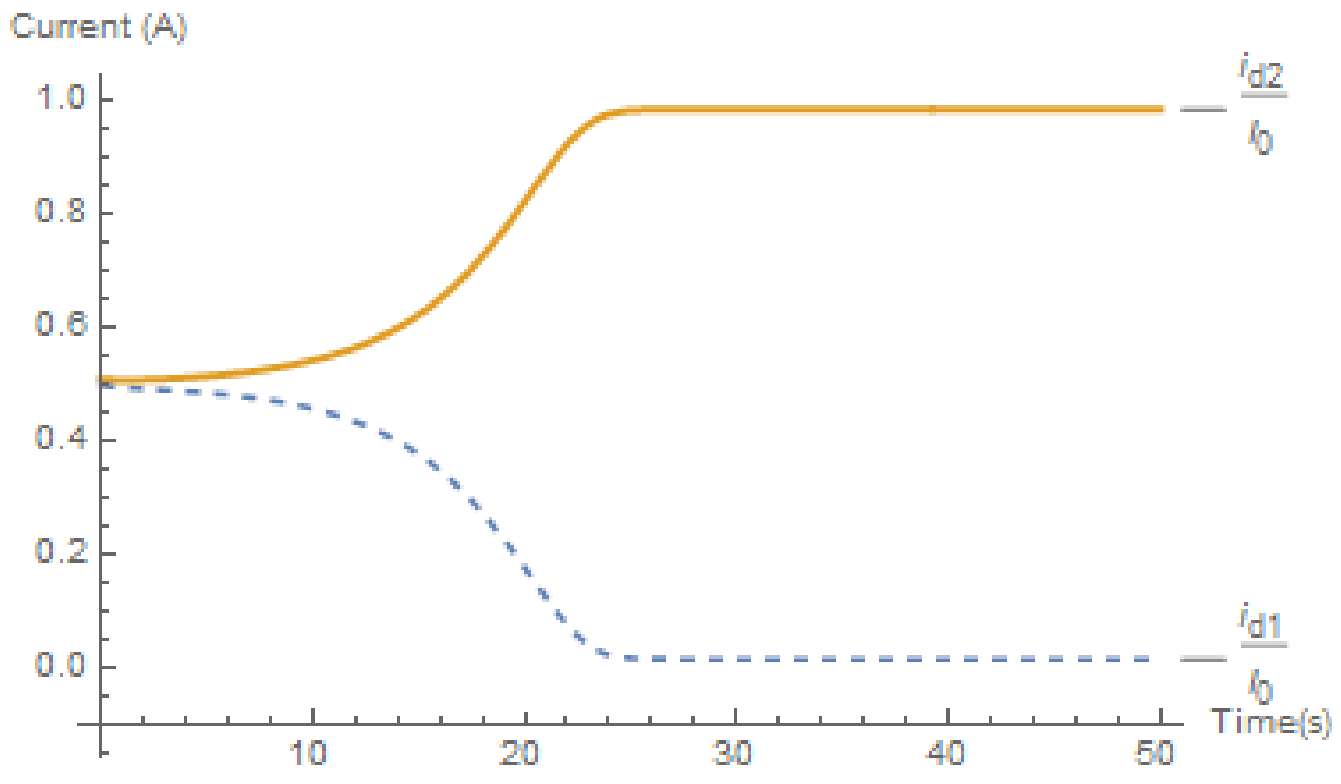

Figure 23. TIME EVOLUTION OF CURRENT, $\frac{i_{d1}}{I_0}$ AND $\frac{i_{d2}}{I_0}$ VS TIME

## APPENDIX B ENERGY LEVELS IN MOMENTUM AND POSITION SPACE

Figure 24 describes the $E-k/E-x$ diagram of standalone transistor $M_1$ (or $M_2$), with $V_{ds}=0$, starting with $V_{gs}=0$ (intrinsic, then doped), then $V_{gs} = V_t$; then $V_{gs} > V_t$, where $V_t$ is the threshold voltage. . The eigenenergies of excess charge carriers are $E(k)=\frac{\hbar^2k^2}{2m^*}$ where $m^*$ is effective mass and $k$ is the wave number, leading to the discrete distribution E(k) in Figure 24. Note $E_f - E_c$ corresponds to $V_{gs}-V_t$ rather than $V_{gs}$, as $V_t$ accounts for the depletion ions, and does not occupy the energy level in $E_f - E_c$.

With an applied $V_{ds}$, Figure 25 describes the $E-k$ diagram at the source and pinch off of transistor $M_1$ as a standalone transistor (i.e. not part of a cross-coupled pair).[38] Furthermore with the capacitor connected to a heat bath or reservoir in the form of the resistor $R$, temperature $T$ is established. The electrons in the semiconductor are distributed across these energy levels, with distribution described by Fermi Dirac statistics centered around the Fermi energy $E_f$. The excess charge is then accounted for by a shift $E_f$ to the new Fermi level, with the shift determined by $n$, the amount of excess charge in the inversion layer.

At source, $E_c$, the energy of the electron, is set to 0 via the inversion layer to source connection.

When $V_{ds} \neq 0$, there is a resultant electric field $\mathcal{E}$, giving rise to drift velocity $v_d$, drift momentum $k_d$. There is no interaction energy between the different electrons occupying the different energy levels.

Around drain end, and at pinch off, $E_c$, the energy of the electron, is set by $V_{gs}-V_t$, as shown in Figure 25.

Meanwhile distance between the Fermi level and $E_c$, goes from $V_{gs}-V_t$ at source end to 0 at pinch off, and so the average number of carriers is proportional to $\frac{1}{2}(V_{gs}-V_t)$. The current density is given by equation (119), with electric field denoted by $\mathcal{E}$.

$$J=\mu q n \mathcal{E} \tag{119}$$

[38] This basically assumes excess charge starts occupation at $E_c$

Focusing at pinch off end, $\mathcal{E} \propto (V_{gs} - V_t)$. Thus $J \propto \frac{1}{2}(V_{gs} - V_t)(V_{gs} - V_t)$, which is consistent for current equation at saturation.

In a more microscopic view, we do the following elaboration.

First, starts from a piece of doped semiconductor, whose $E - k$ diagram at one position is considered. When $V_{ds}$ is non-zero, referring to Figure 26, an electric field $\mathcal{E}$ is applied, which tilts the $E - k$ diagram, resulting in asymmetry; thus there is net momentum, resulting in drift velocity, and hence current.

$E_f$ is then separated into quasi Fermi levels $E_f^+$ and $E_f^-$ given by equations (120) and (121). It is the electrons in these states that carry the current. These excess charge normally they are spread a few $kT$ around $E_f$ [30] ).

$$E_f^+ = \frac{\hbar^2 (k_f + k_d)^2}{2m^*} \tag{120}$$

$$E_f^- = \frac{\hbar^2 (k_f - k_d)^2}{2m^*} \tag{121}$$

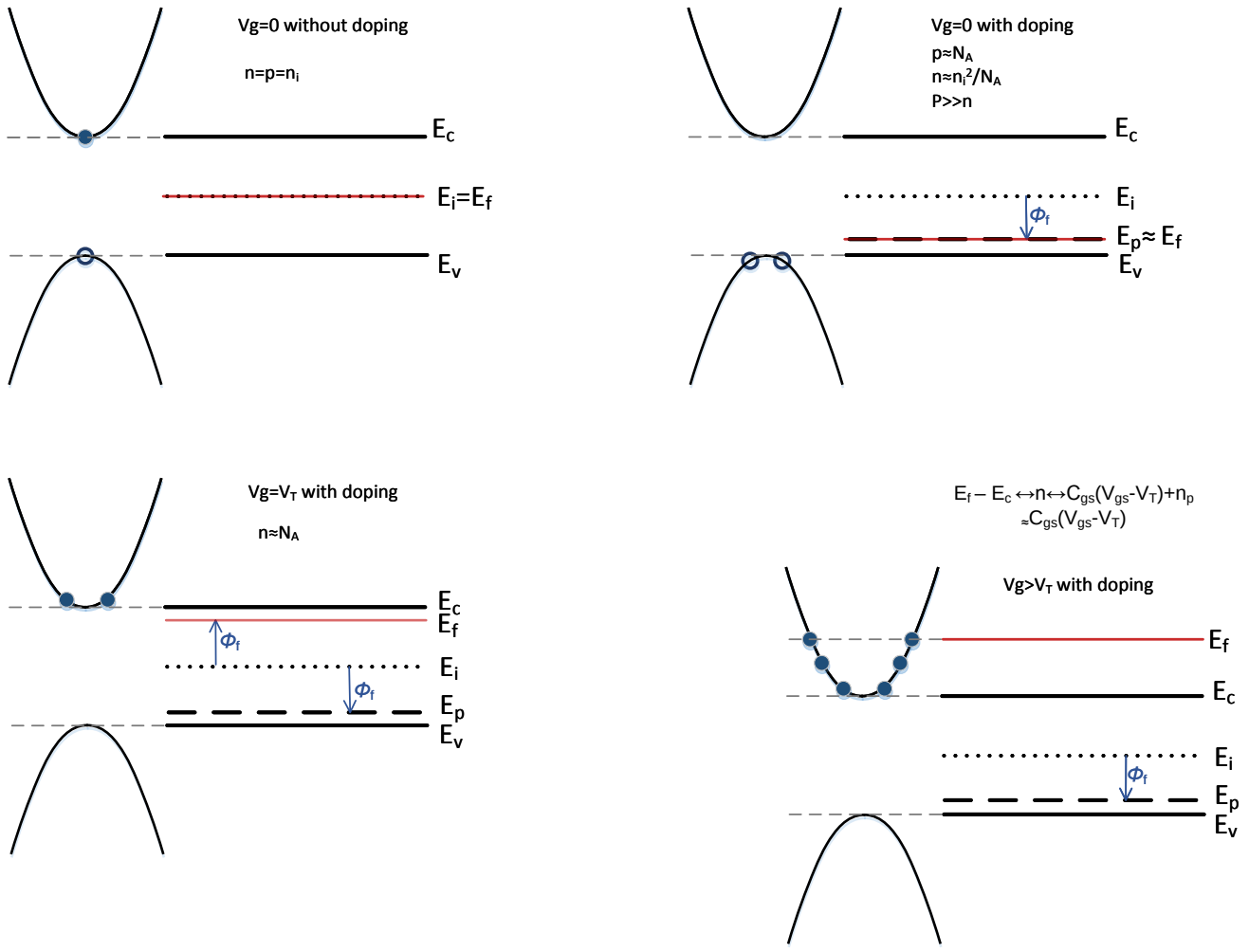

Figure 24. A BAND DIAGRAM SHOWING THE ORIGIN OF ENERGY LEVELS IN A SEMICONDUCTING SYSTEM.

Finally we incorporate this piece of semiconductor in our MOS system. The change is that, due to field effect, the number of carriers changes, and the distance between the Fermi level and conduction band changes to be proportional to $V_{gs} - V_t$, following Figure 25. Also $\mathcal{E}$ is now $V_{gs} - V_t$. Thus again from (119) $J \propto (V_{gs} - V_t)^2$, which is consistent for I-V equation at saturation (no channel length modulation).

Now visualize this capacitor as the capacitor $C_{gs}$ of a transistor, standalone, and the situation remains the same i.e. no interaction energy.

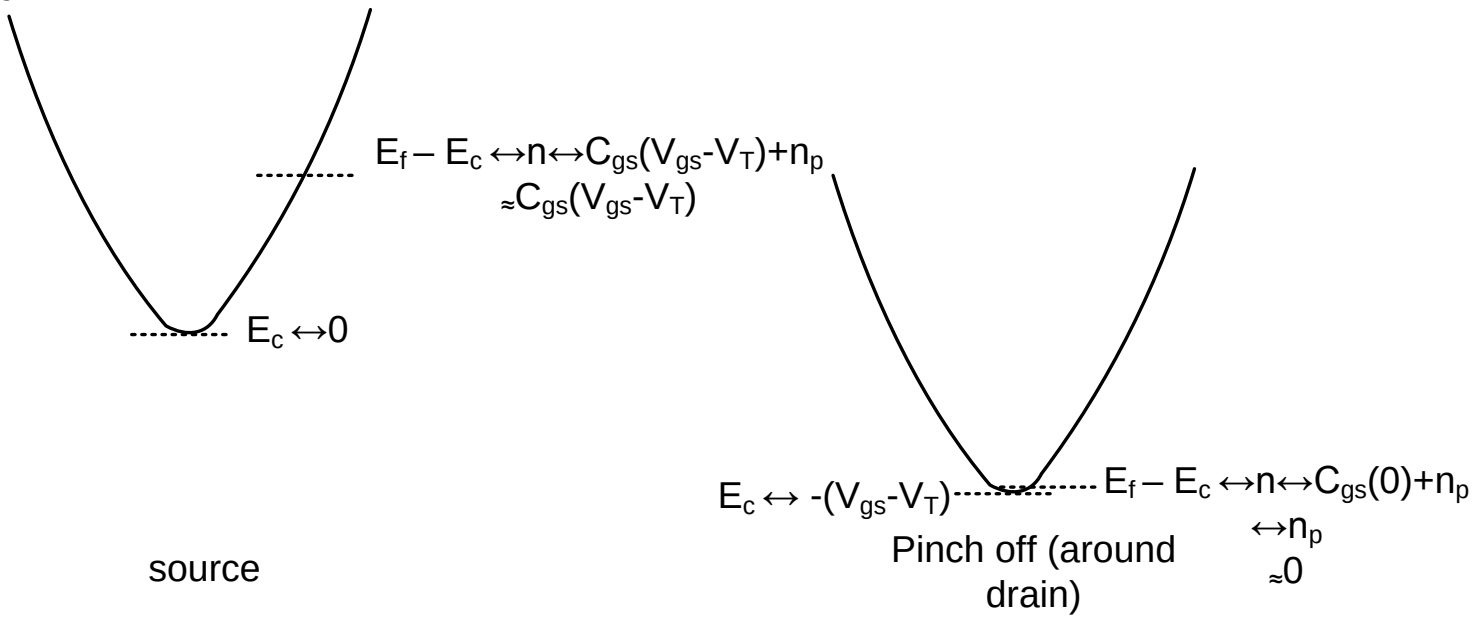

Figure 25. A BAND DIAGRAM SHOWING THE RELATION BETWEEN THE MICROSCOPIC VARIABLE $E_f - E_C$ AND MACROSCOPIC $V_{gs} - V_t$.

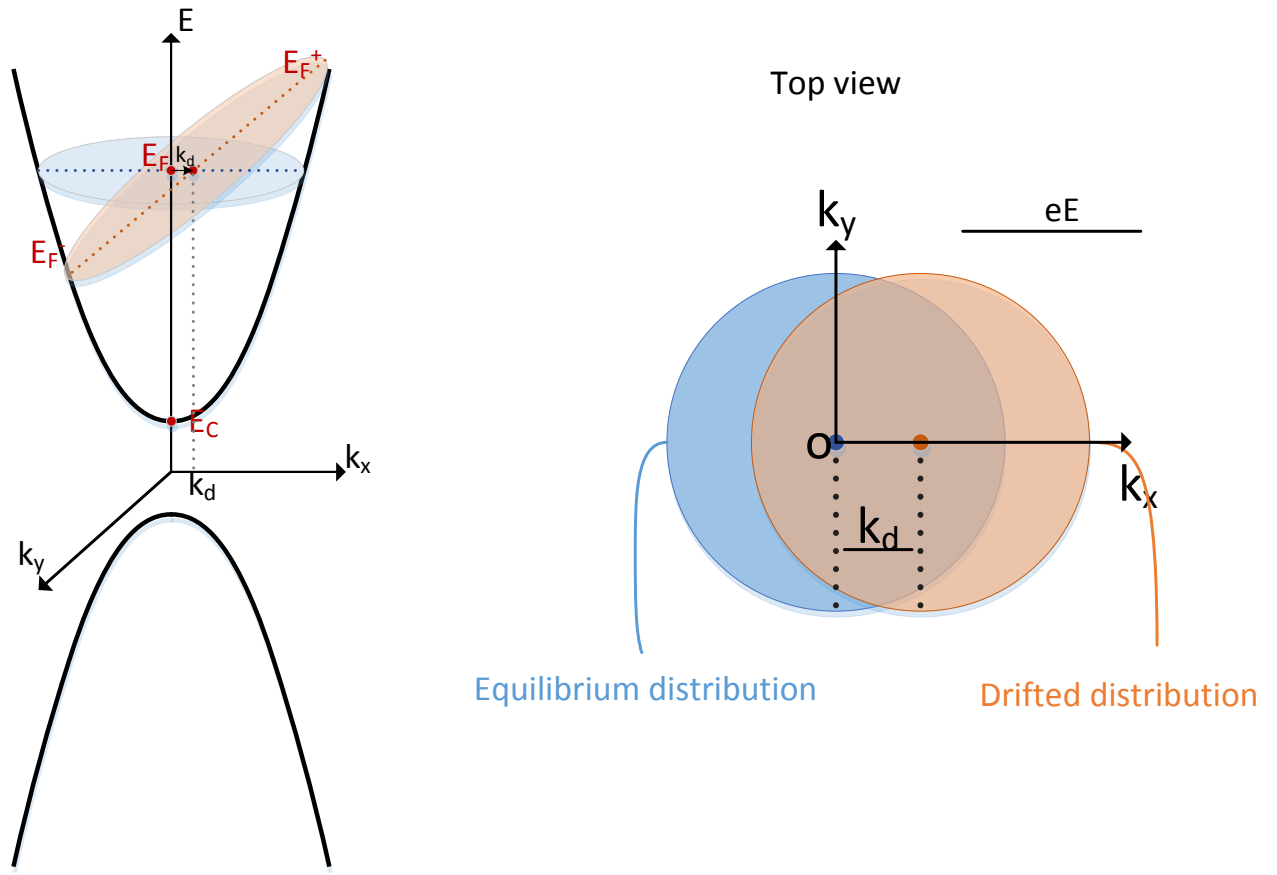

Figure 26. THE EFFECT OF APPLYING AN ELECTRIC FIELD IN THE $x$ DIRECTION ON THE CONDUCTION BAND ON THE WAVENUMBER DISTRIBUTION.

Specifically, the Fermi level on + and – side of momentum ($E_f^+$,$E_f^-$) is different, resulting in number of carriers having opposite momentum being different, resulting in a overall net momentum in the direction opposite to the field, as shown in (122).

$$E_f^+ - E_f^- \propto \mathcal{E} \tag{122}$$

This means the separation of quasi Fermi level is proportional to the energy that the electron gain from $\mathcal{E}$ (in mean free path). Next we apply this argument from one position to throughout, and if further take $E_f \sim \frac{(E_f^+ + E_f^-)}{2}$, then if the situation is as shown in Figure 27, $\mathcal{E}$ results from voltage applied, which in Figure 27 is set to $V_{ds} - V_t$. Thus the $E_C$ changes by the same amount.[39] For now the number of carriers remains unchanged, and so the distance between the Fermi level and conduction band remains the same, and so the Fermi level also drops by $V_{ds} - V_t$.

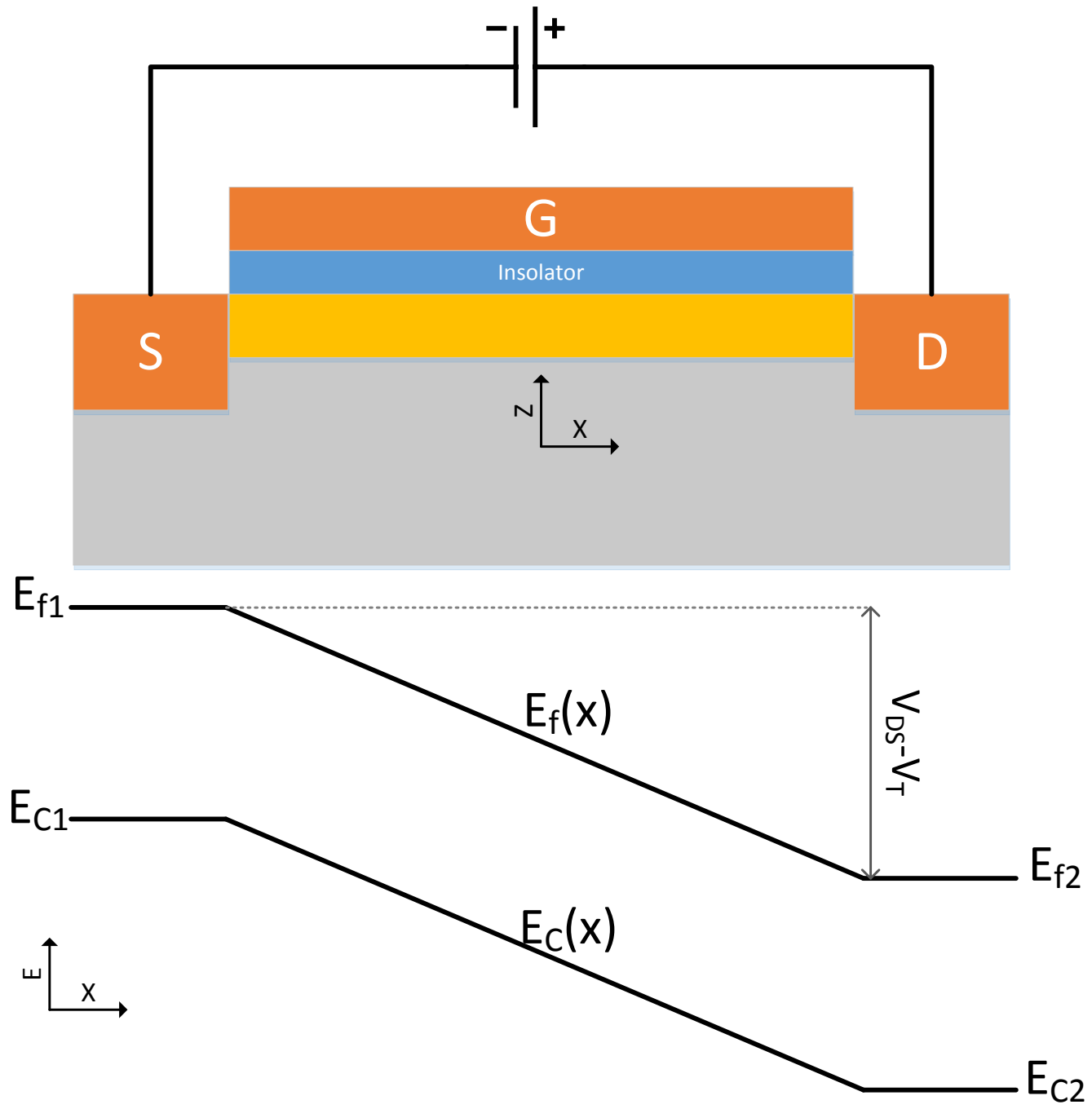

Figure 27. AN $E - x$ DIAGRAM SHOWING THE TREND OF $E_c$ AND $E_f$ ACROSS THE MOSFET.

[39] See fig.1.73 of [25]. Also this is following [25], which is more sophisticated than the Drude model (that relates $E_c$ change to electrostatic applied voltage, OK for simple conduction in free space,; but not OK for conduction in band, as electron needs vacancy to conduct, i.e. only level few $kT$ around $E_f$ participate in conduction; hence $E_f$, rather than $E_c$ alone, is involved).

## APPENDIX C FAST DYNAMIC AND SLOW DYNAMIC OF RELAXATION OSCILLATOR

The whole cycle of oscillation can be separated into a regeneration (fast) phase, and a charging/discharging (slow) phase. An attempt to connect the modelling the relaxation oscillator as a bistable circuit/charging-discharging timing capacitor $C$ combination, to modelling of the relaxation oscillator as a circuit having time scale separation (slow-fast dynamics), is given below.

The circuit in Figure 1 with $C_{gs1}$ and $C_{gs2}$ explicitly shown and the associated equation describing the circuit is already in [1]. The connection of the circuit dynamics with time scale separation model is shown below in Table 4, where the correspondence between two models is established. In Table 4, for ease of illustration, we use the ground capacitor topology for relaxation oscillator.

The variables in left hand column follow eq. 4-5 of [1] and in right hand column follow eq. 15-26 of [1], repeated in method 1. On the left hand column of Table 4 are the differential equations for the time scale separation model. On the right hand column are the differential equations for the bistable circuit/charging-discharging capacitor combination model. They have similar forms, which establish the connection.

TABLE 4 COMPARSION OF TWO TIME SCALE MODEL AND BISTABLE CIRCUIT/CHARGING-DISCHARGING CAPACITOR COMBINATION MODEL.

| Two time scale model | Bistable circuit/charging-discharging capacitor combination model |
|---|---|
| $\frac{dx}{dt} = y$ <br> $\varepsilon \frac{dy}{dx} = -x + y - y^2$ | $\frac{dv}{dt} = \frac{I_0}{C} z$ <br> $g_m \varepsilon \frac{dz}{dt} = 2(g_m R - 1)z - \frac{1}{4z^2} - v \frac{g_m}{I_0}$ |

It should be noted crudely speaking, in the bistable circuit/charging-discharging timing capacitor combination model, one can identify the two time scale (slow-fast) dynamics as follows: timing capacitor $C$ typically is much larger than $C_{gs1}$ and $C_{gs2}$. Thus, $\varepsilon$ associated with the parasitic capacitors $C_{gs1}$ and $C_{gs2}$ tend to determine the fast dynamics. The timing capacitor $C$ determines the slow dynamics.

The noise is dominated by regeneration/fast phase. As shown in [1] the phase noise is peaked up during the regeneration. Meanwhile phase noise during the charging/discharging is small, (given by $\frac{kT}{C_{gs}}$ ).

## APPENDIX D USE OF ELECTRON STATE KET $|\alpha\rangle$ VERSUS ENERGY EIGENSTATE $|\alpha'\rangle$ AND DISCUSSION OF SUPERPOSITION OF $|\alpha\rangle$

### 1. Superposition and Time evolution

a) In Method 3, we consider each electron is in a superposition state, while in Method 2, each electron is considered in an energy/base eigenstate (e.g. in Figure 4 to Figure 11). Method 2 is, therefore, just an approximation. We can handle that by simply using the 'strongest' energy (0.71 $i = 1$, first electron in $M_1$ or $q_1$; 0.71 $i = 2$, first electron in $M_2$ or $q_2$ peak in Figure 14a-c) to represent $|\alpha'\rangle$.

b) Also Pauli Exclusion Principle still works. Here quantum state, instead of an energy eigenenergy, is a state that corresponds to a superposition. Then the $i^{th}$ site (Ising) becomes a "superposition" site; again as one moves from site to site, one moves from superposition state to superposition state (e.g., $q_1$, $q_2$ state ket are different, with their expansion in energy eigenbasis given in Figure 14 having different $c_n$).

### 2. Taking expectation with respect to a superposition:

Concepts: state ket (represent the state, in superposition) $S_z$ or $|\alpha\rangle$; observable $B/S_x^{\pm}$/momentum $P$; Hamiltonian $H$/E (dictate allowable observed eigenstate/energy eigenstate/base ket $|\alpha'\rangle$. Thus $S_z$ or $|\alpha\rangle = \sum c|\alpha'\rangle$, i.e. state ket is expanded in $|\alpha'\rangle$) and so equation (123) holds.

$$\langle B\rangle = \sum\sum c_{\alpha'} c_{\alpha''} \langle\alpha'|B|\alpha''\rangle \exp\left(-i\frac{(E_{a''} - E_{a'})t}{\hbar}\right) \tag{123}$$

A property/observable of state (e.g. momentum of electron): taken by property operator with respect to state ket, but with state ket (e.g. electron at center of system at time=1) expanded in allowable observed eigenstate, as dictated by the underlying system (e.g. electron in a potential)

a) First, like [20], pg. 77, eq. 2.1.58, where observe $S_x$ (property) using $S_z$ (state we are in), here we are observing $B$ (property) using energy eigenstate, and $B$ will be shown to be momentum $P$, closer to our fluctuation of interest, current, $I$, via (124), where J is current density:

$$J|\psi\rangle = \frac{1}{2m^*} p|\psi\rangle \tag{124}$$

b) Next, we want to take against a superposition, rather than just eigenstates, like [20], pg.77, 2.1.59, again for property $S_x$ ($= \langle S_x\rangle$); Thus state ket is, $|S_x, +\rangle$, and it needs a superposition of $S_z$, $|+\rangle$ and $|-\rangle$ and not just $|+\rangle$ or $|-\rangle$. Again, because initial ket can be e.g. $S_x^+$ state, in much the same way, our initial ket ($|\alpha\rangle$) can be in a superposition of energy eigenstate. We cannot assume that it is in the stationary energy eigenstate. Specifically energy eigenstate does not undergo time evolution.

## APPENDIX E Describing Quantum-Mechanical Fluctuations using the Density Operator

As with any microscopic description/many degree of freedom (internal degree of freedom), we start with the ensemble concept (see IV.A). Remember classically to describe such ensemble, we use configuration/state and partition function to characterize it. In the generalization to quantum mechanics the configuration $\nu$ becomes the quantum state $\nu$, with the accompanying Boltzmann factor $\exp(-\beta E_\nu)$ and partition function $\sum \exp(-\beta E_\nu)$. The quantum mechanical description uses the density matrix, whose diagonal element is just the Boltzmann factor (quantum), normalized by the partition function (quantum).

In general ensemble is a mixed ensemble which has density operator given by (125).

$$\rho \equiv \sum_i w_i \, |\alpha^{(i)}\rangle\langle\alpha^{(i)}| \tag{125}$$

Making a measurement on a mixed ensemble with observable $B$, the ensemble average $\langle B\rangle$ is given by (126). The quantum-mechanical expectation value of $O$ taken with respect to the state $|\alpha^{(i)}\rangle$ is $\langle\alpha^{(i)}|B|\alpha^{(i)}\rangle$.

$$\langle B\rangle = \sum_i w_i \, \langle\alpha^{(i)}|B|\alpha^{(i)}\rangle \tag{126}$$

In (126), $|\alpha^{(i)}\rangle$ is the state ket of a pure ensemble $(i)$, and $w_i$ is the fraction of the pure ensemble i. meanwhile when they turned on, system has a new Hamiltonian (which classically is described by (25)), a new set of energy eigenstates (described by (28)). As time evolves, there is time evolution (or superposition can change $\exp\left(-\frac{iEt}{\hbar}\right)$, since $w_i$ change is small, $\exp\left(-\frac{iEt}{\hbar}\right)$ is quick, $\langle B\rangle$ changes is dictated by $\exp\left(-\frac{iEt}{\hbar}\right)$. After $t = t_{reg}$, $w_i$ starts to change, and the change in $\exp\left(-\frac{iEt}{\hbar}\right)$ starts to cancel out, so the change in $\langle B\rangle$ is dictated by $w_i$. There is, however, not a whole lot of change, since by then $w_i$ relaxes to (127).

$$\rho_{kk} = \frac{\exp(-\beta E_k)}{\sum_l^N \exp(-\beta E_l)} \tag{127}$$

Returning back to (125) the expectation values are weighed by the corresponding fractional populations $w_i$. The concept of probabilities manifests twice, with the first in $|\langle\alpha'|\alpha^{(i)}\rangle|^2$ as the quantum mechanical probability for state $|\alpha^{(i)}\rangle$ to be found in $B$ eigenstate $|\alpha'\rangle$; and the second in $w_i$ as the probability factor for finding in the ensemble of a quantum-mechanical state characterized by $|\alpha^{(i)}\rangle$. An element of the density matrix is given by (128), where $b'$ represent the dimensionality of the ket space.

$$\langle b''|\rho|b'\rangle = \sum_i w_i \, \langle b''|\alpha^{(i)}\rangle\langle\alpha^{(i)}|b'\rangle \tag{128}$$

The time evolution of ensembles is now discussed. Assume at time $t_0$ the density matrix is given by (125). If the ensemble is left undisturbed, we cannot change the fractional population $w_i$. So the change in $\rho$ is governed solely by the time evolution of state ket $\left|\alpha^{(i)}\right\rangle$. From the fact that this satisfies Schrodinger equation we obtain the von-Neumann equation as shown in (129).

$$i\hbar\frac{\partial\rho}{\partial t} = -[\rho, H] \tag{129}$$

The diagonal element of the density matrix $\rho_{kk}$ stands for the fractional population for an energy eigenstate with energy eigenvalue $E_k$. Once thermal equilibrium is established, $\frac{\partial\rho}{\partial t}$ in (129) is 0, and we have the diagonal element given in (127). It is recognized that the denominator is the partition function. The exponential of $\beta E_k$ represents a distribution of $\rho_{kk}$ at energy eigenvalue $E_k$ which is a function of $\beta = \frac{1}{kT}$.

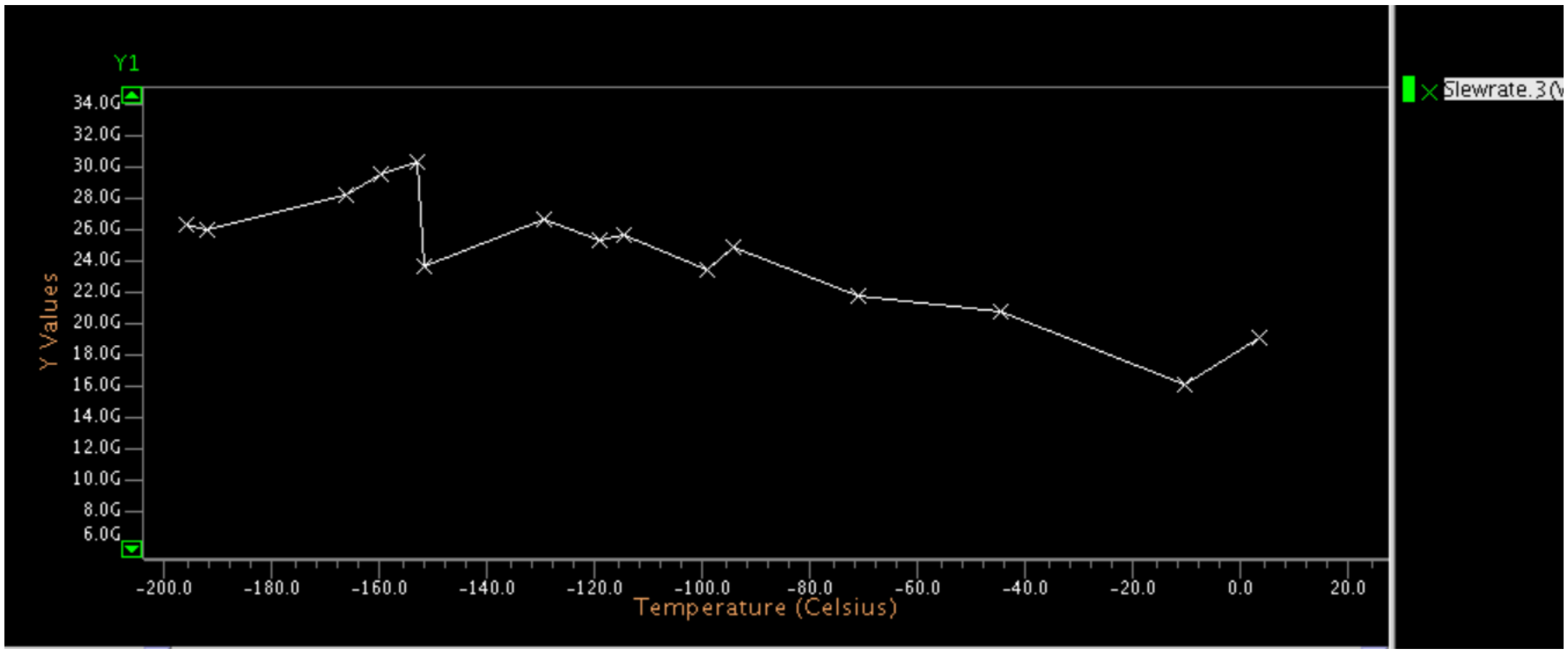

Figure 28. The Measured Slew Rate as a function of Temperature for $RO_1$.

## APPENDIX F The Slew Rate Simulation of Relaxation Oscillator with Temperature Drift

Figure 28 shows the change of slew rate in a function of temperature. The timing jitter is related to the noise amplitude such that with $t_j = \frac{amplitude\ noise}{slew\ rate}$ where $t_j$ is the timing jitter. The slew rate does not change much with temperature, thus the noise amplitude variation with temperature follows the same trend as jitter/phase noise variation. Specifically, the trend in the temperature dependency in the phase noise remains.

## References

[1] B. H. Leung, "Noise spike model in relaxation oscillators based on physical phase change," *IEEE Trans. Circuits Syst. I, Reg. Papers*, vol. 63, no. 6, pp. 871–882, 2016.

[2] Y. Cao *et al.*, "A 63,000 Q-factor relaxation oscillator with switched-capacitor integrated error feedback," in *2013 IEEE Int. Solid-State Circuits Conf.*, San Francisco, pp. 186–187.

[3] S. Hashemi and B. Razavi, "Analysis of Metastability in Pipelined ADCs," *IEEE J. Solid-State Circuits*, vol. 49, no. 6, pp. 1198–1209, May 2014.

[4] P. M. Figueiredo, "Comparator metastability in the presence of noise," *IEEE Trans. Circuits Syst. I, Reg. Papers*, vol. 60, no. 5, pp. 1286–1299, May 2013.

[5] P. Nuzzo *et al.*, "Noise Analysis of Regenerative Comparators for Reconfigurable ADC Architectures," *IEEE Trans. Circuits Syst. I, Reg. Papers*, vol. 55, no. 6, pp. 1441–1454, July 2008.

[6] S. K. Mathew *et al.*, "$\mu$ RNG: A 300-950 mV, 323 Gbps/W All-Digital Full-Entropy True Random Number Generator in 14 nm FinFET CMOS," *IEEE J. Solid-State Circuits*, vol. 51, no. 7, pp. 1695–1704, July 2016.

[7] Y. Cao, P. Leroux, W. De Cock, and M. Steayeart, "A 1.7mW 11b 1-1-1 MASH ΔΣ time-to-digital converter," in *2011 IEEE Int. Solid-State Circuits Conf.*, San Francisco, pp. 480-482.

[8] R. Smunyahirun and E. L. Tan, "Derivation of the Most Energy-Efficient Source Functions by Using Calculus of Variations," *IEEE Trans. Circuits Syst. I, Reg. Papers*, vol. 63, no. 4, pp. 494–502, April 2016.

[9] H. Homulle, S. Visser, and E. Charbon, "A Cryogenic 1 GSa/s, Soft-Core FPGA ADC for Quantum Computing Applications," *IEEE Trans. Circuits Syst. I, Reg. Papers*, vol. 63, no. 11, pp. 1854–1865, November 2016.

[10] A. Buonomo, "Analysis of emitter (source)-coupled multivibrators," *IEEE Trans. Circuits Syst. I, Reg. Papers*, vol. 53, no. 6, pp. 1193 – 1202. June 2006.

[11] A. A. Abidi and R. G. Meyer, “Noise in relaxation oscillators,” *IEEE J. Solid-State Circuits*, vol. 18, no. 6, pp. 794–802, December 1983.
[12] A. Hajimiri and T. H. Lee, “A general theory of phase noise in electrical oscillators,” *IEEE J. Solid-State Circuits*, vol. 33, no. 2, pp. 179–194, February 1998.
[13] B. Razavi, “A study of phase noise in CMOS oscillators,” *IEEE J. Solid-State Circuits*, vol. 31, no. 3, pp. 331–343, March 1996.
[14] B. Patra *et al.*, “Cryo-CMOS Circuits and Systems for Quantum Computing Applications,” *IEEE J. Solid-State Circuits*, vol. 53, no. 1, pp. 309-321, January 2018.
[15] S. S. Sastry, “The effects of small noise on implicitly defined nonlinear dynamical systems,” IEEE Transactions on Circuits & Systems, vol. 30, no. 9, pp. 651–663, 1982.
[16] R. Shankar, *Principles of Quantum Mechanics*, 2nd ed., New York, NY, USA: Kluwer Academic / Plenum Publishers, 1994.
[17] T. Liu et.al, “A temperature compensated triple-path PLL with $K_{VCO}$ Non-linearity Desensitization Capable of Operating at 77 K”, *IEEE Trans. Circuits Syst. I, Reg. Papers*, vol. 64, no. 11, pp. 2835-2843, November 2017.
[18] M. T. Rahman, and T. Lehmann, “A Self-Calibrated Cryogenic Current Cell for 4.2 K Current Steering D/A Converters,” *IEEE Trans. Circuits Syst. II, Exp. Briefs*, vol. 64, no. 10, pp. 1152-1156, October 2017.
[19] R. K. Pathria, *Statistical Mechanics*, 1st ed. Oxford, UK: Pergamon Press Ltd., 1972.
[20] D. Chandler, *Introduction to Modern Statistical Mechanics*, New York, NY, US: Oxford University Press, 1987.
[21] S. H. Strogatz, *Nonlinear Dynamics and Chaos*, Reading, MA, US: Addison-Wesley Publishing Company, 1994.
[22] C. E. Smith, “Lagrangians and Hamiltonians with friction,” *J. Phys.: Conf. Ser.*, vol. 237, no. 1, 2010.
[23] B. H. Leung, “Novel dissipative Lagrange-Hamilton formalism for LC/van der pol oscillator with new implication on phase noise dependency on quality factor,” in *2014 IEEE 57th Intl. Midwest Symp. on Circuits and Systems*, College Station, TX, US, pp. 507-510.
[24] V. V. Schmidt, *The Physics of Superconductors*, Berlin, DE: Springer-Verlag Berlin Heidelberg, 1997.
[25] S. Datta, *Electronic Transport in Mesoscopic Systems*, Cambridge, UK: Cambridge University Press, 1995.
[26] D. Griffiths, *Introduction to Electrodynamics*, 2nd ed. Englewood Cliffs, NJ, US: Prentice Hall Inc., 1989.
[27] R. Smunyahirun and E. L. Tan, “Derivation of the Most Energy-Efficient Source Functions by Using Calculus of Variations,” *IEEE Trans. Circuits Syst. I, Reg. Papers,* vol. 63, no. 4, pp. 494-502, April 2016.
[28] R. Wickham, “Notes on Statistical Mechanics Physics I, Physics 704, lec20, U. of Waterloo”, unpublished.
[29] Y. C. Chen, M. P. A. Fisher, and A. J. Leggett, “The return of a hysteretic Josephson junction to the zero-voltage state: I-V characteristic and quantum retrapping,” *J. Appl. Phys.*, vol. 64, no. 6, pp. 3119–3142, 1988.
[30] J. J. Sakurai and S. F. Tuan, *Modern Quantum Mechanics*, 2nd ed., Reading, MA, USA: Addison-Wesley Publishing Company, 1994.
[31] B. H. Leung, “Noise/jump phenomenon of relaxation oscillators based on phase change using path integral/lagrangian formulation in quantum mechanics,” in *2012 IEEE 55th Intl. Midwest Symp. on Circuits and Systems*, Boise, ID, US, pp. 246-249.
[32] P. Allen, *CMOS Analog Circuit Design*, 2nd ed, Oxford U. Press